\newif\ifsingle
\newif\ifFullVersion
\FullVersiontrue % comment out for full proofs version

\ifsingle
\documentclass[11pt,draftclsnofoot, onecolumn]{IEEEtran}		
\else		
\documentclass[10pt,final, twocolumn]{IEEEtran}
\fi

\usepackage{times}
\usepackage{amsmath,dsfont}
\usepackage{amssymb,amsthm}
\usepackage{epsfig,verbatim}
\usepackage{subcaption}
\usepackage{setspace}
\usepackage{color}
\usepackage{cite}
\usepackage{epstopdf}
\usepackage{graphics}
\usepackage{accents}
\usepackage{acronym}
\usepackage[bookmarks,colorlinks]{hyperref}
\usepackage{booktabs}
\usepackage{mathtools}
\usepackage[ruled,linesnumbered]{algorithm2e}  %,algochapter
\SetKwInput{KwData}{\textbf{Init}} 
\usepackage{enumitem}

\definecolor{NewColor}{rgb}{0.2,0,0.5}

\newcommand{\myVec}[1]{{\boldsymbol{#1}}}
\newcommand{\myMat}[1]{{\boldsymbol{#1}}}
\newcommand{\mySet}[1]{\mathcal{#1}}
\newcommand{\E}{\mathds{E}}		 			% Stochastic expectation
\newcommand{\myI}{{\myMat{I}}}			 		% Identity matrix
\newcommand{\Input}{\myVec{x}}
\newcommand{\InputSpace}{\mySet{X}}

\newcommand{\Label}{\myVec{s}}
\newcommand{\LabelSpace}{\mySet{S}}

\newcommand{\myS}{\Label}			 		% Unknown object 
\newcommand{\Nusers}{K}
\newcommand{\NusersSet}{\mySet{K}}
\newcommand{\Nantennas}{N}
\newcommand{\Niter}{Q}

\newcommand{\Ntraining}{n_t}
\newcommand{\Pdf}[1]{p_{ { #1}} }
\newcommand{\PdfNew}[1]{p}

\newcommand{\SigW}{\sigma_w^2}

\newcommand{\CovMat}[1]{\myMat{\Sigma}_{#1}}			% covariance matrix
\newcommand{\Mem}{l}			 			% observations length
\newcommand{\Blklen}{t}			 			% observations length
\newcommand{\myState}{{\myS}} 
\newcommand{\myStateR}{{\myVec{s}}}
\newcommand{\FwdMsg}[2]{\overrightarrow{\mu}_{#2}}
\newcommand{\BwdMsg}[2]{\overleftarrow{\mu}_{#2}}%{\mu_{#1 \rightarrow #2}}
\newcommand{\Blkset}{\mySet{T}}

\newcommand{\dnnParam}{\myVec{\theta}}
\newcommand{\dnnFunc}{f_{\dnnParam}}
\newcommand{\dnnLatent}{\myVec{z}}

\newcommand{\dnnEnc}{f_{\rm enc}}
\newcommand{\dnnDec}{f_{\rm dec}}
\newcommand{\csSignal}{\myVec{s}}
\newcommand{\csObs}{\Input}
\newcommand{\csLatent}{\myVec{z}}
\newcommand{\csMatrix}{\myMat{H}}
\newcommand{\csNoise}{\myVec{w}}

\newcommand{\csLoss}{\mySet{L}_{\text{CS}}}
\newcommand{\csLasso}{\mySet{L}_{\rm LASSO}}
\acrodef{cnn}[CNN]{convolutional neural network} 
\acrodef{relu}[ReLU]{rectified linear unit}

\acrodef{adc}[ADC]{analog-to-digital convertor}
\acrodef{cs}[CS]{compressed sensing}
\acrodef{bp}[BP]{belief propagation}
\acrodef{bpsk}[BPSK]{binary phase shift keying}
\acrodef{dtft}[DTFT]{discrete-time Fourier transform}
\acrodef{dnn}[DNN]{deep neural network} 
\acrodef{gan}[GAN]{generative adversarial network} 
\acrodef{gnn}[GNN]{graph neural network} 
\acrodef{gru}[GRU]{gated recurrent unit} 
\acrodef{csi}[CSI]{channel state information}
\acrodef{map}[MAP]{maximum a-posteriori probability}
\acrodef{snr}[SNR]{signal-to-noise ratio}
\acrodef{bs}[BS]{base station} 
\acrodef{em}[EM]{expectation maximization} 
\acrodef{iot}[IOT]{Internet of Things}
\acrodef{mimo}[MIMO]{multiple-input multiple-output}
\acrodef{mse}[MSE]{mean-squared error}
\acrodef{pdf}[PDF]{probability density function}
\acrodef{rv}[RV]{random variable}
\acrodef{fec}[FEC]{forward error correction} 
\acrodef{lti}[LTI]{linear time-invariant}
\acrodef{wss}[WSS]{wide-sense stationary}
\acrodef{psd}[PSD]{power spectral density}
\acrodef{ser}[SER]{symbol error rate} 
\acrodef{ber}[BER]{bit error rate} 
\acrodef{sgd}[SGD]{stochastic gradient descent}  
\acrodef{awgn}[AWGN]{additive white Gaussian noise} 
\acrodef{ut}[UT]{user terminal}  
\acrodef{ml}[ML]{machine learning}  
\acrodef{rnn}[RNN]{recurrent neural network} 
\acrodef{fc}[FC]{fully-connected}
\acrodef{sic}[SIC]{soft interference cancellation}
\acrodef{pmf}[PMF]{probability mass function}
\acrodef{sp}[SP]{sum-product} 
\acrodef{ista}[ISTA]{iterative soft thresholding algorithm}
\acrodef{pca}[PCA]{principal component analysis}
\acrodef{admm}[ADMM]{alternating direction method of multipliers}
\acrodef{mri}[MRI]{magnetic resonance imaging}

\definecolor{NewColor}{rgb}{0,0,0}%{0.54, 0.17, 0.89} 

\ifsingle

\else

\fi % ------------------------------------------

\IEEEoverridecommandlockouts

\title{Model-Based Deep Learning
}
\author{  
	\IEEEauthorblockN{Nir Shlezinger, Jay Whang, Yonina C. Eldar, and Alexandros G. Dimakis\\
	} 
	\thanks{
		N. Shlezinger is with the School of ECE, Ben-Gurion University of the Negev, Be'er-Sheva, Israel (e-mail: \mbox{nirshl@bgu.ac.il}).  	
		J. Whang is with the Department of CS,  University of Texas at Austin, Austin, TX (e-mail:  \mbox{jaywhang@cs.utexas.edu}). 	
		Y. C. Eldar is with the Faculty of Math and CS, Weizmann Institute of Science, Rehovot, Israel (e-mail: \mbox{yonina@weizmann.ac.il}).
		A. G. Dimakis is with the Department of ECE, University of Texas at Austin, Austin, TX (e-mail:   \mbox{dimakis@austin.utexas.edu}).
	} 			
	
}
\vspace{-0.75cm}

\begin{document}
	
	\maketitle
	\pagestyle{plain}
	\thispagestyle{plain}
	%----------------------------------------------------------------------------------------
	%	ABSTRACT
	%----------------------------------------------------------------------------------------
	\begin{abstract}
Signal processing, communications, and control have traditionally relied on classical statistical modeling techniques. Such model-based methods utilize mathematical formulations that represent the underlying physics, prior information and additional domain knowledge.
Simple classical models are useful but sensitive to inaccuracies and may lead to poor performance when real systems display complex or dynamic behavior. 
On the other hand, purely data-driven approaches that are model-agnostic are becoming increasingly popular as data sets become abundant and the power of modern deep learning pipelines increases.  Deep neural networks (DNNs) use generic architectures which learn to operate from data, and demonstrate excellent performance, especially for supervised problems.  However, DNNs typically require massive amounts of data and immense computational resources, limiting their applicability for some scenarios.

In this article we present the leading approaches for studying and designing model-based deep learning systems. These are methods  that combine principled mathematical models with data-driven systems to benefit from the advantages of both approaches. Such model-based deep learning methods exploit both partial domain knowledge, via mathematical structures designed for specific problems, as well as learning from limited data.  
Among the applications detailed in our examples for model-based deep learning are compressed sensing, digital communications, and tracking in state-space models. Our aim is to facilitate the design and study of future systems on the intersection of signal processing and machine learning that incorporate the advantages of both domains.
	\end{abstract}
	
	%----------------------------------------------------------------------------------------
	%	Introduction
	%----------------------------------------------------------------------------------------
	%\vspace{-0.4cm}
	\section{Introduction}
	%\vspace{-0.1cm} 
    Traditional signal processing is dominated by algorithms that are based on simple mathematical models which are hand-designed from domain knowledge. Such knowledge can come from statistical models based on measurements and understanding of the underlying physics, or from fixed deterministic representation of the particular problem at hand. These domain-knowledge-based processing algorithms, which we refer to henceforth as {\em model-based methods},  carry out inference based on knowledge of the underlying  model relating the observations at hand and the desired information. Model-based methods do not rely on data to learn their mapping, though data is often used to estimate a small number of parameters. Fundamental techniques like the Kalman filter and message passing algorithms belong to the class of model-based methods. %In some cases these schemes may rely on data to estimate some unknown model parameter, while the overall flow of the algorithm, such as some statistical moment of the measurements, while the inference processing. 
    Classical statistical models rely on simplifying assumptions (e.g., linear systems, Gaussian and independent noise, etc.) that make inference tractable, understandable and computationally efficient. On the other hand, simple models frequently fail to represent nuances of high-dimensional complex data and dynamic variations.

	% Paragraph - deep learning and its challenges

The incredible success of deep learning, e.g., on vision \cite{lecun2015deep},\cite{he2015delving} as well as challenging games such as Go \cite{silver2017mastering} and Starcraft \cite{vinyals2019alphastar}, has initiated a general data-driven mindset. It is currently prevalent to replace simple principled models with purely data-driven pipelines, trained with massive labeled data sets.  In particular, \acp{dnn} can be trained in a supervised way end-to-end to map inputs to predictions.
The benefits of data-driven methods over  model-based approaches are twofold: 
First, purely-data-driven techniques do not rely on analytical
approximations and thus can operate in scenarios where
analytical models are not known.
Second, for complex systems, data-driven algorithms are able to recover features from observed data which are needed to carry out inference \cite{Bengio09learning}. This is sometimes difficult to achieve analytically, even when complex models are perfectly known. 
%Finally, the main complexity in utilizing \ac{ml} methods is in the training stage, which is typically carried out offline. Once trained, they tend to implement inference at a  lower  delay compared to their analytical model-based counterparts \cite{gregor2010learning}. 

%	\begin{figure}
%	\centering
%	\includefig{fig/ML_triangle.png} 
%	\caption{Key machine learning ingredients.} 
%	\label{fig:ML_triangle}
% \end{figure}

%I think its clear that ML needs data and compute- i dont think we need a figure for that. 

The  computational burden of training and utilizing  highly-parametrized \acp{dnn}, as well as the fact that massive data sets are typically required to train such \acp{dnn} to learn a desirable mapping, may constitute major drawbacks in various signal processing, communications, and control applications. This is particularly relevant for hardware-limited  devices, such as mobile phones, unmanned aerial vehicles, and \ac{iot} systems, which are often limited in their ability to utilize highly-parametrized \acp{dnn} \cite{chen2019deep}, and require adapting to dynamic conditions.   Furthermore, \acp{dnn} are commonly utilized as black-boxes; {understanding how their predictions are obtained and characterizing confidence intervals tends to be quite challenging. As a result, deep learning does} not yet offer the interpretability, flexibility, versatility, and reliability of model-based methods~\cite{monga2019algorithm}.

% \begin{figure}
% 	\centering
% 	\includefig{fig/SymDetComp3.png} 
% 	\caption{Model-based methods versus deep learning for symbol detection: $a)$ a receiver uses its knowledge of the statistical channel, denoted $\Pdf{\Input|\myS}$ to detect the transmitted symbols in a model-based manner; $b)$ a receiver utilizes a \ac{dnn} trained using the data set $\{\myVec{s}_t, \Input_t\}$ for recovering the symbols.} 
% 	\label{fig:SymDetComp1}
% \end{figure}

% Paragraph - the fusion of deep learning and model-based algorithms. Emphasize that the prolifiration of techniques on the interface of signal processing and deep learning motivates the formulation of a unified framework for such techniques. 
The limitations associated with model-based methods and black-box deep learning systems gave rise to a multitude of techniques for combining signal processing and machine learning to benefit from both approaches. These methods are application-driven, and are thus designed and studied in light of a specific task. For example, the combination of \acp{dnn} and model-based \ac{cs} recovery algorithms was shown to facilitate sparse recovery \cite{gregor2010learning, wu2018learning} as well as enable \ac{cs} beyond the domain of sparse signals \cite{bora2017compressed, whang2020compressed}; Deep learning was used to empower regularized optimization methods \cite{gilton2019neumann,venkatakrishnan2013plug}, while model-based optimization contributed to the design of \acp{dnn} for such tasks \cite{aggarwal2018modl}; Digital communication receivers used \acp{dnn} to learn to carry out and enhance symbol detection and decoding algorithms in a data-driven manner \cite{shlezinger2019viterbinet, shlezinger2019deepSIC, nachmani2018deep}, while symbol recovery methods enabled the design of model-aware deep receivers \cite{samuel2019learning, he2018model,khani2020adaptive, pratik2020re}. The proliferation of  hybrid model-based/data-driven systems, each designed for a unique task, motivates establishing a concrete systematic framework for combining domain knowledge in the form of model-based methods and deep learning, which is the focus of this article. 
 	
\begin{figure}
 	    \centering
 	    \includegraphics[width=\columnwidth]{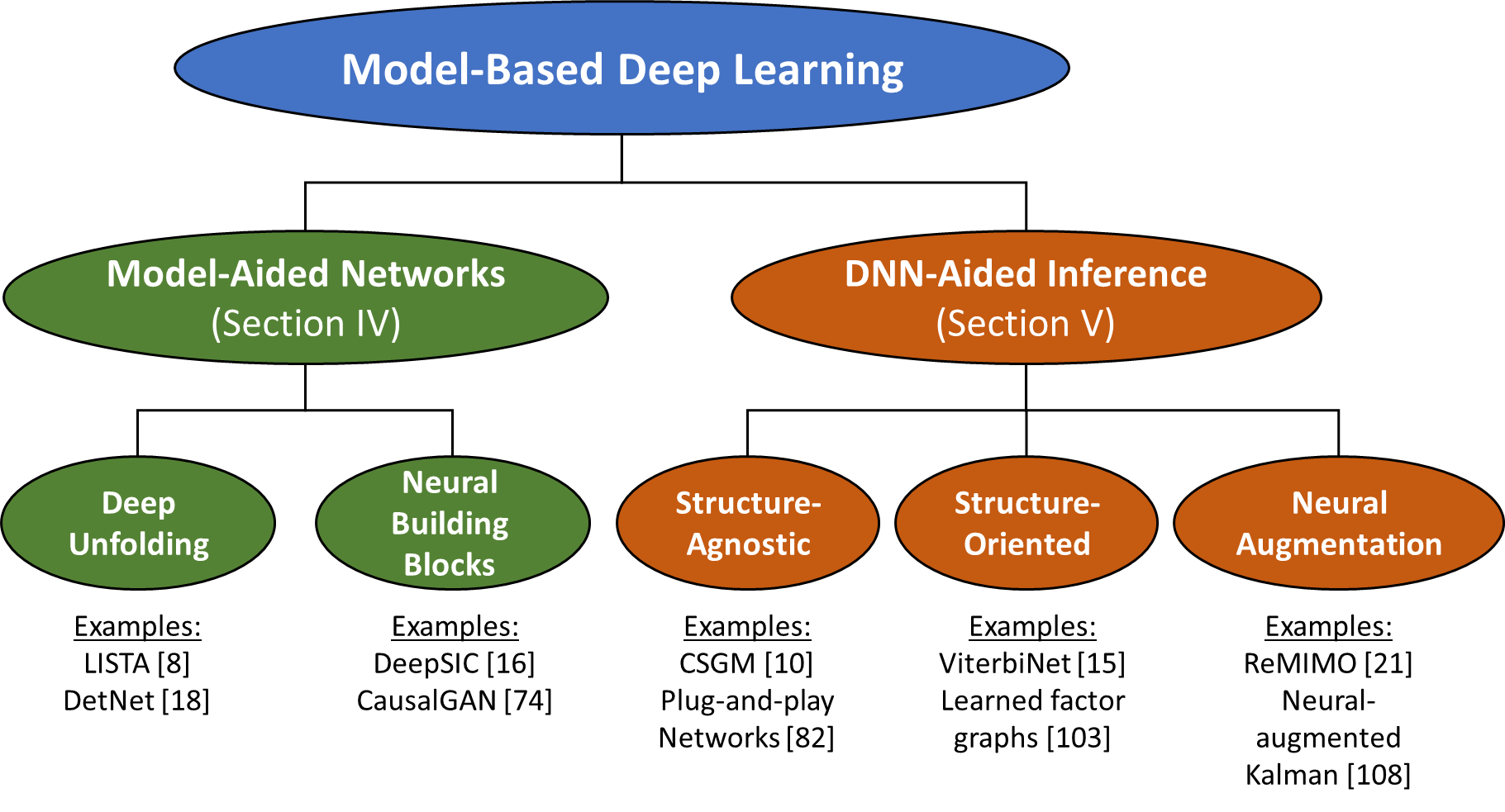}
 	    \caption{
 	    Division of model-based deep learning techniques into categories and sub-categories. }
 	    \label{fig:Division1}
 	\end{figure}

 	In this article we review leading strategies for designing systems whose operation combines domain knowledge and data via model-based deep learning in a tutorial fashion. To that aim, we present a unified framework for studying   hybrid model-based/data-driven systems, without focusing on a specific application, while being geared towards families of problems typically studied in the signal processing literature. The proposed framework divides  systems combining model-based signal processing and deep learning into two main strategies: The first category includes \acp{dnn} whose architecture is specialized to the specific problem using model-based methods, referred to here as {\em model-aided networks}. The second one, which we refer to as {\em \ac{dnn}-aided inference}, consists of techniques in which  inference is carried out by a model-based algorithm whose operation is augmented with deep learning tools. This integration of model-agnostic deep learning tools allows one to use model-based inference algorithms while having access only to partial domain knowledge. Based on this division, we provide concrete guidelines for studying, designing, and comparing model-based deep learning systems. {An illustration of the proposed division   into categories and sub-categories  is depicted in Fig.~\ref{fig:Division1}.}
 	
 	We begin by discussing the high level concepts of model-based, data-driven, and hybrid  schemes. Since we focus on \acp{dnn} as the current leading data-driven technique, we briefly review basic concepts in deep learning,  ensuring that the tutorial is accessible to readers without background in deep learning. We then elaborate on the fundamental strategies for combining model-based methods with deep learning. 
 	\label{txt:unfold1}
 	For each such strategy, we present a few concrete implementation approaches in a systematic manner, including established approaches such as deep unfolding, which was originally proposed in 2010 by Gregor and LeCun \cite{gregor2010learning}, as well as recently proposed model-based deep learning paradigms such as \ac{dnn}-aided inference \cite{farsad2020data} and neural augmentation \cite{satorras2020neural}. For each approach we formulate system design guidelines  for a given problem; provide detailed examples from the recent literature; and discuss its properties and use-cases. Each of our detailed examples focuses on a different application in signal processing, communications, and control, demonstrating the breadth and the wide variety of applications that can benefit from such hybrid designs. We conclude the article with a summary and a qualitative comparison of model-based deep learning approaches, along with a description of some  future research topics and  challenges. We aim to encourage future researchers and practitioners with a signal processing background to study and design model-based deep learning. % from both a theoretical and an application-oriented perspectives. 

 	This overview article focuses on strategies for designing architectures whose operation combines deep learning with model-based methods, as illustrated in Fig.~\ref{fig:Division1}. These strategies can also be integrated into existing mechanisms for incorporating model-based domain knowledge in the selection of the task for which data-driven systems are applied, as well as in the generation and manipulation of the data. An example of a family of such mechanisms for using model-based knowledge in the selection of the application and the data is the learning-to-optimize framework, which is the focus of growing attention in the context of wireless networks design \cite{zappone2019model,zappone2019wireless,liang2019deep}; this framework advocates the usage of pre-trained \acp{dnn} for realizing fast solvers for complex optimization problems which rely on objectives and constraints formulated based on domain knowledge, along with the usage of model-based generated data for offline training. An additional related family is that of channel autoencoders, which  integrate mathematical modelling of random communication channels as layers of deep autoencoders to design channel codes \cite{oshea2017introduction, kim2018communication} and compression mechanisms \cite{mashhadi2020distributed}.
	
	The rest of this article is organized as follows:
	Section~\ref{sec:MBvsDL} discusses the concepts of model-based methods as compared to data-driven schemes, and how they give rise to the model-based deep learning paradigm. 
	Section~\ref{sec:DL} reviews some basics of deep learning. % that are used in our presentation of model-based deep learning strategies in the sequel. 
	The main strategies for designing model-based deep learning systems, i.e., model-aided networks and  \ac{dnn}-aided inference, are detailed in Sections~\ref{sec:Networks}-\ref{sec:Inference}, respectively. Finally, we provide a summary and discuss some future research challenges in Section~\ref{sec:Conclusions}.

% 	Throughout the paper, we use boldface lower-case letters for vectors, e.g., ${\myVec{x}}$;
% 	%	the $i$th element of ${\myVec{x}}$ is written as $({\myVec{x}})_i$. 
% 	Matrices are denoted with boldface upper-case letters,  e.g., 
% 	$\myMat{M}$;   %$(\myMat{M})_{i,j}$   is its $(i,j)$th element. 
% 	calligraphic letters, such as $\mySet{X}$, are used for sets.
% 	% 
% 	%
% 	The $\ell_2$ norm and vectorization operator are denoted by $\| \cdot \|$ and ${\rm vec}(\cdot)$, respectively.
% 	Transpose,  Hermitian transpose, trace, and stochastic expectation are written as  $(\cdot)^T$,  $(\cdot)^H$, ${\rm {Tr}}\left(\cdot\right)$, and $\E\{ \cdot \}$,  respectively. 
% 	Finally, $\mySet{R}$ and $\mySet{Z}$ are the sets of real numbers and integers, respectively.
% 	%$\delta_{k,l}$ is the Kronecker delta, i.e., $\delta_{k,l}\! =\! 1$ when $k\!=\!l$ and $\delta_{k,l}\! =\! 0$ otherwise. 
	
	%----------------------------------------------------------------------------------------
	%	Model-Based versus Data-Driven 
	%----------------------------------------------------------------------------------------
%	\vspace{-0.2cm}
	\section{Model-Based versus Data-Driven Inference}
	\label{sec:MBvsDL}
%	\vspace{-0.1cm}	
    We begin by reviewing the main conceptual differences between model-based and data-driven inference. To that aim, we first present a mathematical formulation of a generic inference problem. Then we discuss how this problem is tackled from a purely model-based perspective as well as from a purely data-driven one, where for the latter we focus on deep learning as a family of generic data-driven approaches. We then formulate the  notion of model-based deep learning based upon these distinct strategies. 
    
      \begin{figure*}
	\centering
	\includegraphics[width=\linewidth]{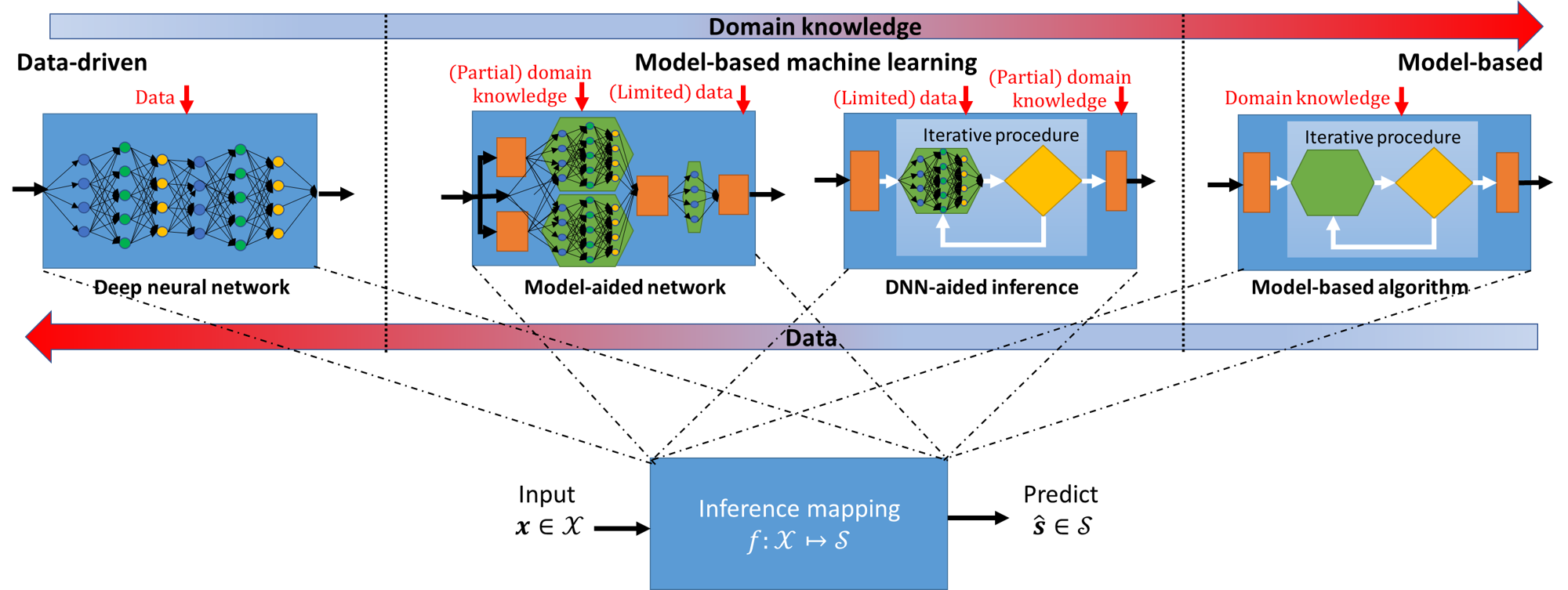} 
	\caption{Illustration of model-based versus data-driven inference. The red arrows correspond to computation performed before the particular inference data is received.  
	} 
	\label{fig:MBvsDL1}
\end{figure*}  

     %----------------------------------------------------------------------------------------
    %	{Inference System
    %----------------------------------------------------------------------------------------      
    \subsection{Inference Systems}
    \label{subsec:InferenceSystem}
    The term {\em inference} refers to the ability to conclude based on evidence and reasoning.
    While this generic definition can refer to a broad range of tasks, we focus in our description on systems which estimate or make predictions based on a set of observed variables.  In this wide family of problems, the system is required to   map an input variable $\Input \in \InputSpace$  into a prediction of a label variable $\Label \in \LabelSpace$, denoted $\hat{\myS}$, where $\InputSpace$ and $\LabelSpace$ are referred to as the {\em input space} and the {\em label space}, respectively.  
    An inference rule can thus be expressed as
    \begin{equation}
        \label{eqn:mapping}
        f:\InputSpace \mapsto  \LabelSpace
    \end{equation}
   and the space of inference mappings is denoted by $\mySet{F}$. We use  $l(\cdot)$ to denote a cost measure defined over  $\mySet{F}\times\InputSpace\times\LabelSpace$, dictated by the specific task \cite[Ch. 2]{shalev2014understanding}. The fidelity of an inference mapping is measured by the {\em risk function}, also known as the generalization error, given by  $\E_{\Input,\Label\sim p_{\Input,\Label}}\{l(f,\Input,\Label)\}$, where $p_{\Input,\Label}$ is the underlying statistical model relating the input and the label.  The goal of both model-based methods and data-driven schemes is to design the inference rule $f(\cdot)$ to minimize the risk for a given problem. The main difference between these strategies is what information is utilized to tune $f(\cdot)$. 

     %----------------------------------------------------------------------------------------
    %	Model-Based Methods
    %----------------------------------------------------------------------------------------      
    \subsection{Model-Based Methods}
    \label{subsec:ModelMethods}
    Model-based algorithms, also referred to as {\em hand-designed methods} \cite{metzler2017learned}, set their inference rule, i.e., tune $f$ in \eqref{eqn:mapping} to minimize the risk function, based on domain knowledge.  The term {\em domain knowledge} typically refers to  prior knowledge of the underlying statistics relating the input $\Input$ and the label $\Label$. %, such as  the generative conditional distribution $\Pdf{\Input | \Label}$. Alternatively, domain knowledge can also accommodate deterministic modelling, as commonly used in \ac{cs} applications \cite{eldar2012compressed,eldar2015sampling}, where, e.g., $\Input$ represents linear projections of a sparse $\Label$. 
    %
    %Regardless of the nature of the domain knowledge at hand, the main characteristic of model-based methods is that their mapping is dictated by this  knowledge. 
    In particular, an analytical mathematical expression describing the underlying model, i.e., $p_{\Input,\Label}$, %e.g., the  distribution of $\Input$ conditioned on $\Label$ denoted $\Pdf{\Input | \Label}$, 
    is required. Model-based algorithms can provably implement the risk minimizing inference mapping, e.g., the \ac{map} rule. While computing the risk minimizing rule is often computationally prohibitive, various model-based methods approximate this rule at controllable complexity, and in some cases also provably approach its performance. This is typically achieved  using iterative methods comprised of multiple stages, where each stage involves generic mathematical manipulations and model-specific computations.

    Model-based methods do not rely on data to learn their mapping, as illustrated in the right part of Fig.~\ref{fig:MBvsDL1}, though data is often used to estimate unknown model parameters. In practice, accurate knowledge of the statistical model relating the observations and the desired information is typically unavailable, and thus applying such techniques commonly requires imposing some assumptions on the underlying statistics, which  in some cases  reflects the actual behavior, but may also constitute a crude approximation of the true dynamics. In the presence of inaccurate model knowledge,   either as a result of estimation errors or due to enforcing a model which does not fully capture the environment, the performance of model-based techniques tends to degrade. This limits the applicability of model-based schemes in scenarios where, e.g.,  $\Pdf{\Input , \Label}$ is unknown, costly to estimate accurately, or too complex to express analytically.
    
     %----------------------------------------------------------------------------------------
    %	Data-Driven Schemes
    %----------------------------------------------------------------------------------------      
    \subsection{Data-Driven Schemes}
    \label{subsec:DataSchemes}
    Data-driven systems learn their mapping from data. In a supervised setting, data is comprised of a training set consisting of $\Ntraining$ pairs of inputs and their corresponding labels, denoted $\{(\Input_t, \Label_t)\}_{t=1}^{\Ntraining}$. 
    Data-driven schemes do not have access to the underlying distribution, and thus cannot compute the risk function. As a result, the inference mapping is typically tuned based on an empirical risk function, referred henceforth as {\em loss function}, which for an inference mapping $f$ is given by
    \begin{equation} 
    \label{eqn:EmpRisk}
    \mySet{L}(f)=\frac{1}{\Ntraining}\sum_{t=1}^{\Ntraining}l(f,\Input_t,\Label_t).
    \end{equation} 
    
    Since one can usually form an inference rule which minimizes the empirical loss \eqref{eqn:EmpRisk} by memorizing the data, i.e., overfit, data-driven schemes often restrict the domain of feasible inference rules \cite[Ch. 2]{shalev2014understanding}.
    A leading strategy in data-driven systems, upon which deep learning is based, is to assume some highly-expressive generic parametric model on the mapping in \eqref{eqn:mapping}, while incorporating optimization mechanisms to avoid overfitting and allow the resulting system to infer reliably with new data samples. In such cases, the inference rule is dictated by a set of parameters denoted $\myVec{\theta}$, and thus the system mapping is written as $f_{\myVec{\theta}}$.

    The conventional application of deep learning implements $f_{\myVec{\theta}}$ using a \ac{dnn} architecture, where $\myVec{\theta}$ represent the weights of the network.  Such highly-parametrized networks can effectively approximate any Borel measurable mapping, as follows from the universal approximation theorem \cite[Ch. 6.4.1]{goodfellow2016deep}. Therefore, by properly tuning their parameters using sufficient training data, as we elaborate in Section~\ref{sec:DL}, one should be able to obtain the desirable inference rule. 

 Unlike model-based algorithms, which are specifically tailored to a given scenario, purely-data-driven methods are model-agnostic, as illustrated in the left part of Fig.~\ref{fig:MBvsDL1}. The unique characteristics of the specific scenario are encapsulated in 
 the learned weights. The parametrized inference rule, e.g., the \ac{dnn} mapping, is generic and can be applied to a broad range of different problems.  
While standard \ac{dnn} structures are highly model-agnostic and are commonly treated as black boxes, one can still incorporate some level of domain knowledge in the selection of the specific network architecture. For instance, when the input is known to exhibit temporal correlation, architectures based on \acp{rnn} \cite{hochreiter1997long} or attention mechanisms \cite{vaswani2017attention} are often preferred. Alternatively, in the presence of spatial patterns, one may   utilize convolutional layers \cite{lecun1995convolutional}. An additional method to incorporate domain knowledge into a black box \ac{dnn} is by pre-processing of the input via, e.g., feature extraction. 

The generic nature of data-driven strategies induces some drawbacks. Broadly speaking,  learning a large number of parameters requires a massive data set to train on. Even when a sufficiently large data set is available, the resulting training procedure is typically lengthy and involves high computational burden. Finally, the black-box nature of the resulting mapping implies that data-driven systems in general lack interpretability, making it difficult to provide performance guarantees and insights into the system operation. 
    
     %----------------------------------------------------------------------------------------
    %	Model-Based Deep Learning
    %----------------------------------------------------------------------------------------   
    \subsection{Model-Based Deep Learning} 
    \label{subsec:MBDL}
    
    Completely separating existing literature into model-based versus data-driven is a subjective and debatable task. Instead, we focus on some approaches which clearly lie in the middle ground to give a useful overview of the landscape. The considered families of methods incorporate domain knowledge in the form of an established model-based algorithm which is suitable for the problem at hand, while combining capabilities to learn from data via deep learning techniques.
    
    Model-based deep learning schemes tune their mapping of the input $\Input$ based on both data, e.g., a labeled training set  $\{(\Input_t, \Label_t)\}_{t=1}^{\Ntraining}$, as well as some domain knowledge, such as partial knowledge of the underlying distribution $\Pdf{\Input , \Label}$. Such hybrid data-driven model-aware systems can typically learn their mappings from smaller training sets compared to purely model-agnostic \acp{dnn}, and commonly operate without full accurate knowledge of the underlying model upon which model-based methods are based. 
    
    \textcolor{NewColor}{Most existing techniques for implementing inference rules in a hybrid model-based/data-driven fashion are designed for a specific application, i.e., to solve a given problem rather than formulate a systematic methodology. Nonetheless, one can identify a common rationale for categorizing existing schemes in a systematic manner that is not tailored to a specific scenario. In particular, model-based deep learning techniques}
     can be divided into two main strategies, as illustrated in Fig.~\ref{fig:MBvsDL1}. These strategies may each be further specialized to various different tasks, as we show in the sequel. The first of the two, which we refer to as  {\em model-aided networks}, utilizes \acp{dnn} for inference; however, rather than using conventional \ac{dnn} architectures, here a specific \ac{dnn} tailored for the problem at hand is designed by following the operation of suitable  model-based methods. 
	The second strategy, which we call {\em \ac{dnn}-aided inference systems},  uses conventional  model-based methods for inference; however, unlike purely model-based schemes, here specific parts of the model-based algorithm are augmented with deep learning tools, allowing the resulting system to implement the algorithm while learning to overcome partial or mismatched domain knowledge from data.  
	
	\textcolor{NewColor}{The systematic categorization of model-based deep learning methodologies can facilitate the study and design of future techniques in different and diverse application areas. One may also propose schemes which combine aspects from both categories, building upon the understanding of the characteristics and gains of each approach, discussed in the sequel.} 
	Since both strategies rely on deep learning tools, we first provide a brief overview of key concepts in  deep learning in the following section, after which we elaborate on model-aided networks and \ac{dnn}-aided inference in Sections~\ref{sec:Networks} and \ref{sec:Inference}, respectively.

	%----------------------------------------------------------------------------------------
	%	Basics of Deep Learning 
	%----------------------------------------------------------------------------------------
%	\vspace{-0.2cm}
	\section{Basics of Deep Learning}
	\label{sec:DL}
%	\vspace{-0.1cm}
	Here, we cover the basics of deep learning required to understand the \ac{dnn}-based components in the model-based/data-driven approaches discussed later.  Our aim is to equip the reader with necessary foundations upon which our formulations of model-based deep learning systems are presented. %, and also to clarify pros and cons of black-box \ac{dnn} techniques.
	
	As discussed in Subsection~\ref{subsec:DataSchemes}, in deep learning, the target mapping is constrained to take the form of a parametrized function $f_{\myVec{\theta}}: \InputSpace \to \LabelSpace$. In particular, the inference mapping belongs to a fixed family of functions $\mySet{F}$  specified by a predefined \ac{dnn} architecture, which is represented by a specific choice of the parameter vector $\myVec{\theta}$.   Once the function class $\mySet{F}$ and loss function $\mySet{L}$ are defined, where the latter is dictated by the training data  \eqref{eqn:EmpRisk} while possibly including some regularization on $\myVec{\theta}$, one may attempt to find the function which minimizes the loss within $\mySet{F}$, i.e., 
	\begin{equation}
	\label{eqn:Loss1}
	 {\myVec{\theta}}^{\ast} = \mathop{\arg\min}\limits_{f_{\myVec{\theta}}\in \mySet{F}} \mySet{L}(f_{\myVec{\theta}}).
	\end{equation}
	A common challenge in optimizing based on \eqref{eqn:Loss1} is to guarantee that the inference mapping learned using the data-based loss function rather than the model-based risk function will not overfit and be able to generalize, i.e., infer reliably from new data samples.
	Since the optimization in \eqref{eqn:Loss1} is carried out over $\myVec{\theta}$, we write the loss as $\mySet{L}(\myVec{\theta})$ for brevity.

	The above formulation naturally gives rise to three fundamental components of deep learning: the \ac{dnn} \textit{architecture} that defines the function class $\mySet{F}$; the task-specific \textit{loss} function $\mySet{L}(\myVec{\theta})$; and the \textit{optimizer} that dictates how to search for the optimal $f_{\myVec{\theta}}$ within $\mySet{F}$. Therefore, our review of the basics of deep learning commences with a description of the fundamental architecture and optimizer components in Subsection~\ref{subsec:DL_Pre}. We then present several representative tasks along with their corresponding typical loss functions in Subsection~\ref{subsec:DL_Tasks}. %The formulation of these common tasks is exploited in our presentation of the incorporation of deep learning into hybrid model-based/data-driven schemes  in Sections~\ref{sec:Networks}-\ref{sec:Inference}.
	%We will cover each of the three components in more details below.
  
    %----------------------------------------------------------------------------------------
    %	Deep Learning Preliminaries
    %----------------------------------------------------------------------------------------
 %   \vspace{-0.2cm}
    \subsection{Deep Learning Preliminaries}
	\label{subsec:DL_Pre}
%	\vspace{-0.1cm}	
	The formulation of the parametric empirical risk in \eqref{eqn:Loss1} is not unique to deep leaning, and is in fact common to numerous machine learning schemes. The strength of deep learning, i.e., its ability to learn accurate complex mappings from large data sets, is due to its use of \acp{dnn} to enable a highly-expressive family of function classes $\mySet{F}$, along with dedicated optimization algorithms for tuning the parameters from data. In the following we discuss the high level notion of \acp{dnn}, followed by a description of how they are optimized.
	
	\paragraph{Neural Network Architecture}
	\acp{dnn} implement parametric functions comprised of  a sequence of differentiable transformations called \textit{layers}, whose composition maps the input to a desired output. Specifically, a \ac{dnn} $\dnnFunc$ consisting of $k$ layers $\{h_1, \ldots, h_k\}$ maps the input $\Input$ to the output $\hat{\Label} = \dnnFunc (\Input) = h_k \circ \cdots \circ h_1 (\Input)$, where $\circ$ denotes function composition. % Here $\Input$ and $\Label$ are assumed to be vectors. 
	Since each layer $h_i$ is itself a parametric function, the parameters set of the entire network $\dnnFunc$ is the union of all of its layers' parameters, and thus $\dnnFunc$ denotes a \ac{dnn} with parameters $\myVec{\theta}$.  The \textit{architecture} of a \ac{dnn} refers to the specification of its layers $\{h_i\}_{i=1}^k$.
	
%    \jw{Which layers and activation functions should we cover here? And with how much detail? I'd imagine things like pooling layer are not very relevant.}
	%As layers are one of the fundamental building blocks of \acp{dnn}, 	much of deep learning research has been devoted to coming up with layers that perform well in various scenarios. 
	A generic formulation which captures various parametrized layers is that of an affine transformation, i.e., $h(\myVec{x}) = \myMat{W}\myVec{x} + \myVec{b}$ whose parameters are $\{\myMat{W}, \myVec{b}\}$. For instance, in \textit{\ac{fc}} layers, also referred to as \textit{dense} layers, one can optimize $\{\myMat{W}, \myVec{b}\}$ to take any value. Another extremely common affine transform layer is  \textit{convolutional} layers. Such layers apply a set of discrete convolutional kernels to signals that are possibly comprised of multiple channels, e.g., tensors. The vector representation of their output can be written as an affine mapping of the form $\myMat{W}\myVec{x} + \myVec{b}$, where $\myVec{x}$ is the vectorization of the input, and  $\myMat{W}$ is constrained to represent multiple channels of discrete convolutions \cite[Ch. 9]{goodfellow2016deep}. These \acp{cnn} are known to yield a highly parameter-efficient mapping that captures important invariances such as translational invariance in image data.
	 
    % TODO NIR CONTINUE PROOFREADING FROM HERE
    While many commonly used layers %including \ac{fc} and convolutional layers 
    are affine, \acp{dnn} rely on the inclusion of \textit{non-linear} layers.  If all the layers of a \ac{dnn} were affine, the composition of all such layers would also be affine, and thus the resulting network would only represent affine functions.  For this reason, layers in a \ac{dnn} are interleaved with \textit{activation functions}, which are simple non-linear functions applied to each dimension of the input separately. Activations are often fixed, i.e., their mapping is not parametric and is thus not optimized in the learning process.  Some notable examples of widely-used activation functions include the \ac{relu} defined as ${\rm ReLU}(x) = \max\{x,0\}$ and the sigmoid $\sigma(x) = (1+\exp(-x))^{-1}$.
    
 %   \acp{dnn} allow the function space $\mathcal{F}$ to capture a broad range of functions. In fact, by making the network sufficiently large, one can approximate any Borel measurable mapping, i.e., $\mathcal{F}$ is the space of all Borel measurable functions from $\InputSpace$ to $\LabelSpace$, as follows from the universal approximation theorem \cite[Ch. 6.4.1]{goodfellow2016deep}. The expressiveness of \acp{dnn} combined with the fact that the parameters $\myVec{\theta}$ can be learned from data allows trained \acp{dnn} to operate reliably in a model-agnostic manner. %In fact, highly-parametrized \acp{dnn} have demonstrated the ability to disentangle  semantic information in complex environments \cite{Bengio09learning}.     Optimizing the large amount of parameters of a \ac{dnn} based on massive volumes of data is feasible, though computationally expensive, due to its sequential structure, as detailed next.  
    
    \paragraph{Choice of Optimizer}
    Given a \ac{dnn} architecture and a loss function $\mySet{L}(\myVec{\theta})$, finding a globally optimal $\myVec{\theta}$ that minimizes $\mySet{L}$ is a hopelessly intractable task, especially at the scale of millions of parameters or more.  Fortunately, recent success of deep learning has demonstrated that gradient-based optimization methods work surprisingly well despite their inability to find global optima. The simplest such method is \textit{gradient descent}, which iteratively updates the parameters:
    \begin{equation}
        \myVec{\theta}_{q+1} = \myVec{\theta}_{q} - \eta_{q} \nabla_{\myVec{\theta}} \mySet{L}(\myVec{\theta}_{q})
    \end{equation}
    where $\eta_{q}$ is the \textit{step size} that may change as a function of the step count $q$. Since the gradient $\nabla_{\myVec{\theta}} \mySet{L}(\myVec{\theta}_{q})$ is often too costly to compute over the entire training data, it is estimated from a small number of randomly chosen samples (i.e., a mini-batch). The resulting optimization method is called \textit{mini-batch stochastic gradient descent} and belongs to the family of stochastic first-order optimizers.
    
    Stochastic first-order optimization techniques are well-suited for training \acp{dnn} because their memory usage grows only linearly with the number of parameters, and they avoid the need to process the entire training data at each step of optimization. Over the years, numerous variations of stochastic gradient descent have been proposed. Many modern optimizers such as RMSProp \cite{tieleman2012lecture} and Adam \cite{kingma2014adam} use statistics from previous parameter updates to adaptively adjust the step size for each parameter separately (i.e., for each dimension of $\myVec{\theta}$). %In general, the interplay between a \ac{dnn}, an optimizer, and the downstream performance is not yet well-understood and is an active area of research.

    \subsection{Common Deep Learning Tasks}
	\label{subsec:DL_Tasks}
    % \vspace{-0.1cm}
    As detailed above, the data-driven nature of deep learning is encapsulated in the dependence of the loss function on the training data. %based on which the optimizer tunes the parametric architecture. The loss function thus reflects both the task of the system, as well as the nature of the data available. 
    Thus, the loss function not only implicitly defines the task of the resulting system, but also dictates what kind of data is required.
	Based on the requirements placed on the training data, problems in deep learning largely fall under three different categories: supervised, semi-supervised, and unsupervised.  Here, we define each category and list some example tasks as well as their typical loss functions.
	
	\paragraph{Supervised Learning}
In supervised learning, the training data consists of a set of input-label pairs $\{(\Input_t, \Label_t)\}_{t=1}^{\Ntraining}$, where each pair takes values in $\InputSpace\times\LabelSpace$. As discussed in Subsection~\ref{subsec:DataSchemes}, the goal is to recover a mapping  $\dnnFunc$ which minimizes the risk function, i.e., the generalization error. This is done by optimizing the \ac{dnn} mapping    $\dnnFunc$ using the data-based empirical loss function   $\mySet{L}(\dnnParam)$ \eqref{eqn:EmpRisk}. This setting encompasses a wide range of problems including regression, classification, and structured prediction, through a judicious choice of the loss function.  Below we review commonly used loss functions for classification and regression tasks.
	
%	\begin{enumerate}
	
    	% --- Classification ---
	    $1)$ \textit{Classification}: Perhaps one of the most widely-known success stories of \acp{dnn},  classification (image classification in particular) has remained a core benchmark since the introduction of AlexNet \cite{krizhevsky2017imagenet}. In this setting, we are given a training set $\{(\Input_t, \Label_t)\}_{t=1}^{\Ntraining}$ containing input-label pairs, where each $\Input_t$ is a fixed-size input, e.g., an image, and $\Label_t$ is the one-hot encoding of the class. Such one-hot encoding of class $c$ can be viewed as a probability vector for a $K$-way categorical distribution, with $K = |\LabelSpace|$, with all probability mass placed on class $c$.
	    
    	The \ac{dnn} mapping $\dnnFunc$ for this task  is appropriately designed to map an input  $\Input_t$ to the probability vector $\hat{\Label}_t \triangleq f(\Input_t) = \langle \hat{s}_{t,1}, ..., \hat{s}_{t,K} \rangle$, where $\hat{s}_{t,c}$ denotes the $c$-th component of $\hat{\Label}_t$.  This parametrization allows for the model to return a soft decision in the form of a categorical distribution over the classes.
	
    	A natural choice of loss function for this setting is the \textit{cross-entropy} loss, defined as
    	\begin{equation}
    	\label{eqn:CEloss}
    	    \mySet{L}_{\rm CE}(\dnnParam) = \frac{1}{\Ntraining} \sum_{t=1}^{\Ntraining} \sum_{c=1}^K s_{t,c} (-\log \hat{s}_{t,c}).
    	\end{equation}
    	For a sufficiently large set of i.i.d. training pairs, the empirical cross entropy loss approaches the expected cross entropy measure, which is minimized when the \ac{dnn} output matches the true conditional distribution $\Pdf{\myS|\Input}$. Consequently, minimizing the cross-entropy loss encourages the \ac{dnn} output to match the ground truth label, and its mapping closely approaches the true underlying posterior distribution when properly trained.

    	%One technical but crucial trick in classification setting is the \textit{softmax} function. 
    	The formulation of the cross entropy loss \eqref{eqn:CEloss} implicitly assumes that the \ac{dnn} returns a valid probability vector, i.e., $\hat{s}_{t,c} \ge 0$ and $\sum_{c=1}^{K} \hat{s}_{t,c} = 1$.  However, there is no guarantee that this will be the case, especially at the beginning of  training when the parameters of the \ac{dnn} are more or less randomly initialized.  To guarantee that the \ac{dnn} mapping yields a valid probability distribution, classifiers typically employ the softmax function (e.g., on top of the output layer), given by:
    	\begin{equation*}
    	    \textrm{Softmax}(\myVec{x}) = \left\langle\frac{\exp(x_1)}{\sum_{i=1}^{d} \exp(x_i)}, \ldots, \frac{\exp(x_d)}{\sum_{i=1}^{d} \exp(x_i)}\right\rangle
    	\end{equation*}
    	where $x_i$ is the $i$th entry of $\myVec{x}$.
    	Due to the exponentiation followed by normalization, the output of the softmax function is guaranteed to be a valid probability vector. In practice, one can compute the softmax function of the network outputs when evaluating the loss function, rather than using a dedicated output layer. %In effect, the softmax trick removes the burden of outputting a valid probability distribution from the network and improves the stability and  speed of the training process.
    	
    	\smallskip
    	% --- Regression ---
        $2)$ \textit{Regression}: Another task where \acp{dnn} have been successfully applied is regression, where one attempts to predict  continuous variables instead of categorical ones.  Here, the labels $\{\Label_t\}$ in the training data represent some continuous value, e.g., in $\mathbb{R}$ or some specified range $[a,b]$. 
        % One simple \ac{dnn} architecture for regression is a stack of \ac{fc} layers with activation functions in between as described in Section \ref{subsec:DL_Pre}:
        % \begin{equation}
        %     \hat{\Label}_t \triangleq \dnnFunc(\Input_t) = \sigma \circ h_k \circ \phi \circ \cdots \circ \phi \circ h_1(\Input_t)
        % \end{equation}
        % where each $h_i$ is an affine layer and $\phi$ is some activation function. %Notably, the last layer $h_k$ must output a single scalar rather than a vector.
        
        Similar to the usage of softmax layers for classification, an appropriate final activation function $\sigma$ is needed, depending on the range of the variable of interest. For example, when regressing on a strictly positive value, a common choice is $\sigma(x) = \exp(x)$ or the softplus activation $\sigma(x)=\log(1+\exp(x))$, so that the range of the network $\dnnFunc$ is constrained to be the positive reals. When the output is to be limited to an interval $[a,b]$, then one may use the mapping $\sigma(x) = a + (b-a)(1+\tanh(x))/2$.
        
        Arguably the most common loss function for regression tasks is the empirical \ac{mse}, i.e.,
        \begin{equation}
    	    \mySet{L}_{\rm MSE}(\dnnParam) = \frac{1}{\Ntraining} \sum_{t=1}^{\Ntraining} (\Label_t - \hat{\Label}_t)^2.
        \end{equation}
%	\end{enumerate}
	
	\paragraph{Unsupervised Learning}
	In unsupervised learning, we are only given a set of examples $\{\Input_t\}_{t=1}^{\Ntraining}$  without labels.  Since there is no label to predict, unsupervised learning algorithms are often used to discover interesting patterns present in the given data.  Common tasks in this setting include clustering, anomaly detection, generative modeling, and compression. %A popular type of \ac{dnn}-based generative model is a \textit{\ac{gan}} \cite{goodfellow2014generative}, which has recently shown remarkable success in numerous domains.
	
	%\begin{enumerate}
    % --- GAN --- 
    $1)$ \textit{Generative models}: One goal in unsupervised learning of a generative model is to train a \textit{generator} network $G_{\myVec{\theta}}(\dnnLatent)$ such that the \textit{latent variables} $\dnnLatent$, which follow a simple distribution such as standard Gaussian, are mapped into samples obeying  a distribution similar to that of the training data \cite[Ch. 20]{goodfellow2016deep}. For instance, one can train a generative model to map Gaussian vectors into images of human faces.
    A popular type of \ac{dnn}-based generative model that tries to achieve this goal is \textit{\ac{gan}} \cite{goodfellow2014generative}, which has  shown remarkable success in many domains.
    
   % Since the training data does not provide the ground truth latent variable $\dnnLatent_t$ associated with each example $\Input_t$, 
   \color{NewColor}
    \acp{gan} learn the generative model by employing a  \textit{discriminator} network $D_\myVec{\varphi}$ to assess the generated samples, thus avoiding the need to mathematically handcraft a loss measure quantifying their quality. The parameters $\{\myVec{\theta}, \myVec{\varphi}\}$ of the two networks are learned via \textit{adversarial} training, where $\myVec{\theta}$ and $\myVec{\varphi}$ are updated in an alternating manner. The two networks $G_{\myVec{\theta}}$ and $D_\myVec{\varphi}$ ``compete'' against each other to achieve opposite goals: $G_{\myVec{\theta}}$ tries to fool the discriminator, whereas $D_\myVec{\varphi}$ tries to reliably distinguish real examples from the fake ones made by the generator. 
    
    Various methods have been proposed to train generative models in this adversarial fashion, including, e.g., the Wasserstein \ac{gan} \cite{arjovsky2017wasserstein, gulrajani2017improved}; the least-squares \ac{gan} \cite{mao2017least}; the Hinge \ac{gan} \cite{lim2017geometric}; and the relativistic average \ac{gan} \cite{jolicoeur2018relativistic}. For simplicity, in the following we describe the original \ac{gan} formulation of \cite{goodfellow2014generative}. Here, 
    $D_\myVec{\varphi}: \InputSpace \to [0,1]$ is a binary classifier trained to distinguish real examples $\Input_t$ from the fake examples generated by $G_{\myVec{\theta}}$, and the \ac{gan} loss function is the minmax loss. 
   \color{black}
    The loss is optimized in an alternating fashion by tunning the discriminator $\myVec{\varphi}$ to minimize $\mySet{L}_{\rm D}(\cdot)$ for a given generator $\myVec{\theta}$, followed by a corresponding optimization of the generator based on its loss $\mySet{L}_{\rm G}(\cdot)$. These loss measures are   given by
    \begin{align*}
        \mySet{L}_{\rm D}(\myVec{\varphi} | \myVec{\theta}) &= \frac{-1}{2\Ntraining}\sum_{t=1}^{\Ntraining}\log D_\myVec{\varphi}(\Input_t)  
        + \log \left(1 \!-\!  D_\myVec{\varphi}\big(G_\myVec{\theta}(\dnnLatent_t) \big) \right), \\
        \mySet{L}_{\rm G}(\myVec{\theta} | \myVec{\varphi}) &= \frac{-1}{\Ntraining}\sum_{t=1}^{\Ntraining}\log  \log   D_\myVec{\varphi}\big(G_\myVec{\theta}(\dnnLatent_t) \big).
    \end{align*}
    \color{black}   
        Here, the latent variables $\{\dnnLatent_t\}$ are drawn from its known prior distribution for each mini-batch. 
    
    Among currently available deep generative models, GANs achieve  the best sample quality at an unprecedented resolution.  For example, the current state-of-the-art model StyleGAN2 \cite{karras2020analyzing} is able to generate high-resolution ($1024 \times 1024$) images that are nearly indistinguishable from real photos to a human observer.  That said, GANs do come with several disadvantages as well. The adversarial training procedure is known to be unstable, and many tricks are necessary in practice to train a large GAN.  Also because GANs do not offer any probabilistic interpretation, it is difficult to objectively evaluate the quality of a GAN.

\smallskip
    % --- Autoencoders --- 
    $2)$ \textit{Autoencoders}: %\ns{Jay - can you please add a short description here of basic autoencoders along with their corresponding loss measure?}
    Another well-studied task in unsupervised learning is the training of an \textit{autoencoder}, which has many uses such as dimensionality reduction and representation learning.  An autoencoder  consists of two neural networks: an \textit{encoder} $\dnnEnc: \InputSpace \mapsto \mySet{Z}$ and a \textit{decoder} $\dnnDec: \mySet{Z} \mapsto \InputSpace$, where $\mySet{Z}$ is some predefined latent space. The primary goal of an autoencoder is to reconstruct a signal $\Input$ from itself by mapping it through $\dnnDec \circ \dnnEnc$.  
    %The output of the encoder $\dnnEnc(\Input)$ is often called the \textit{code} or the \textit{latent variable}, and the output of the decoder $\dnnDec(\dnnEnc(\Input))$ is called the \textit{reconstruction} of $\Input$. 
    
    The task of autoencoding may seem pointless at first; indeed one can trivially recover $\Input$ by setting $\mySet{Z} = \InputSpace$ and $\dnnEnc$, $\dnnDec$ to be identity functions.  The interesting case is when one imposes constraints which limit the ability of the network to learn the identity mapping \cite[Ch. 14]{goodfellow2016deep}. One way to achieve this is to form an undercomplete autoencoder, where the {latent space} $\mySet{Z}$ is restricted to be lower-dimensional than $\InputSpace$, e.g., $\InputSpace = \mathbb{R}^n$ and $\mySet{Z} = \mathbb{R}^m$ for some $m < n$. This constraint forces the encoder to map its input into a more compact representation, while retaining enough information so that the reconstruction is as close to the original input as possible. Additional mechanisms for preventing an autoencoder from learning the identity mapping include imposing a regularizing term on the latent representation, as done in sparse autoencoders and contractive autoencoders, or alternatively, by distorting the input to the system, as carried out by denoising autoencoders \cite[Ch. 14.2]{goodfellow2016deep}.  A common metric used to measure the quality of reconstruction is the \ac{mse} loss. Under this setting, we obtain the following loss function for training
    \begin{equation}
    \!\mySet{L}_{\rm MSE}\left(\dnnEnc, \dnnDec \right) \!=\! \frac{1}{\Ntraining} \sum_{t=1}^{\Ntraining} \left\| \Input_t\! -\! \dnnDec(\dnnEnc(\Input_t)) \right\|_2^2.
    \end{equation}
    
    %While the above objective is simple, an antoencoder has several uses.
%    Due to the constraint on $\mySet{Z}$, the latent variable $\dnnLatent = \dnnEnc(\Input)$ can be seen as a continuous low-dimensional representation of $\Input$, even when $\InputSpace$ is discrete or high-dimensional.  When $\dnnLatent$ is appropriately quantized, it can also be used to implement a learned lossy compressor of samples from $\InputSpace$.
    
%    Many variations of the above autoencoder formulation have been proposed and extensively studied in the deep learning literature. For example, a \textit{denoising} autoencoder  \cite{vincent2008extracting} receives a noisy input example and tries to reconstruct the noiseless version of it, leading to a latent representation that is robust to input noise.  Another popular architecture is the variational autoencoder (VAE) \cite{kingma2013auto}, in which the latent variable $\dnnLatent$ is stochastic, and thus the decoder can be used as a learned generative model.

%    \end{enumerate}

	\paragraph{Semi-Supervised Learning}
	Semi-supervised learning lies in the middle ground between the above two categories, where one typically has access to a large amount of unlabeled data and a small set of labeled data.  The goal is to leverage the unlabeled data to improve performance on some supervised task to be trained on the labeled data.  As labeling data is often a very costly process, semi-supervised learning provides a way to quickly learn desired inference rules without having to label all of the available unlabeled data points.

	Various approaches have been proposed in the literature to utilize unlabeled data for a supervised task, see the detailed survey \cite{van2020survey}. One such common technique is to guess the missing labels, while integrating dedicated mechanisms to boost confidence \cite{lee2013pseudo}. This can be achieved by, e.g., applying the \ac{dnn} to various augmentations of the unlabeled data \cite{laine2016temporal}, while combining multiple regularization terms for encouraging consistency and low-entropy of the guessed labels \cite{Berthelot2019mixmatch}, as well as training a teacher \ac{dnn} using the available labeled data to produce guessed labels \cite{xie2020self}.

     %----------------------------------------------------------------------------------------
    %	Challenges
    %----------------------------------------------------------------------------------------
%     \vspace{-0.2cm}
%     \subsection{Strengths and Challenges of Deep Learning}
% 	\label{subsec:DL_Challenges}
% 	\vspace{-0.1cm}	
	
% 	TODO - Jay 
	
% 	\ns{we can remove this section. I have added in Section~\ref{sec:MBvsDL} a discussion on the individual pros and cons of model-based methods and deep learning, so we can omit this part}
	
% 	\jw{Sounds good.}

% 	 Elborate on main strengths - model agnostic nature, ability to operate reliably and disentangle semantic information in complex systems, inference complexity which only grows linearly with the number of layers (though training is computationally heavy).
	 
% 	 Drawbacks - plenty of data needed to train and adapt, as well as immense computational burden. Non-interpretable nature. Highly-parametrized models may be difficult to apply on hardware limited devices. 
	 
	%----------------------------------------------------------------------------------------
	%	Model-Aided Networks
	%----------------------------------------------------------------------------------------
	% \vspace{-0.2cm}
	\section{Model-Aided Networks}
	\label{sec:Networks}
	% \vspace{-0.1cm}
  %----------------------------------------------------------------------------------------
    %	Rationale
    %----------------------------------------------------------------------------------------
%    \vspace{-0.2cm}
%     \subsection{Rationale}
%	\label{subsec:Networks_Rationale}
%	\vspace{-0.1cm}	
	Model-aided networks implement model-based deep learning by using model-aware algorithms to design deep architectures.  
	Broadly speaking, model-aided networks implement the inference system using a \ac{dnn}, similar to conventional deep learning. Nonetheless, instead of applying generic off-the-shelf \acp{dnn}, the rationale here is to tailor the architecture specifically for the scenario of interest, based on a suitable model-based method. 	
	By converting a model-based algorithm into a model-aided network, that learns its mapping from data, one typically achieves improved inference speed, as well as overcome partial or mismatched domain knowledge. In particular, model-aided networks can learn missing model parameters, such as channel matrices \cite{he2018model}, dictionaries \cite{tolooshams2019convolutional}, and noise covariances \cite{xu2021ekfnet}, as part of the learning procedure. \textcolor{NewColor}{Alternatively, it can be used to learn a surrogate model for which the resulting inference rule best matches the training data~\cite{shlezinger2022model}.}

\begin{figure*}
	    \centering
	    \includegraphics[width=0.8\linewidth]{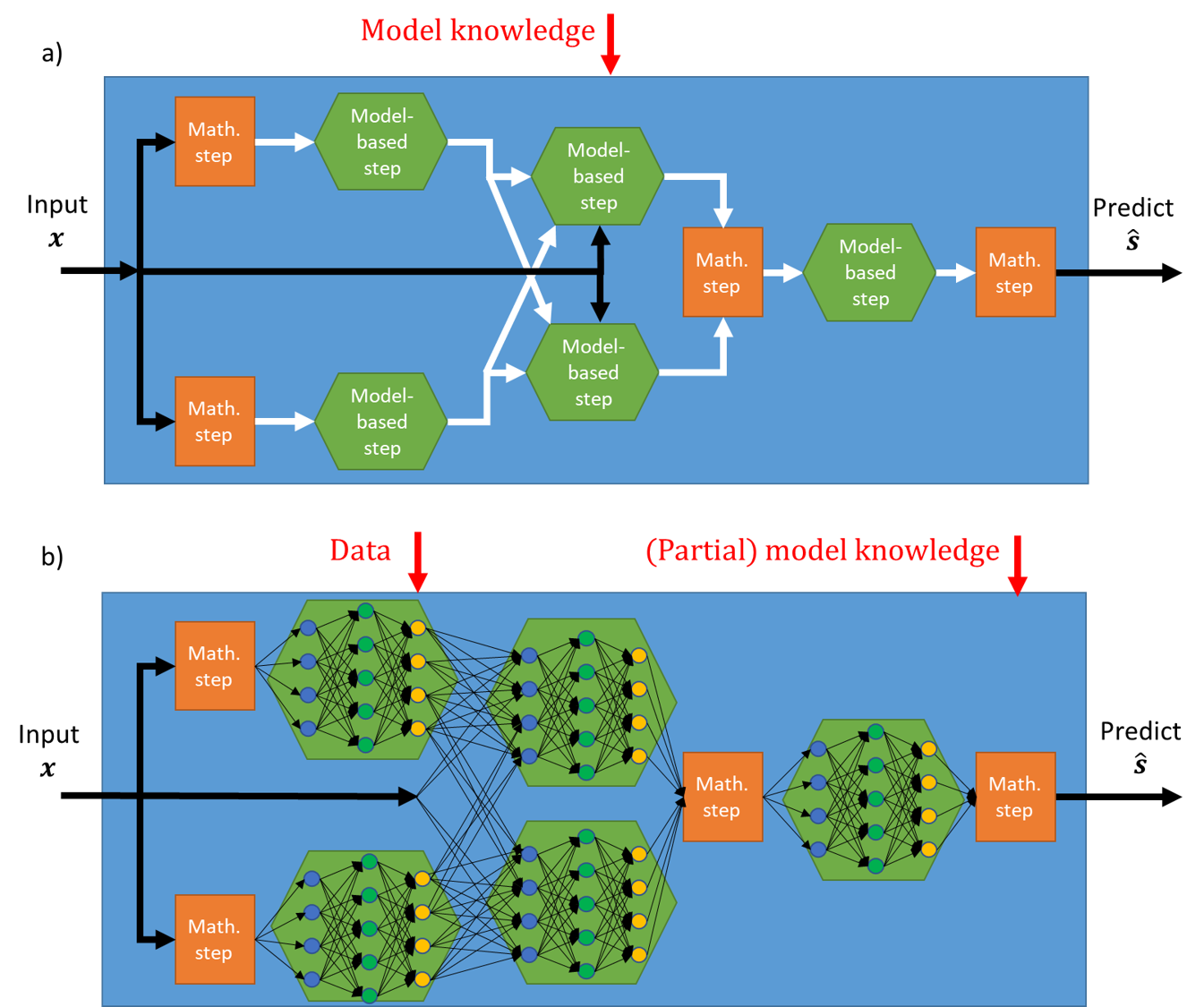}
	    \caption{Model-aided DNN illustration: $a)$ a model-based algorithm comprised of a series of model-aware computations and generic mathematical steps; $b)$ A DNN whose architecture and  inter-connections are designed based on the model-based algorithm. \textcolor{NewColor}{Here, data can be used to train the overall architecture end-to-end, typically requiring the intermediate mathematical steps to be either differentiable or well-approximated by a differentiable mapping.}}
	    \label{fig:Model-Aided1}
	\end{figure*}

	Model-aided networks obtain dedicated \ac{dnn} architectures by identifying structures in a model-based algorithm one would have utilized for the  problem  given full domain knowledge and sufficient computational resources. Such structures can be given in the form of an iterative representation of the model-based algorithm, as exploited by \textit{deep unfolding} detailed in Subsection~\ref{subsec:Networks_Unfolding}, or via a block diagram algorithmic representation, which \textit{neural building blocks} rely upon, as presented in Subsection~\ref{subsec:Networks_Blocks}. The dedicated neural network is then formulated as a discriminative architecture~\cite{ng2001discriminative,shlezinger2022discriminative} whose trainable parameters, intermediate mathematical manipulations, and interconnections follow the operations of the model-based algorithm, as illustrated in Fig.~\ref{fig:Model-Aided1}.

	\textcolor{NewColor}{In the following we describe these methodologies in a systematic manner. In particular, our presentation of each approach commences with a high level description and generic design outline, followed by one or two concrete model-based deep learning examples from the literature, and concludes with a summarizing discussion. For each example, we first detail the system model and model-based algorithm from which it originates. Then, we describe  the hybrid model-based/data-driven system by detailing its architecture and training procedure, as well as present some representative quantitative results.  }

	%We next elaborate on the main approaches for designing model-aided networks, namely, deep unfolding and neural building blocks, in Subsection~\ref{subsec:Networks_Unfolding} and \ref{subsec:Networks_Blocks}, respectively. 
	%and conclude with a discussion and summary in Subsection~\ref{subsec:Networks_Summary}.
	
 %   The main rationale in model-aided networks is to design the network to imitate the operation of a model-based iterative optimization algorithm applicable for the considered scenario. In particular, each iteration of the model-based algorithm is replaced with a dedicated layer with trainable parameters whose structure is based on the iterative procedure.

     %----------------------------------------------------------------------------------------
    %	Deep Unfolding
    %----------------------------------------------------------------------------------------
    % \vspace{-0.2cm}
     \subsection{Deep Unfolding}
	\label{subsec:Networks_Unfolding}
	% \vspace{-0.1cm}	
    Deep unfolding \cite{hershey2014deep}, also referred to as {\em deep unrolling}, converts an iterative algorithm into a \ac{dnn} by designing each layer to resemble a single iteration. 
    Deep unfolding was originally proposed by Greger and LeCun in \cite{gregor2010learning}, where  a deep architecture was designed to learn to carry out the \ac{ista} for sparse recovery.  Deep unfolded networks have since been applied in various applications in image denoising \cite{li2019deep, solomon2019deep}, sparse recovery \cite{wu2018learning,cui2020jointly,metzler2017learned}, dictionary learning \cite{chang2019randnet,tolooshams2019convolutional}, communications \cite{balatsoukas2019deep, samuel2019learning,he2018model,takabe2019deep,hu2020iterative,khobahi2021model}, ultrasound \cite{mischi2020deep}, and super resolution \cite{dardikman2020learned,zhang2020deep,huang2020unfolding}. A recent review can be found in \cite{monga2019algorithm}.
    
    \subsubsection*{Design Outline} 
    The application of deep unfolding to design a model-aided deep network is based on the following steps:
    \begin{enumerate}
    	\item Identify an iterative optimization algorithm which is useful for the problem at hand. 
    	For instance, recovering a sparse vector from its noisy projections can be tackled using   \ac{ista}, unfolded into LISTA in \cite{gregor2010learning}. 
    	\item Fix a number of iterations in the optimization algorithm.
    	\item Design the layers to imitate the free parameters of each iteration in a trainable fashion. 
    	\item Train the overall resulting network end-to-end.
    \end{enumerate} 
    \textcolor{NewColor}{The selection of the free parameters to learn in the third step determines the resulting trainable architecture. One can set these parameters to be the hyperparameters of the iterative optimizer (such as step-size), thus leveraging data to automatically determine parameters typically selected by hand~\cite{shlezinger2022model}. Alternatively, the architecture may be designed to learn the parameters of the objective optimized in each iteration, thus achieving a more abstract family of inference rules compared with the original iterative algorithm, or even convert the operation of each iteration into a trainable neural architecture.} 
We next demonstrate how this rationale is translated into concrete architectures, 
using two examples: the first is the DetNet system of \cite{samuel2019learning} which unfolds projected gradient descent optimization;  the second is the unfolded dictionary learning for Poisson image denoising proposed in \cite{tolooshams2019convolutional}.

%\textcolor{red}{Explain also about the different forms of unfolding - what exactly is to be learned.}

% ---- DetNet -------
\subsubsection*{Example 1: Deep Unfolded Projected Gradient Descent}
Projected gradient descent is a simple and common iterative algorithm for tackling constrained optimization. While the projected gradient descent method is quite generic and can be applied in a broad range of constrained optimization setup, in the following we focus on its implementation for symbol detection in linear memoryless \ac{mimo} Gaussian channels. In such cases, where the constraint follows from the discrete nature of digital communication symbols, the iterative projected gradient descent gives rise to the DetNet architecture proposed in \cite{samuel2019learning} via deep unfolding. 

%refers to the problem of recovering a sparse matrix where the measurements are corrupted by low-rank clutter as well as noise. Such scenarios are commonly encountered, e.g., in medical imaging. In the following, we first formulate the robust \ac{pca} system model and objective, after which we discuss the generalization of the \ac{ista} algorithm for such setups, and show how it is unfolded into the deep unfolded robust \ac{pca} architecture proposed in \cite{solomon2019deep}. 

\paragraph{System Model}
Consider the problem of symbol detection in  linear memoryless \ac{mimo} Gaussian channels. The task is to recover the $\Nusers$-dimensional vector $\myS$ from the $\Nantennas\times 1$ observations $\Input$, which are related via:
\begin{equation}
\label{eqn:Gaussian}
\Input = \myMat{H}\myS + \myVec{w}.
\end{equation}
Here, $\myMat{H}$ is a known deterministic $\Nantennas\times \Nusers$ channel matrix, and $\myVec{w}$ consists of $\Nantennas$ i.i.d Gaussian \acp{rv}. For our presentation we consider the case in which the entries of $\myS$ are symbols  generated from a \ac{bpsk} constellation in a uniform i.i.d. manner, i.e., $\mathcal{S}=\{\pm 1\}^K$. In this case, the \ac{map} rule given an observation $\Input $ becomes the minimum distance estimate, given by 
\begin{equation}
\label{eqn:DetNetObj}
\hat{\myVec{s}} = \mathop{\arg \min}\limits_{\myVec{s}\in\{\pm 1\}^\Nusers} \|\Input-\myMat{H}\myVec{s}\|^2.
\end{equation}

\paragraph{Projected Gradient Descent}

While directly solving \eqref{eqn:DetNetObj} involves an exhaustive search over the $2^\Nusers$ possible symbol combinations, it can be tackled with affordable computational complexity using the iterative projected gradient descent algorithm. 
\color{NewColor}
This method, whose derivation is detailed in Appendix~\ref{app:DetNet_Exm}, is summarized as Algorithm~\ref{alg:PGD}, where  $\mySet{P}_{\mySet{S}}(\cdot)$ denotes the projection operator into $\mySet{S}$, which for \ac{bpsk} constellations is the element-wise sign function.

\begin{algorithm}  
		\caption{Projected gradient descent for system model~\eqref{eqn:Gaussian}}
		\label{alg:PGD}
		\KwData{Fix step-size  $\eta>0$. Set initial guess $\hat{\myVec{s}}_0$} 
		\For{$q=0,1,\ldots$}{
	        Update $\hat{\myVec{s}}_{q + 1}  =\mySet{P}_{\mySet{S}}\left(\hat{\myVec{s}}_{q} - \eta\myMat{H}^T\Input + \eta\myMat{H}^T\myMat{H}\hat{\myVec{s}}_{q} \right)$.\label{stp:ProjGrad} 
		}
		\KwOut{Estimate $\hat{\myVec{s}} = \myVec{s}_q$.}
\end{algorithm}

 \color{black}

\paragraph{Unfolded DetNet}
DetNet unfolds the projected gradient descent iterations, \textcolor{NewColor}{repeated until convergence in Algorithm~\ref{alg:PGD}}, into a \ac{dnn}, which learns to carry out this optimization procedure from data. To formulate DetNet, we first fix a number of iterations $\Niter$. Next, a \ac{dnn} with $\Niter$ layers is designed, where each layer imitates a single iteration of Algorithm~\ref{alg:PGD} in a trainable manner.

\begin{figure*}
	\centering
	\includegraphics[width=0.9\linewidth]{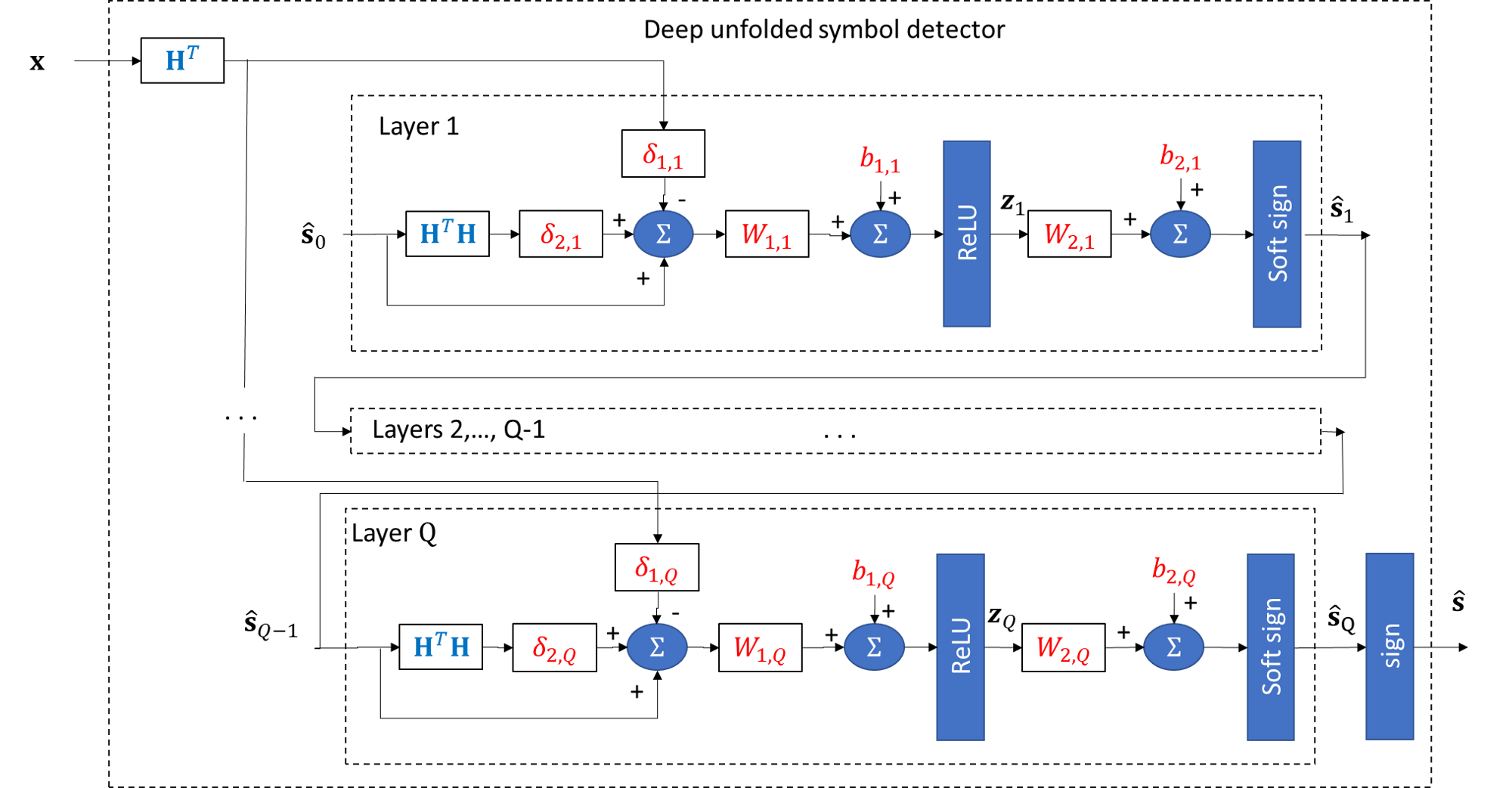} 
	\caption{DetNet illustration. Parameters in red fonts are learned in training, while those in blue fonts are externally provided.} 
	\label{fig:DetNet}
\end{figure*}

{\bf Architecture:} DetNet builds upon the observation that the update rule in Step~\ref{stp:ProjGrad} of Algorithm~\ref{alg:PGD} consists of two stages: gradient descent computation, i.e., gradient step $\hat{\myVec{s}}_{q} - \eta\myMat{H}^T\Input + \eta\myMat{H}^T\myMat{H}\hat{\myVec{s}}_{q}$; and projection, namely, applying $ \mySet{P}_{\mySet{S}}(\cdot)$. Therefore, each unfolded iteration is represented as two sub-layers: The first sub-layer learns to compute the gradient descent stage by treating the step-size as a learned parameter and applying an \ac{fc} layer with ReLU activation to the obtained value. For iteration index $q$, this results in 
\begin{equation*}
\myVec{z}_{q} \!=\! {\rm ReLU}\left(\myMat{W}_{1,q}\left((\myMat{I}\! +\! \delta_{2,q}\myMat{H}^T\myMat{H})\hat{\myVec{s}}_{q-1} \!-\! \delta_{1,q}\myMat{H}^T\Input  \right) \!+\! \myVec{b}_{1,q}  \right)
%\label{eqn:Layer1}
\end{equation*}
in which $\{\myMat{W}_{1,q}, \myVec{b}_{1,q}, \delta_{1,q}, \delta_{2,q}\}$ are learnable parameters. 
The second sub-layer learns the projection operator by approximating the sign operation with a soft sign activation preceded by an \ac{fc} layer, leading to
\begin{equation}
\hat{\myVec{s}}_q = {\rm soft~sign}\left( \myMat{W}_{2,q} \myVec{z}_{q} +\myVec{b}_{2,q} \right).
\label{eqn:Layer2}
\end{equation}
Here, the learnable parameters are $\{\myMat{W}_{2,q}, \myVec{b}_{2,q}\}$. 
The resulting deep network is depicted in Fig. \ref{fig:DetNet}, in which the output after $\Niter$ iterations, denoted $\hat{\myVec{s}}_{\Niter}$, is used as the estimated symbol vector by taking the sign of each element.

 {\bf Training:} Let  $\myVec{\theta} = \{(\myMat{W}_{1,q}, \myMat{W}_{2,q}, \myVec{b}_{1,q}, \myVec{b}_{2,q}, \delta_{1,q}, \delta_{2,q})\}_{q=1}^Q$ be the trainable parameters of DetNet\footnote{The formulation of DetNet in \cite{samuel2019learning} includes an additional sub-layer in each iteration intended to further lift its input into higher dimensions and introduce additional trainable parameters, as well as reweighing of the outputs of subsequent layers. As these operations do not follow directly from unfolding  projected gradient descent, they are not included in the description here.}.  To tune $\myVec{\theta}$, the overall network is trained end-to-end  to minimize the empirical weighted $\ell_2$ norm loss over its intermediate layers, given by
 %In particular,by letting $\{\myVec{s}_t, \Input_t\}_{t=1}^{\Ntraining}$ denote the training set consisting of channel outputs and their corresponding transmitted symbols,  the loss function used for training  DetNet is given by
\begin{equation}
\label{eqn:LossDetNet}
\mySet{L}(\myVec{\theta}) = \frac{1}{\Ntraining}\sum_{t=1}^{\Ntraining}\sum_{q=1}^{\Niter} \log(q)\|\myVec{s}_t - \hat{\myVec{s}}_q(\Input_t; \myVec{\theta}) \|^2
\end{equation}
where $\hat{\myVec{s}}_q(\Input_t; \myVec{\theta})$ is the output of the $q$th layer of DetNet with parameters $\myVec{\theta}$ and input $\Input_t$. This loss measure accounts for the interpretable nature of the unfolded network, in which the output of each layer is a further refined estimate of $\myVec{s}$.

	{\bf Quantitative Results:} The experiments reported in \cite{samuel2019learning} indicate that, when provided sufficient training examples, DetNet outperforms leading \ac{mimo} detection algorithms based on approximate message passing and semi-definite relaxation. It is also noted in \cite{samuel2019learning} that the unfolded network requires an order of magnitude less layers compared to the number of iterations required by the model-based optimizer to converge. This gain is shown to be translated into reduced run-time during inference, particularly when processing batches of data in parallel. In particular, it is reported in \cite[Tbl. 1]{samuel2019learning} that DetNet successfully successfully detects a batch of $1000$ channel outputs in a $60\times 30$ static \ac{mimo} channel at run-time which is three times faster than that required by approximate message passing to converge, and over 80 times faster than semi-definite relaxation. 
	
%	EMPHASIZE THAT MUCH LESS LAYERS THAN ITERATIONS ARE REQUIRED,
	
%	An experimental study of deep unfolded robust \ac{pca}, demonstrating the  ability of deep unfolding to carry out inference both quickly and reliably, is presented in Fig.~\ref{fig:DeepRPCA_Res2}. In particular, Fig.~\ref{fig:DeepRPCA_Res2}(c) shows the recovered  ultrasound (contrast agents) image from a cluttered image (Fig.~\ref{fig:DeepRPCA_Res2}(a)) achieved using deep unfolded robust \ac{pca} with $\Niter=10$ layers trained using $\Ntraining = 4800$ images. Comparing the recovered image to the ground-truth in  Fig.~\ref{fig:DeepRPCA_Res2}(b) demonstrates the accuracy in using a \ac{dnn} to imitate the operations of the generalized \ac{ista} algorithm in a learned fashion. Furthermore, the fact that the unfolded network learns its parameters from data for each layer allows it to infer with a notably reduced number of layers compared to the corresponding number of iterations required by the model-based algorithm, which utilizes its full domain knowledge in applying the hard-coded iterative procedure. This is illustrated in Fig.~\ref{fig:DeepRPCA_Res2}(d) which demonstrates that the trained unfolded network can achieve with only a few layers an \ac{mse} accuracy which the model-based fast \ac{ista} of \cite{beck2009fast,palomar2010convex} does not approach even in $50$ iterations. 

\setcounter{paragraph}{0}
% ---- DCEA -------
\subsubsection*{Example 2: Deep Unfolded Dictionary Learning} DetNet exemplifies how deep unfolding can be used to realize rapid implementations of exhaustive optimization algorithms that typically require a very large amount of iterations to converge. However, DetNet requires full domain knowledge, i.e., it assumes the system model follows \eqref{eqn:Gaussian} and that the channel parameters $\myMat{H}$ are known. An additional benefit of deep unfolding is its ability to learn missing model parameters along with the overall optimization procedure, as we illustrate in the following example proposed in \cite{tolooshams2019convolutional}, which focuses on dictionary learning learning for Poisson image denoising. Similar examples where channel knowledge is not required in deep unfolding can be found in, e.g., \cite{khobahi2021model,li2019deep, he2018model}

\paragraph{System Model}
Consider the problem of reconstructing an image $\myVec{\mu} \in \mathbb{R}^N$ from its noisy measurements $\Input \in \mathbb{R}^N$. The image is corrupted by Poisson noise, namely, $p_{\Input | \myVec{\mu}}$ is a multivariate Poisson distribution with mutually independent entries and mean  $\myVec{\mu}$. 
Furthermore, it is assumed that for the clean image $\myVec{\mu}$, it holds that $\log (\myVec{\mu})$ (taken element-wise) 
\color{NewColor}
can be written as 
\begin{equation}
    \label{eqn:ConvGenModel}
    \log (\myVec{\mu}) =  \myMat{H}\myVec{s}.
\end{equation}
In~\eqref{eqn:ConvGenModel},  $\myMat{H}$, referred to as the {\em dictionary}, is an unknown  block-Toeplitz matrix (representing a convolutional dictionary), while $\myVec{s}$ is an unknown sparse vector.

\paragraph{Proximal Gradient Mapping}
The recovery of the clean image $\myVec{\mu}$ is tackled by alternating optimization \cite{agarwal2016learning}. In each iteration, one first recovers $\myVec{s}$ for a fixed $\myMat{H}$, after which $\myVec{s} $ is set to be fixed and $\myMat{H}$ is estimated. The resulting iterative algorithm, whose detailed derivation is given in Appendix~\ref{app:DCEA_Exm}, is summarized as Algorithm~\ref{alg:DictLearn}. Here, $\eta >0$ is the step size; $\myVec{1}$ is the all-ones vector; $b$ is a threshold parameter; and  $\mySet{T}_{b}$ is the soft-thresholding operator, also referred to as the shrinkage operator, applied element-wise and is given by $\mySet{T}_{b}(x) = {\rm sign}(x)\max\{|x|-b, 0\}$. Furthermore, the optimization variable $\myMat{H}$ in Step~\ref{stp:DictLearn} is constrained to be block-Toeplitz.
		
\begin{algorithm}  
		\caption{Alternating image recovery and dictionary learning for system model~\eqref{eqn:ConvGenModel}}
		\label{alg:DictLearn}
		\KwData{Fix step-size  $\eta>0$. Set initial guess ${\myVec{s}}_0$} 
		\For{$l=0,1,\ldots$}{
		    Update  $\hat{\myVec{H}}_{l} = \mathop{\arg \min}\limits_{ \myVec{H}  } \myVec{1}^T\exp\left(\myVec{H} {\myVec{s}}_l\right)-\Input^T \myVec{H}{\myVec{s}}_l$. \label{stp:DictLearn} \\
    		    Set $\hat{\myVec{s}}_0 = {\myVec{s}}_l$.\\
    		\For{$q=0,1,\ldots$}{ 
    	        Update 
    	        \begin{equation*}
    	          \hat{\myVec{s}}_{q+1} = \mySet{T}_{b}\big( \hat{\myVec{s}}_{q} + \eta \myMat{H}^T\left(\Input - \exp\left(\myVec{H} \hat{\myVec{s}}_{q}\right) \big)  \right).  
    	        \end{equation*}
    	        \label{stp:Proxmapping}
    		}
    		Set ${\myVec{s}}_{l+1} = \hat{\myVec{s}}_{q}$.
		}
		\KwOut{Estimate clean image via  $\hat{\myVec{\mu}} = \exp\big(\hat{\myMat{H}}_l{\myVec{s}}_l\big)$.}
\end{algorithm}

\color{black}

\paragraph{Deep Convolutional Exponential-Family Autoencoder} 
The hybrid model-based/data-driven architecture entitled deep convolutional exponential-family autoencoder (DCEA) architecture proposed in \cite{tolooshams2019convolutional} unfolds the proximal gradient iterations in Step~\ref{stp:Proxmapping} of Algorithm~\ref{alg:DictLearn}. By doing so, it avoids the need to learn the dictionary $\myMat{H}$ by alternating optimization, as it is implicitly learned from data in the training procedure of the unfolded network.

    {\bf Architecture:}  DCEA treats the two-step convolutional sparse coding problem as an autoencoder, where the encoder computes the sparse vector $\myVec{s}$ by unfolding $Q$ proximal gradient iterations  as in Step~\ref{stp:Proxmapping} of Algorithm~\ref{alg:DictLearn}. The decoder then converts $ \hat{\myVec{s}}$ produced by the encoder into a recovered clean image $\hat{\myVec{\mu}}$. 
    
\begin{figure*}
	\centering
	\includegraphics[width=0.8\linewidth]{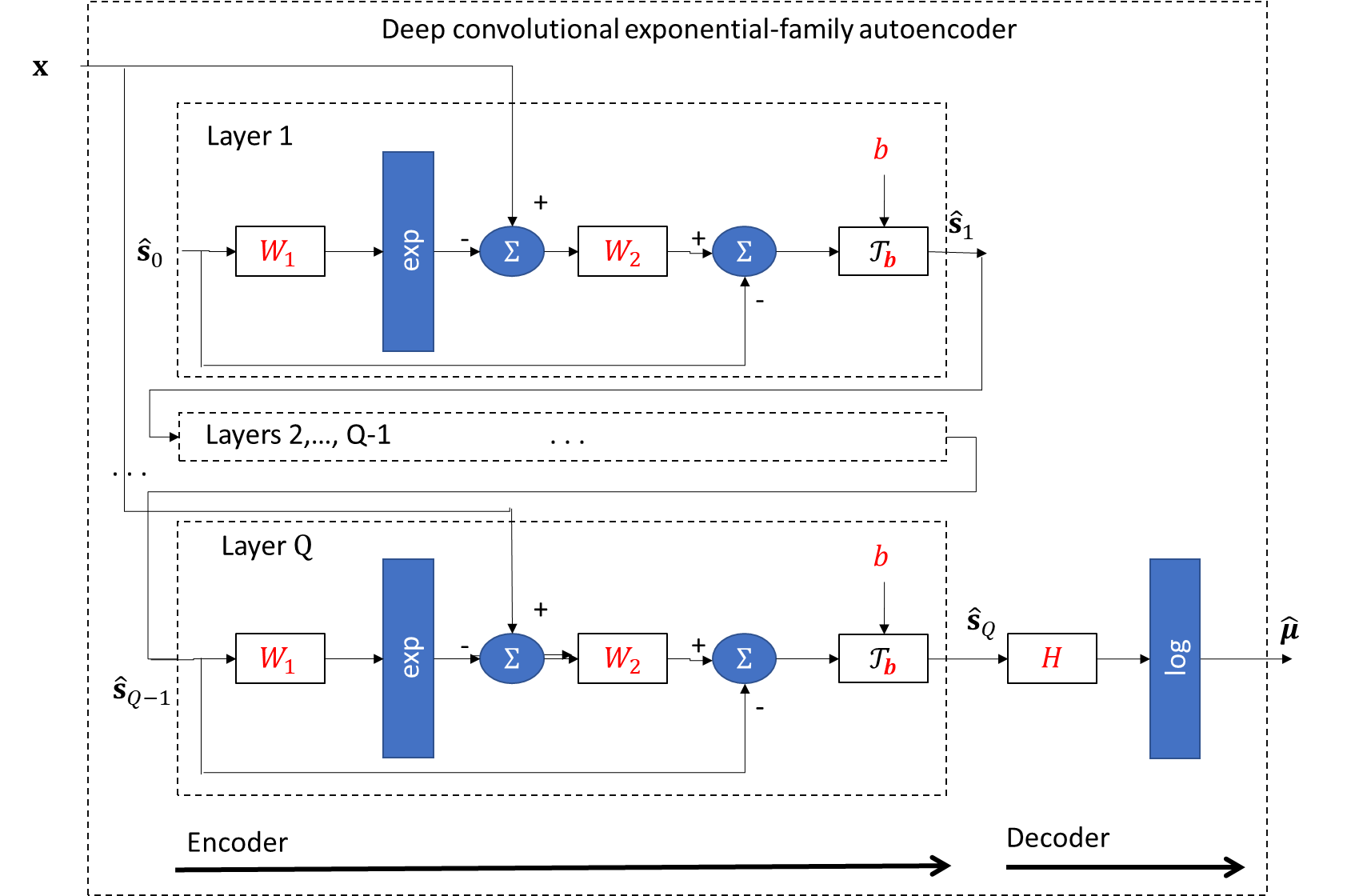} 
	\caption{DCEA illustration. Parameters in red fonts are learned in training, while those in blue fonts are externally provided.} 
	\label{fig:DCEA}
\end{figure*}

    In particular, \cite{tolooshams2019convolutional} proposed two implementations of DCEA. The first, referred to as DCEA-C, directly implements $Q$  proximal gradient iterations followed by the decoding step which computes $\hat{\myVec{\mu}}$, where both the encoder and the decoder use the same value of the dictionary matrix $\myMat{H}$.  This is replaced with a convolutional layer and is learned via end-to-end training along with the thresholding parameters, bypassing the need to explicitly recover the dictionary for each image, as in Step~\ref{stp:DictLearn} of Algorithm~\ref{alg:DictLearn}. The second implementation, referred to DCEA-UC, decouples the convolution kernels of the encoder and the decoder, and lets the encoder carry out $Q$ iterations of the form
    \begin{equation}
         \hat{\myVec{s}}_{q+1} = \mySet{T}_{b}\left( \hat{\myVec{s}}_{q} + \eta \myMat{W}_2^T\left(\Input - \exp\left(\myVec{W}_1 \hat{\myVec{s}}_{q}\right) \right)  \right). \label{eqn:Proxmapping2}
    \end{equation}  
    Here, $\myVec{W}_1$ and $\myVec{W}_2$ are convolutional kernels which are not constrained to be equal to $\myMat{H}$ used by the decoder\footnote{The architecture proposed in \cite{tolooshams2019convolutional} is applicable for various exponential-family noise signals. Particularly for Poisson noise, an additional exponential linear unit was applied to $\Input - \exp\left(\myVec{W}_1 \hat{\myVec{s}}_{q}\right)$ which was empirically shown to improve the convergence properties of the network.}. An illustration of the resulting architecture is depicted in Fig.~\ref{fig:DCEA}.

    {\bf Training:} The parameters of  DCEA  are $\myVec{\theta}=\{\myMat{H}, \myVec{b}\}$ for DCEA-C, and $\myVec{\theta}=\{\myMat{W}_1,\myMat{W}_2,\myMat{H}, \myVec{b}\}$ for DCEA-UC. The vector $\myVec{b}\in\mathbb{R}^C$ is comprised of the thresholding parameters used at each channel. 
    When  applied for Poisson image denoising, DCEA is trained in a supervised manner using the \ac{mse} loss, namely, a set of $\Ntraining$ clean images $\{\myVec{\mu}_t\}_{t=1}^{\Ntraining}$ are used along with their Poisson noisy version $\{\Input_t\}_{t=1}^{\Ntraining}$. By letting $f_{\myVec{\theta}}(\cdot)$ denote the resulting mapping of the unfolded network, the loss function is formulated as
    \begin{equation}
    \label{eqn:DCEALoss}
        \mySet{L}(\myVec{\theta}) = \frac{1}{\Ntraining}\sum_{t=1}^{\Ntraining} \|\myVec{\mu}_t - f_{\myVec{\theta}}(\Input_t)\|^2. 
    \end{equation}

\begin{figure*}
	    \centering
	    \includegraphics[width=0.7\linewidth]{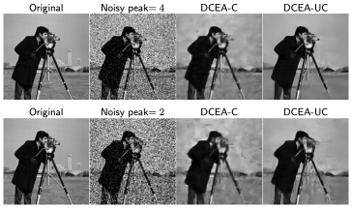}
	    \caption{Illustration of an image corrupted by different levels of Poisson noise and the resulting denoised images produced by the unfolded DCEA-C and DCEA-UC. {Figure reproduced from \cite{tolooshams2019convolutional} with authors' permission.}}
	    \label{fig:DCEA_res1}
	\end{figure*}

	{\bf Quantitative Results:} The experimental results reported in \cite{tolooshams2019convolutional} evaluated the ability of the unfolded DCEA-C and DCEA-UC in recovering images corrupted with different levels of Poisson noise. An example of an image denoised by the unfolded system is depicted in Fig.~\ref{fig:DCEA_res1}. It was noted in \cite{tolooshams2019convolutional} that the proposed approach allows to achieve similar and even improved results to those of purely data-driven techniques based on black-box \acp{cnn} \cite{remez2018class}. However, the fact that the denoising system is obtained by unfolding the model-based optimizer in Step~\ref{stp:Proxmapping} of Algorithm~\ref{alg:DictLearn} allows this performance to be achieved while utilizing $3\%-10\%$ of the overall number of trainable parameters as those used by the conventional \ac{cnn}.
	
 \color{black}
	
\subsubsection*{Discussion}
Deep unfolding incorporates model-based domain knowledge to obtain a dedicated \ac{dnn} design which resembles an iterative optimization algorithm. Compared to conventional \acp{dnn}, unfolded networks are typically interpretable, and tend to have a smaller number of parameters, and can thus be trained more quickly \cite{balatsoukas2019deep,monga2019algorithm}. %Nonetheless, these deep networks are still quite highly parametrized and require a large volume of training data. %For instance, DetNet is trained in \cite{samuel2019learning} using approximately $250$ million labeled samples. 
A key advantage of deep unfolding over model-based optimization is in inference speed.  For instance,  unfolding projected gradient descent iterations into DetNet allows to infer with much fewer layers  compared to the number of iterations required by the model-based algorithm to converge. Similar observations have been made in various unfolded algorithms  \cite{dardikman2020learned,solomon2019deep}.

One of the key properties of unfolded networks is their reliance on knowledge of the model describing the setup (though not necessarily on its parameters). For example, one must know that the image is corrupted by Poisson noise to formulate the iterative procedure in Algorithm~\ref{alg:DictLearn} unfolded into DCEA, or that the observations obey a linear Gaussian model to unfold the projected gradient descent iterations into DetNet. However, the parameters of this model, e.g., the matrix $\myMat{H}$ in \eqref{eqn:Gaussian} and \eqref{eqn:ConvGenModel}, can be either provided based on domain knowledge, as done in DetNet, or alternatively, learned in the training procedure, as carried out by DCEA. 
The model-awareness of deep unfolding has its advantages and drawbacks. When the model is accurately known, deep unfolding essentially incorporates it into the \ac{dnn} architecture, as opposed to conventional black-box \acp{dnn} which must learn it from data. However, this approach does not exploit the model-agnostic nature of deep learning, and thus may lead to degraded performance when the true relationship between the measurements and the desired quantities deviates from the model assumed in design. Nonetheless, training an unfolded network designed with a mismatched model using data corresponding to the true underlying scenario typically yields more accurate inference compared to the model-based iterative algorithm with the same model-mismatch, as the unfolded network can learn to compensate for this mismatch \cite{khobahi2021model}.

     \subsection{Neural Building Blocks}
	\label{subsec:Networks_Blocks}
	% \vspace{-0.1cm}	
    Neural building blocks is an alternative approach to design model-aided networks, which can be treated as a generalization of deep unfolding. It is based on representing a model-based algorithm, or alternatively prior knowledge of an underlying statistical model, as an interconnection of distinct building blocks. Neural building blocks implement a \ac{dnn} comprised of multiple sub-networks. Each module learns to carry out the specific computations of the different building blocks constituting the model-based algorithm, as done in \cite{duan2019vs, shlezinger2019deepSIC,merkofer2022deep,van2022deep}, or to capture a known statistical relationship, as in~\cite{kocaoglu2017causalgan}.

   Neural building blocks are designed for scenarios which are tackled using algorithms with flow diagram representations, that can be captured as a sequential and parallel interconnection of building blocks. In particular, deep unfolding can be obtained as a special case of neural building blocks, where the original algorithm is an iterative optimizer, such that the building blocks are interconnected in a sequential fashion, and implemented using a single layer. However, the generalization of neural building blocks compared to deep unfolding is not encapsulated merely in its ability to implement non-sequential interconnections between algorithmic building  blocks in a learned fashion, but also in the identification of the specific task of each block, as well as the ability to convert known statistical relationships such as causal graphs into dedicated \ac{dnn} architectures.

\subsubsection*{Design Outline}     The application of neural building blocks to design a model-aided deep network is based on the following steps:
\begin{enumerate}
	\item Identify an algorithm or a flow-chart structure  which is useful for the problem at hand, and can be decomposed into multiple building blocks.  
	\item Identify which of these building blocks should be learned from data, and what is their concrete task.
	\item Design a dedicated neural network for each building block capable of learning to carry out its specific task.
	\item Train the overall resulting network, either in an end-to-end fashion or by training each building block network individually.
\end{enumerate}
%Several unfolding-based deep symbol detectors have been proposed recently, mostly tackling symbol detection in flat Gaussian \ac{mimo} channels \cite{samuel2019learning,he2018model,takabe2019deep}, but also for alternative models such as one-bit quantized receivers \cite{khobahi2019deep}. 
We next demonstrate how one can design a model-aided network comprised of neural building blocks. Our example focuses on symbol detection in flat \ac{mimo} channels, where we consider the data-driven implementation of the iterative \ac{sic} scheme of \cite{choi2000iterative}, which is the DeepSIC algorithm  proposed in \cite{shlezinger2019deepSIC}.
	
% ---- DeepSIC -------	
\subsubsection*{Example 3: DeepSIC for MIMO Detection}
Iterative \ac{sic}  \cite{choi2000iterative} is a \ac{mimo} detection method  suitable for linear Gaussian channels, i.e., the same channel models as that described in the example of DetNet in Subsection~\ref{subsec:Networks_Unfolding}. DeepSIC is a hybrid model-based/data-driven implementation of the iterative \ac{sic} scheme \cite{shlezinger2019deepSIC}. However, unlike its model-based counterpart, and alternative deep \ac{mimo} receivers \cite{samuel2019learning,he2018model,balatsoukas2019deep}, DeepSIC is not particularly tailored for linear Gaussian channels, and can be utilized in various flat MIMO channels.  
We formulate DeepSIC by first reviewing the model-based iterative \ac{sic}, and present DeepSIC as its data-driven implementation.

% \paragraph{System Model}
% Iterative \ac{sic} is a method for symbol detection in  linear memoryless \ac{mimo} Gaussian channels. The task is to recover the $\Nusers$-dimensional vector $\myS$ from the $\Nantennas\times 1$ observations $\Input$, which are related via:
% \begin{equation}
% \label{eqn:Gaussian}
% \Input = \myMat{H}\myS + \myVec{w}.
% \end{equation}
% Here, $\myMat{H}$ is a known deterministic $\Nantennas\times \Nusers$ channel matrix, and $\myVec{w}$ consists of $\Nantennas$ i.i.d Gaussian \acp{rv}. For our presentation we consider the case in which the entries of $\myS$ are symbols  generated from a \ac{bpsk} constellation in a uniform i.i.d. manner, i.e., $\mathcal{S}=\{\pm 1\}$. In this case the \ac{map} rule given an observation $\Input $ becomes the minimum distance estimate, given by 
% \begin{equation}
% \label{eqn:DetNetObj}
% \hat{\myVec{s}} = \mathop{\arg \min}\limits_{\myVec{s}\in\{\pm 1\}^\Nusers} \|\Input-\myMat{H}\myVec{s}\|^2.
% \end{equation}

\paragraph{Iterative Soft Interference Cancellation}
The iterative  \ac{sic} algorithm proposed in \cite{choi2000iterative} is a \ac{mimo}  detection method that combines multi-stage interference cancellation with soft decisions.  
The detector operates iteratively,  where in each iteration, an estimate of the conditional \ac{pmf} of $s_k$, which is the $k$th entry of $\myS$, given the observed $\Input $, is generated for every symbol $k \in \{1,2,\ldots, \Nusers\} :=\NusersSet$. \textcolor{NewColor}{Each \ac{pmf}, which is an $|\mySet{S}|\times 1$ vector denoted $\hat{\myVec{p}}_k^{(q)}$ at the $q$th iteration, is computed using the corresponding estimates of the interfering symbols $\{s_l\}_{l \neq k}$ obtained in the previous iteration.} Iteratively repeating this procedure refines the \ac{pmf} estimates, allowing to accurately recover each symbol from the output of the last iteration.  
\color{NewColor}
This iterative procedure is illustrated in Fig.~\ref{fig:SoftIC1}(a) and summarized as Algorithm~\ref{alg:Algo1SIC}, whose derivation is detailed in Appendix~\ref{app:DeepSIC_Exm}. Algorithm~\ref{alg:Algo1SIC} is detailed for linear Gaussian models as in \eqref{eqn:Gaussian}, assuming that the noise $\myVec{w}$ has variance $\SigW$. We use $\myVec{h}_l$ to denote the $l$th column of $\myMat{H}$, while $\mathcal{N}(\myVec{\mu},\myMat{\Sigma})$ is the Gaussian distribution with mean $\myVec{\mu}$ and covariance $\myMat{\Sigma}$.

\begin{algorithm}  
		\caption{Iterative \ac{sic} for system model~\eqref{eqn:Gaussian}}
		\label{alg:Algo1SIC}
		\KwData{Set initial \acp{pmf} guess $\{\hat{\myVec{p}}_k^{(0)}\}_{k =1}^{\Nusers}$} 
		\For{$q=0,1,\ldots$}{
		    For each $k \in \NusersSet$, compute expected values $e_k^{(q)}$ and variance  $v_k^{(q)}$ from $\hat{\myVec{p}}_k^{(q)}$.	\label{stp:MF1a} \\
    		 {\em Interference cancellation:} For each $k \in \NusersSet$ compute
    		 \begin{equation*}
    		  \myVec{z}_k^{(q+1)}=\Input - \sum\limits_{l \neq k} \myVec{h}_l e_l^{(q)}.   
    		 \end{equation*}
    		 \label{stp:IC} \\
		    {\em Soft decoding:} For each $k \in \NusersSet$, estimate   $\hat{\myVec{p}}_k^{(q+1)}$ as the \ac{pmf} of $s_k$ given $\myVec{z}_k^{(q+1)}$,  assuming that 
		    \begin{equation*}
		     \myVec{z}_k^{(q+1)}|s_k \sim \mathcal{N}\Big(\myVec{h}_k s_k, \SigW \myI_{\Nusers} + \sum_{l \neq k}v_l^{(q)}  \myVec{h}_l \myVec{h}_l^T\Big).   
		    \end{equation*}
		    \label{stp:SoftDec}
		}
		\KwOut{Estimate $\hat{\myVec{s}}$ by setting each $\hat{s}_k$ as the symbol maximizing the estimated \ac{pmf} $\hat{\myVec{p}}_k^{(q)}$.}
\end{algorithm}

\begin{figure*}
	\centering
	\includegraphics[width = 0.75\linewidth]{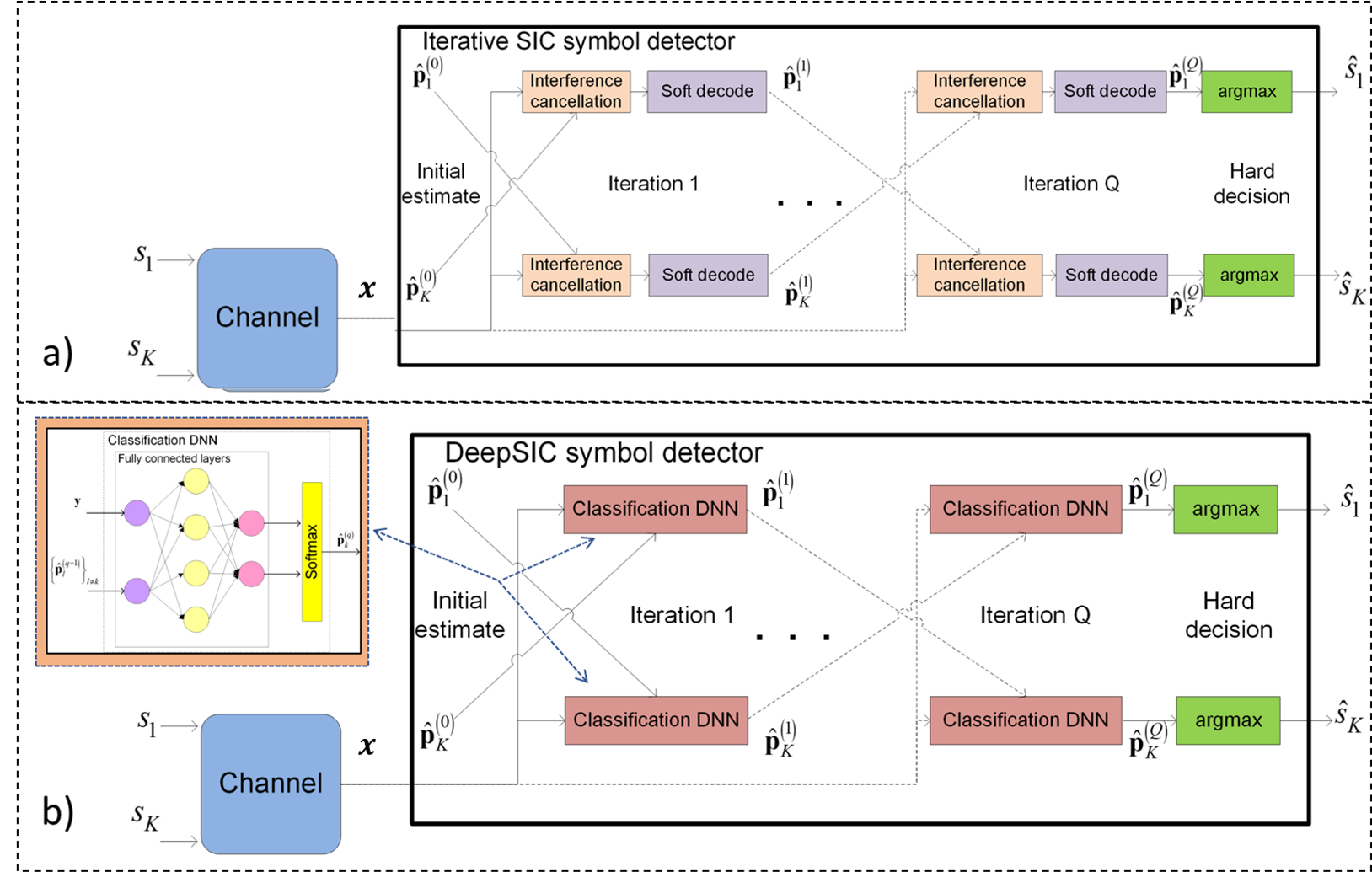}
%	\vspace{-0.4cm}
	\caption{Iterative \ac{sic} illustration: $a)$ model-based method; $b)$ DeepSIC.}
	%	\vspace{-0.4cm}
	\label{fig:SoftIC1}
\end{figure*}

\color{black}

\paragraph{DeepSIC}
Iterative \ac{sic} is specifically designed for linear channels of the form \eqref{eqn:Gaussian}. In particular, the interference cancellation Step~\ref{stp:IC} of Algorithm~\ref{alg:Algo1SIC} requires the contribution of the interfering symbols to be additive. Furthermore, it requires accurate complete knowledge of the underlying statistical model, i.e., of \eqref{eqn:Gaussian}.  DeepSIC propsoed in \cite{shlezinger2019deepSIC} learns to implement the iterative \ac{sic} from data as a set of neural building blocks, thus circumventing these limitations of its model-based counterpart.

{\bf Architecture:}
The  iterative \ac{sic} algorithm can be viewed as a set of interconnected basic building blocks, each implementing the two stages of interference cancellation and soft decoding, as illustrated in Fig.~\ref{fig:SoftIC1}(a). While the block diagram in Fig. \ref{fig:SoftIC1}(a) is ignorant of the underlying channel model, the basic building blocks are model-dependent.  
Although each of these basic building blocks consists of two sequential procedures which are completely channel-model-based, the purpose of these computations is to carry out a classification task. In particular, the $k$th building block of the $q$th iteration, $k \in \NusersSet$, produces $\hat{\myVec{p}}_k^{(q)}$, which is an estimate of the conditional \ac{pmf} of $s_k $ given $\Input $ based on $\{\hat{\myVec{p}}_l^{(q-1)}\}_{l\neq k}$. Such computations are naturally implemented by classification \acp{dnn}, e.g., \ac{fc} networks with a softmax output layer. %An illustration of such a network implementing the $k$th basic block of the $q$th iteration is depicted in Fig. \ref{fig:DNN1}. 
Embedding these conditional \ac{pmf} computations into the iterative \ac{sic} block diagram in Fig. \ref{fig:SoftIC1}(a) yields the overall receiver architecture depicted in Fig. \ref{fig:SoftIC1}(b). 
%The initial estimates $\{\hat{\myVec{p}}_k^{(0)}\}_{k=1}^{\Nusers}$ can be set to represent a uniform distribution, i.e., $\big(\hat{\myVec{p}}_k^{(0)}\big)_j \equiv \frac{1}{\CnstSize}$. % for each $j \in \{1,2,\ldots,\CnstSize\}$ and $k \in \NusersSet$.  

A major advantage of using classification \acp{dnn} as the basic building blocks in Fig. \ref{fig:SoftIC1}(b) stems from their ability to accurately compute conditional distributions in   complex non-linear setups without requiring a-priori knowledge of the channel model and its parameters. Consequently, when these building blocks are trained to properly implement their classification task, the  receiver essentially realizes iterative \ac{sic} for arbitrary channel models in a data-driven fashion.

{\bf Training:} 
In order for DeepSIC to  reliably implement symbol detection, its building block classification \acp{dnn} must be properly trained. Two possible training approaches are considered based on a labeled set of $\Ntraining$ samples $\{(\myVec{s}_t, \Input_t) \}_{t=1}^{\Ntraining}$:  

$(i)$ {End-to-end training}: The first approach jointly trains the entire network, i.e., all the building block \acp{dnn}. Since the output of the deep network is the set of \acp{pmf} $\{\hat{\myVec{p}}_k^{(\Niter)}\}_{k=1}^{\Nusers}$,   the sum cross entropy loss is used. Let 
$\myVec{\theta}$ be the network parameters, and 
$\hat{\myVec{p}}_k^{(\Niter)}(\Input, \alpha; \myVec{\theta} )$ be the entry of $\hat{\myVec{p}}_k^{(\Niter)}$ corresponding to $s_k  = \alpha$ when the input to the network parameterized by $\myVec{\theta}$ is $\Input$.
The sum cross entropy loss is 
\begin{equation}
\label{eqn:SumCE}
\mathcal{L} (\myVec{\theta})= \frac{1}{\Ntraining}\sum_{t=1}^{\Ntraining}
\sum_{k=1}^{\Nusers} -\log \hat{\myVec{p}}_k^{(\Niter)}\big(\Input_t, (\myVec{s}_t)_k ; \myVec{\theta}\big). 
\end{equation} 

Training the interconnection of \acp{dnn} in Fig. \ref{fig:SoftIC1}(b) end-to-end based on  \eqref{eqn:SumCE} jointly updates the coefficients of all the $\Nusers \cdot \Niter$ building block \acp{dnn}. For a large number of symbols, i.e., large $\Nusers$, training so many parameters simultaneously is expected to require a large labeled set. %, motivating the sequential training approach. 

$(ii)$ {Sequential training}: The fact that DeepSIC is implemented as an interconnection of neural building blocks, implies that each block can be trained with a reduced number of training samples. Specifically,  the goal of each building block \ac{dnn} does not depend on the iteration index: The $k$th building block of the $q$th iteration outputs a soft estimate of $s_k $ for each iteration $q$. Therefore, each building block \ac{dnn} can be trained individually, by minimizing the conventional cross entropy loss. To formulate this objective, let  
$\myVec{\theta}_{k}^{(q)}$ represent the parameters of the $k$th \ac{dnn} at iteration $q$, and write 
$\hat{\myVec{p}}_k^{(q)}\big(\Input, \{\hat{\myVec{p}}_l^{(q-1)}\}_{l\neq k}, \alpha; \myVec{\theta}_{k}^{(q)}\big)$ as the entry of $\hat{\myVec{p}}_k^{(q)}$ corresponding to $s_k = \alpha$ when the \ac{dnn} parameters are $\myVec{\theta}_{k}^{(q)}$ and its inputs are $\Input$ and  $\{\hat{\myVec{p}}_l^{(q-1)}\}_{l\neq k}$. The cross entropy loss is 
\begin{equation}
\label{eqn:CE}
\hspace{-0.2cm}
\mathcal{L}\big(\myVec{\theta}_{k}^{(q)}\big)\!  =\! \frac{-1}{\Ntraining}\sum_{t=1}^{\Ntraining}
\log \hat{\myVec{p}}_k^{(q)}\big(\tilde{\Input}_t, \{\hat{\myVec{p}}_{t,l}^{(q\!-\!1)}\}_{l\neq k}, (\tilde{\myVec{s}}_t)_k ; \myVec{\theta}_{k}^{(q)}\big)
\end{equation}	
where $\{\hat{\myVec{p}}_{t,l}^{(q-1)}\}$ represent the estimated \acp{pmf} associated with $\Input_i$ computed at the previous iteration.
The problem with training each \ac{dnn} individually is that the soft estimates $\{\hat{\myVec{p}}_{t,l}^{(q-1)}\}$ are not provided as part of the training set. This challenge can be tackled by training the \acp{dnn} corresponding to each layer in a sequential manner, where for each layer the outputs of the trained previous iterations are used as the soft estimates fed as training samples. % This process is summarized below as Algorithm \ref{alg:Algo2}.

\begin{figure*}
	\centering
	\begin{subfigure}{0.42\textwidth}
		\centering
		{\includegraphics[width=\columnwidth]{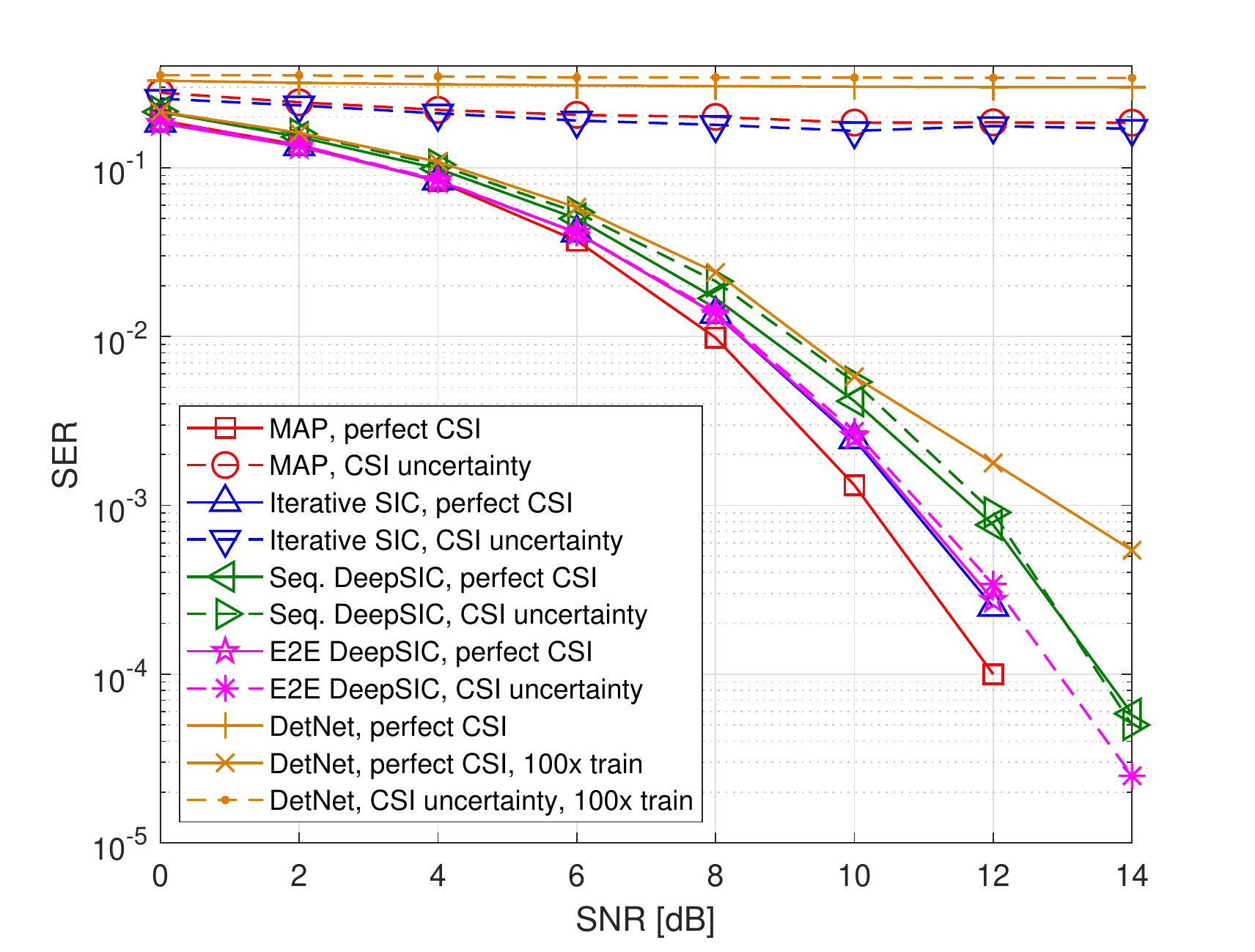}} 
		\caption{ $32\times32$ Gaussian channel.
		}
		\label{fig:AWGN6} 	
	\end{subfigure}
	$\quad$
	\begin{subfigure}{0.42\textwidth}
		\centering
		{\includegraphics[width=\columnwidth]{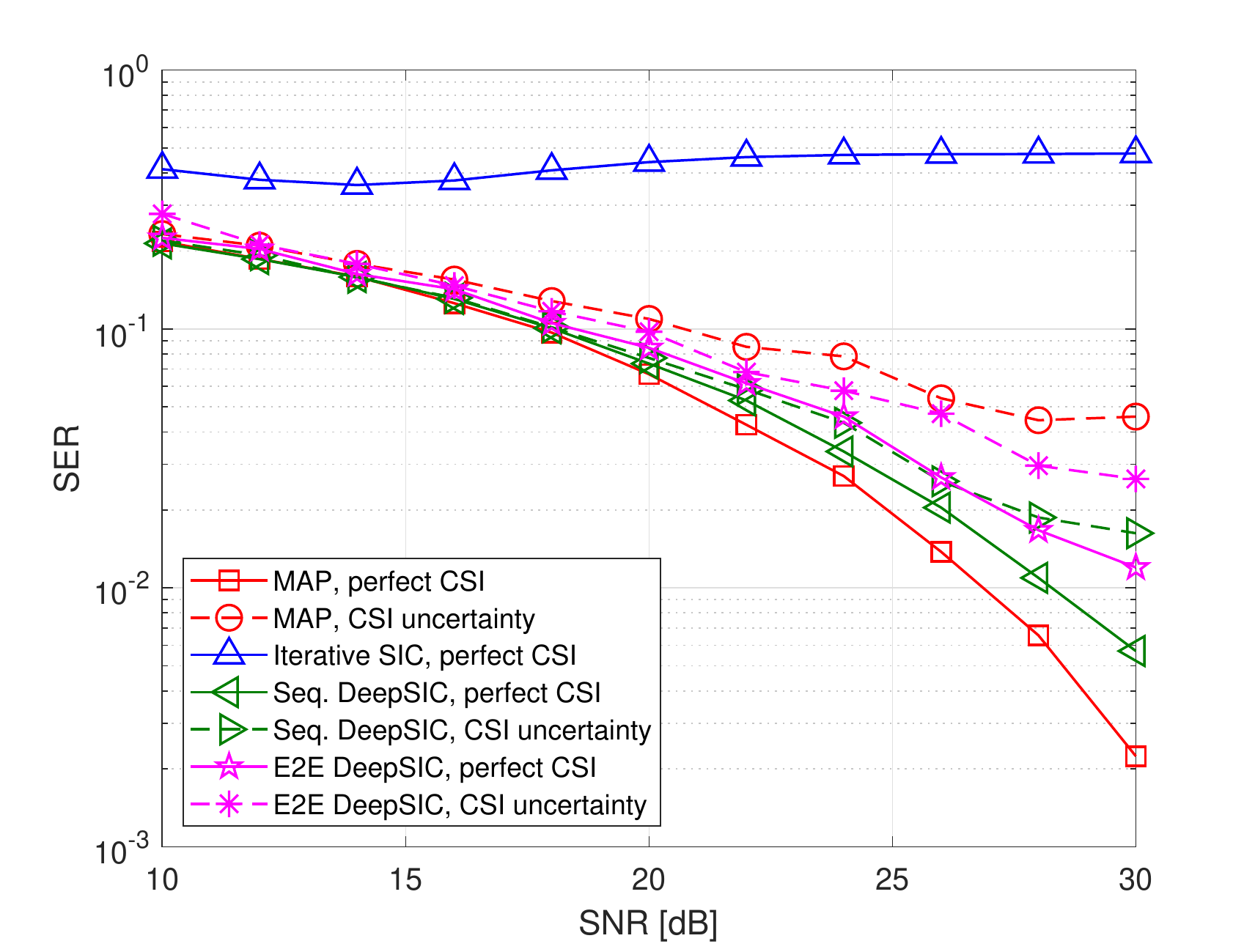}} 
		\caption{$4\times 4$ Poisson channel.
		}
		\label{fig:Poisson4} 
	\end{subfigure}
	\caption{Experimental results from \cite{shlezinger2019deepSIC} of DeepSIC compared to the model-based iterative \ac{sic}, the model-based \ac{map} (when feasible) and the data-driven DetNet of \cite{samuel2019learning} (when applicable). {\em Perfect CSI} implies that the system is trained and tested using
samples from the same channel, while under {\em CSI uncertainty} they are trained using samples from a set of different channels.}
    \label{fig:DeepSICRes}
	%\vspace{-0.2cm}
\end{figure*}

{\bf Quantitative Results:} Two experimental studies of DeepSIC taken from \cite{shlezinger2019deepSIC} are depicted in Fig.~\ref{fig:DeepSICRes}. These results compare the \ac{ser} achieved by DeepSIC which learns to carry out $\Niter=5$ \ac{sic} iterations from $\Ntraining = 5000$ labeled samples. In particular, Fig.~\ref{fig:AWGN6} considers a Gaussian channel of the form \eqref{eqn:Gaussian} with $\Nusers = \Nantennas = 32$, resulting in \ac{map} detection being computationally infeasible, and compares DeepSIC to the model-based iterative \ac{sic} as well as the data-driven DetNet  \cite{samuel2019learning}. Fig.~\ref{fig:Poisson4} considers a Poisson channel, where $\Input$ is related to $\myS$ via a multivariate Poisson distribution, for which schemes requiring a linear Gaussian model such as the iterative \ac{sic} algorithm are not suitable. The ability to use \acp{dnn} as neural building blocks to carry out their model-based algorithmic counterparts in a robust and model-agnostic fashion is demonstrated in  Fig.~\ref{fig:DeepSICRes}. In particular, it is demonstrated that DeepSIC approaches the \ac{ser} values of the iterative \ac{sic} algorithm in linear Gaussian channels, while notably outperforming it in the presence of model mismatch, as well as when applied in non-Gaussian setups. It is also observed in Fig.~\ref{fig:AWGN6} that the resulting architecture of DeepSIC can be trained with smaller data sets compared to alternative data-driven receivers, such as DetNet.

\subsubsection*{Discussion}
The main rationale in designing \acp{dnn} as interconnected neural building blocks is to facilitate learned inference by preserving the structured operation of a model-based algorithm applicable for the problem at hand given full domain knowledge. As discussed earlier, this approach can be treated as an extension of deep unfolding, allowing to exploit additional structures beyond a sequential iterative operation. %For instance, DeepSIC designs its building-block \acp{dnn} to be interconnected in both parallel as well as a sequential manner. 
The generalization of deep unfolding into a set of learned building blocks opens additional possibilities in designing model-aided networks. 

First, the treatment of the model-based algorithm as a set of building blocks with concrete tasks allows a \ac{dnn} architecture designed to comply with this structure not only to learn to carry out the original model-based method from data, but also to robustify it and enable its application in diverse new scenarios. This follows since the block diagram structure of the algorithm may be ignorant of the specific underlying statistical model, and only rely upon a set of generic assumptions, e.g., that the entries of the desired vector $\myS$ are mutually independent. Consequently, replacing these building blocks with dedicated \acp{dnn} allows to exploit their model-agnostic nature, and thus the original algorithm can now be learned to be carried out in complex environments. For instance, DeepSIC can be applied to non-linear channels, owing to the implementation of the building blocks of the iterative \ac{sic} algorithm using generic \acp{dnn}, while the model-based algorithm is limited to setups of the form \eqref{eqn:Gaussian}. % where the contribution of the interference is additive and can be canceled by subtraction. 

In addition, the division into building blocks gives rise to the possibility to train each block separately. 
The main advantage in doing so is that a smaller training set is expected to be required, though in the horizon of a sufficiently large amount of training, end-to-end training is likely to yield a more accurate model as its parameters are jointly optimized. For example, in DeepSIC, sequential training uses the $\Ntraining$ input-output pairs to train each \ac{dnn} individually. Compared to the end-to-end training that utilizes the training samples to learn the complete set of parameters, which can be quite large, sequential training uses the same data set  to learn a significantly smaller number of parameters, reduced by a factor of $\Nusers\cdot \Niter$, multiple times. This indicates that the ability to train the blocks individually is expected to require much fewer training samples, at the cost of a longer learning procedure for a given training set, due to its sequential operation, and possible performance degradation as the building blocks are not jointly trained. 
In addition, training each block separately facilitates adding and removing blocks, when such operations are required in order to adapt the inference rule.

	\section{DNN-Aided Inference}
	\label{sec:Inference}
\begin{figure*}
	    \centering
	    \includegraphics[width=0.7\linewidth]{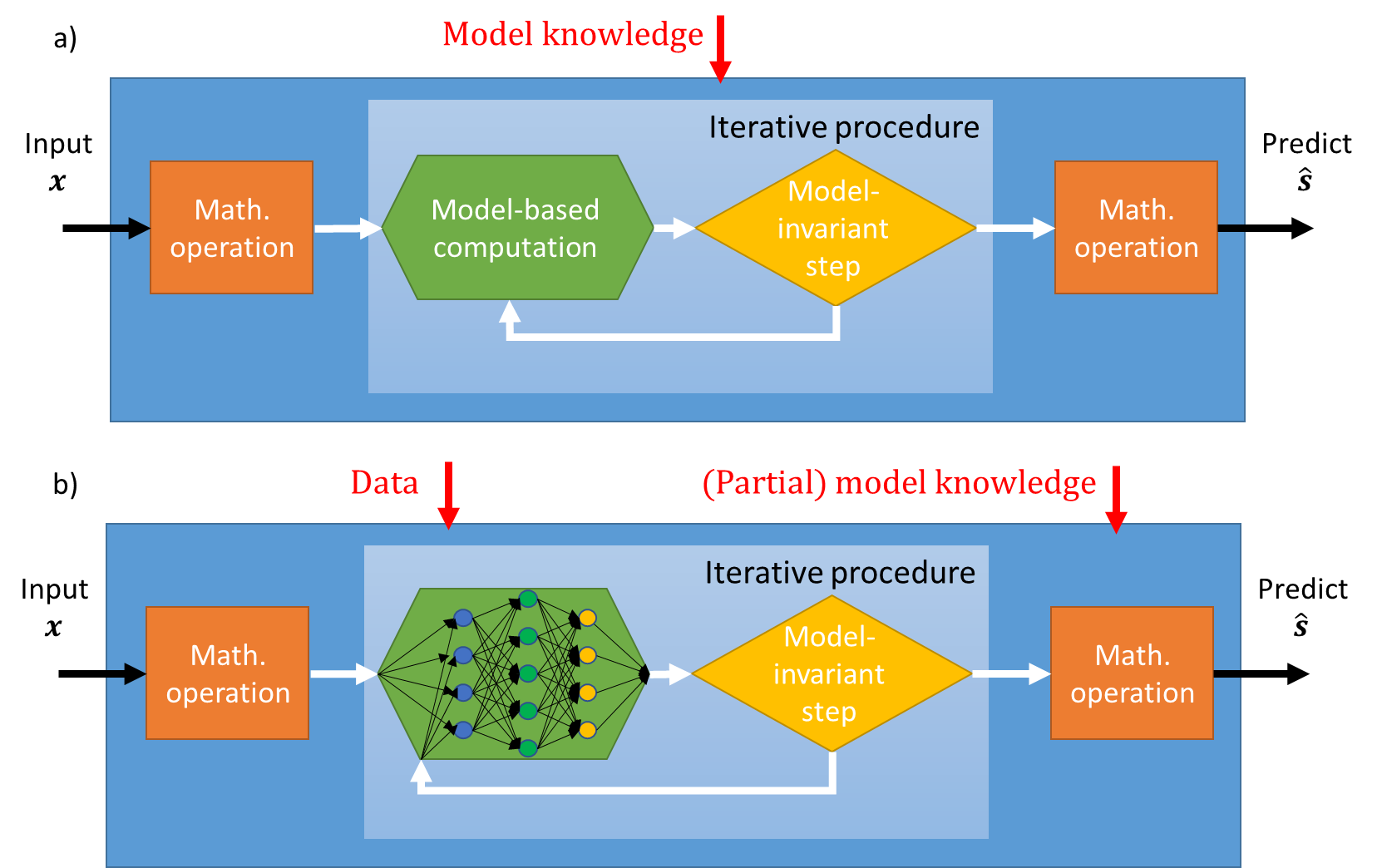}
	    \caption{DNN-aided inference illustration: $a)$ a model-based algorithm comprised of multiple iterations with intermediate model-based computations; $b)$ A data-driven implementation of the algorithm, where the specific model-based computations are replaced with dedicated learned deep models. \textcolor{NewColor}{ Here, one can possibly use data to train the internal \acp{dnn} individually, or to train the overall inference mapping end-to-end as a discriminative learning model~\cite{ng2001discriminative,shlezinger2022discriminative}, typically requiring the intermediate mathematical steps to be either differentiable or well-approximated by a differentiable mapping.}}
	    \label{fig:DNN-Aided1}
	\end{figure*}

	%\vspace{-0.1cm}
    \ac{dnn}-aided inference is a family of model-based deep learning algorithms in which \acp{dnn} are incorporated into model-based methods. As opposed to  model-aided networks  discussed in Section~\ref{sec:Networks}, where the resultant system is a deep network whose architecture imitates the operation of a model-based algorithm, here inference is carried out using a traditional model-based method, while some of the intermediate computations are augmented by \acp{dnn}. 
        The main motivation of \ac{dnn}-aided inference is to exploit the established benefits of model-based methods, in terms of performance, complexity, and suitability for the problem at hand. Deep learning is incorporated to mitigate sensitivity to inaccurate model knowledge, facilitate operation in complex environments, and enable application in new domains. 
    An illustration of a \ac{dnn}-aided inference system %obtained by integrating deep learning into a model-based algorithm 
    is depicted in Fig.~\ref{fig:DNN-Aided1}.

	\ac{dnn}-aided inference is particularly suitable for scenarios in which one only has access to partial domain knowledge. In such cases, the available domain knowledge dictates the algorithm utilized, while the part that is not available or is too complex to model analytically is tackled using deep learning. We divide our description of \ac{dnn}-aided inference schemes into three main families of methods: The first, referred to as {\em structure-agnostic \ac{dnn}-aided inference} detailed in Subsection~\ref{subsec:Agnostic}, utilizes deep learning to capture structures in the underlying data distribution, e.g., to represent the domain of natural images. This \ac{dnn} is then utilized by model-based methods, allowing them to operate in a manner which is invariant to these structures. The family of {\em structure-oriented \ac{dnn}-aided inference} schemes, detailed in Subsection~\ref{subsec:Inference_Oriented}, utilizes model-based algorithms to exploit a known tractable statistical structure, such as an underlying Markovian behavior of the considered signals. In such methods, deep learning is incorporated into the structure-aware algorithm, thereby capturing the remaining portions of the underlying model as well as mitigating sensitivity to uncertainty. Finally, in Subsection~\ref{subsec:Inference_Augmentation}, we discuss {\em neural augmentation} methods, which are tailored to robustify model-based processing in the presence of inaccurate knowledge of the parameters of the underlying model. Here, inference is carried out using a model-based algorithm based on its available domain knowledge, while a deep learning system operating in parallel is utilized to compensate for errors induced by model inaccuracy. \textcolor{NewColor}{Our description of these methodologies in Subsections~\ref{subsec:Agnostic}-\ref{subsec:Inference_Augmentation} follows the same systematic form used in Section~\ref{sec:Networks}, where each approach is detailed by a high-level description; design outline; one or two concrete examples; and a summarizing discussion.}
	
     %----------------------------------------------------------------------------------------
    %	Structure Agnostic DNN-Aided Inference
    %----------------------------------------------------------------------------------------
    % \vspace{-0.2cm}
     \subsection{Structure-Agnostic DNN-Aided Inference}
	\label{subsec:Agnostic}
	% \vspace{-0.1cm}	
	The first family of \ac{dnn}-aided inference utilizes deep learning to implicitly learn structures and statistical properties of the signal of interest, in a manner that is amenable to model-based optimization. These inference systems are particularly relevant for various inverse problems in signal processing, including denoising, sparse recovery, deconvolution, and super resolution \cite{ongie2020deep}. 
	Tackling such problems typically involves imposing some structure on the signal domain. This prior knowledge is then incorporated into a model-based optimization procedure, such as   \ac{admm} \cite{boyd2011distributed},  fast iterative shrinkage and thresholding algorithm \cite{beck2009fast}, and primal-dual splitting \cite{chambolle2011first}, which recover the desired signal with provable performance guarantees. 
	
	Traditionally, the prior knowledge encapsulating the structure and properties of the underlying signal is represented by a handcrafted regularization term or constraint incorporated into the optimization objective. For example, a common model-based strategy used in various inverse problems is to impose sparsity in some given dictionary, which facilitates \ac{cs}-based optimization.  Deep learning brings forth the possibility to avoid such explicit constraint, thereby mitigating the detrimental effects of crude, handcrafted approximation of the true underlying structure of the signal,  % due to the difficulty in providing an analytical formulation of complex domain.
	%Structure-agnostic \ac{dnn} aided inference builds upon the model-agnostic nature of deep learning and its ability to operate reliably in complex and analytically intractable setups, to 
	while enabling optimization with implicit data-driven regularization. This can be implemented by incorporating deep denoisers as learned proximal mappings in iterative optimization, as carried out by plug-and-play networks\footnote{The term {\em plug-and-play} typically refers to the usage of an image denoiser as proximal mapping in regularized optimization \cite{ahmad2020plug}. As this approach can also utilize model-based denoisers, we use the term {\em plug-and-play networks} for such methods with \ac{dnn}-based denoisers.} \cite{venkatakrishnan2013plug, aggarwal2018modl,ahmad2020plug, zhang2017learning,ryu2019plug,ono2017primal,kamilov2017plug, meinhardt2017learning}. \ac{dnn}-based priors can also be used to enable, e.g., \ac{cs} beyond the domain of sparse signals  \cite{bora2017compressed, whang2020compressed}.

		\subsubsection*{Design Outline} %Structure-agnostic \ac{dnn}-aided inference integrates deep learning into model-based optimization in order to faithfully capture complex signal domains without having to manually characterize them. The resulting hybrid model-based/data-driven systems thus carry out model-based optimization over  analytically intractable domains.
		Designing structure-agnostic \ac{dnn}-aided systems can be carried out via the following steps:
		\begin{enumerate}
		    \item Identify a suitable  optimization procedure, given the domain knowledge for the signal of interest. 
		    \item The specific parts of the optimization procedure which rely on complicated and possibly analytically intractable domain knowledge are replaced with a \ac{dnn}. 
		    \item The integrated data-driven module can either be trained separately from the inference system, possibly in an unsupervised manner as in \cite{bora2017compressed}, or alternatively, the complete inference system is trained  end-to-end  \cite{gilton2019neumann}.
		\end{enumerate}
		
		We next demonstrate how these steps are carried out in two examples: \ac{cs} over complicated domains, where deep generative networks are used for capturing the signal domain  \cite{bora2017compressed}; and plug-and-play networks, which augment \ac{admm} with a \ac{dnn} to bypass the need to express a proximal mapping.

	\subsubsection*{Example 4: Compressed Sensing using Generative Models}
   \ac{cs} refers to the task of recovering some unknown signal from (possibly noisy) lower-dimensional observations. The mapping that transforms the input signal into the observations is known as the \textit{forward operator}.  In our example, we focus on the setting where the forward operator is a particular linear function that is known at the time of signal recovery.  
    
%	\jw{Should these sections be written from first-person perspective? In other words, should we say ``We do ....''? A bit confused since these results are from a paper by someone else.} \\
    
    The main challenge in \ac{cs} is that there could be (potentially infinitely) many signals that agree with the given observations.  Since such a problem is underdetermined, it is necessary to make some sort of structural assumptions on the unknown signal to identify the most plausible one.  A classic assumption is that the signal is \textit{sparse} in some known basis.
	
	\paragraph{System Model} We consider the problem of noisy \ac{cs}, where we wish to reconstruct an unknown $N$-dimensional signal $\csSignal^*$ from the following observations
	\begin{equation}
	\label{eqn:CSModel}
	    \csObs = \csMatrix \csSignal^* + \csNoise
	\end{equation}
	where $\csMatrix$ is an $M\times N$ matrix,  modeled as random Gaussian matrix with entries $\csMatrix_{ij} \sim \mathcal{N}(0, 1/M)$, with $M < N$, and $\csNoise$ is an $M\times1$ noise vector.
	
	\paragraph{Sparsity-based \ac{cs}} %Owing to the pioneering result by \cite{candes2006robust,donoho2006compressed},
	\color{NewColor}
	We next focus on a particular technique as a representative example of model-based \ac{cs}. We rely here on the assumption that $\csSignal^*$ is sparse, and seek to recover $\csSignal^*$ from $\csObs$ by solving the $\ell_1$ relaxed LASSO objective
	\begin{equation}
	\label{eqn:Lasso}
	    \csLasso(\csSignal) \triangleq \|\csMatrix\csSignal-\csObs\|_2^2 + \lambda \| \csSignal \|_{1}.
	\end{equation}
	While the derivation above assumes that $\csSignal^*$ is sparse, the LASSO objective can also be used when $\csSignal^*$ is sparse in some dictionary $\myMat{B}$, e.g., in the wavelet domain, and the detailed formulation is given in Appendix~\ref{app:CS_Exm}.
	
	\color{black}	
	
	\paragraph{\ac{dnn}-Aided Compressed Sensing} In a data-driven approach, we aim to replace the sparsity prior with a learned \ac{dnn}.  The following description is based on \cite{bora2017compressed}, which proposed to use a deep generative prior. Specifically, we replace the explicit sparsity assumption on true signal $\csSignal^*$, with a requirement that it lies in the range of a pre-trained generator network $G: \mathbb{R}^l \to \mathbb{R}^N$ (e.g., the generator network of a \ac{gan}). 
	
	{\bf Pre-training: } To implement deep generative priors, one first has to train a generative network $G$ to map a latent vector $\csLatent$  into a signal $\csSignal$ which lies in the domain of interest.  %These generative models are trained in advance using unlabeled data representing the domain of signals one wishes to capture.  	
	A major advantage of employing a \ac{dnn}-based prior in this setting is that  generator networks are agnostic to how they are used and can be \textit{pre-trained} and reused for multiple downstream tasks.  The pre-training thus follows the standard unsupervised training procedure, as discussed, e.g., in Subsection \ref{subsec:DL_Tasks} for \acp{gan}.
	In particular, the work \cite{bora2017compressed} trained a  Deep convolutional GAN \cite{radford2015unsupervised} on the CelebA data set \cite{liu2015faceattributes}, to represent $64\times 64$ color images of human faces, as well as a variational autoencoder (VAE) \cite{kingma2013auto} for representing handwritten digits in $28\times 28$ grayscale form based on the MNIST data set \cite{lecun2010mnist}. 
% 	, We consider using two different types of generator network $G: \csLatent \mapsto \csSignal$, one for each data set.
% 	The first is DCGAN \cite{radford2015unsupervised}, a particular type of GAN designed to work well with image data.  We train a DCGAN on the CelebA data set \cite{liu2015faceattributes}, which consists of $64 \times 64$ color images of human faces.
% 	The second model is a variational autoencoder (VAE) \cite{kingma2013auto}, another type of deep generative model that produces a generator network similar to that of a GAN. We train a VAE on the MNIST data set \cite{lecun2010mnist}, which contains $28 \times 28$ grayscale images of handwritten digits.
% 	Also we assume a Gaussian prior $\mySet{N}(\myVec{0},\myMat{I}_n)$ over latent variable $\csLatent$ for both models.
%	
%	Such domains are extremely complex to represent analytically. For instance,the generator networks for MNIST and CelebA need to capture highly multimodal distributions of dimensions $28^2 = 784$ and $64^2 \cdot 3 = 12288$, respectively. At this scale, manually creating a handcrafted model often becomes prohibitively expensive, if not impossible.
	
    \begin{figure}
    	\centering
    	\includegraphics[width=\columnwidth]{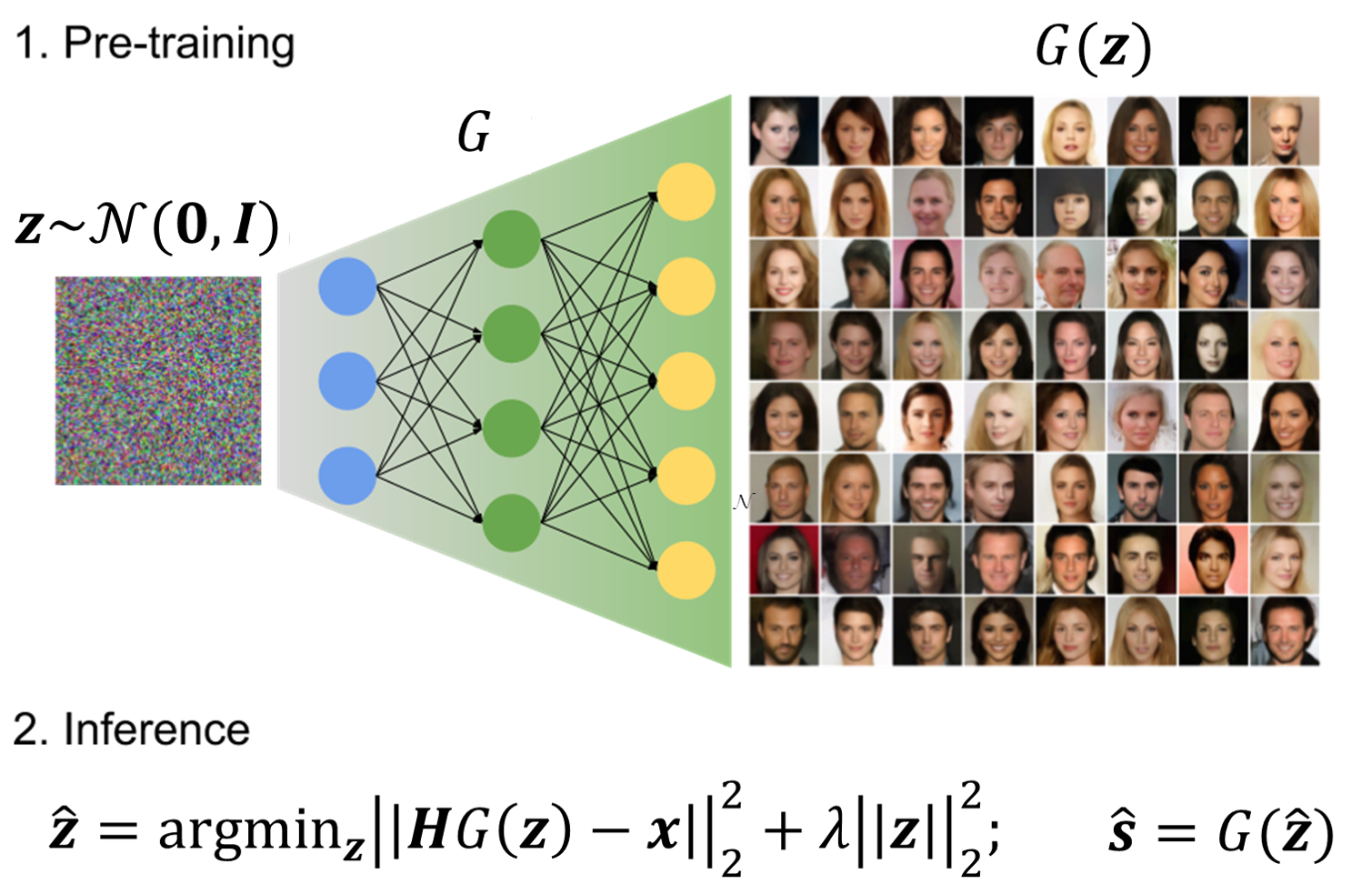}
    	\caption{High-level overview of \ac{cs} with a \ac{dnn}-based prior. The generator network $G$ is pre-trained to map Gaussian latent variables to plausible signals in the target domain. Then signal recovery is done by finding a point in the range of $G$ that minimizes reconstruction error via gradient-based optimization over the latent variable.}
    	%\\    	\jw{Diagram link: https://docs.google.com/drawings/d/1RpO7bxb4XFV1UDB0iH3mX7iQI6ainDnLdGPPj-bUfZQ/edit}
    	\label{fig:CSGM_flow}	 
    \end{figure}

	{\bf Architecture:} Once a pre-trained generator network $G$ is available, it can be incorporated as an alternative prior for the inverse  model in \eqref{eqn:CSModel}. The key intuition behind this approach is that the range of $G$ should only contain \textit{plausible} signals. Thus one can replace the handcrafted sparsity prior with a data-driven \ac{dnn} prior $G$ by constraining our signal recovery to the range of $G$. 
	
	One natural way to impose this constraint is to perform the optimization in the latent space to find $\csLatent$ whose image $G(\csLatent)$ matches the observations. This is carried out by minimizing the following loss function in the latent space of $G$:
	\begin{equation}
	    \mySet{L}(\csLatent) = \| \csMatrix G(\csLatent) - \csObs \|_2^2.
	\end{equation}
	Because the above loss function involves a highly non-convex function $G$, there is no closed-form solution or guarantee for this optimization problem. However the loss function is differentiable with respect to $\csLatent$, so it can be tackled using conventional gradient-based optimization techniques. Once a suitable latent  $\csLatent$ is found, the signal is recovered as $G(\csLatent)$.
	
	In practice, \cite{bora2017compressed} reports that incorporating an $\ell_2$ regularizer on $\csLatent$ helps.  This is possibly due to the Gaussian prior assumption for the latent variable, as the density of $\csLatent$ is proportional to $\exp(-\|\csLatent\|_2^2)$. Therefore, minimizing $\|\csLatent\|_2^2$ is equivalent to maximizing the density of $\csLatent$ under the Gaussian prior. This has the effect of avoiding images that are extremely unlikely under the Gaussian prior even if it matches the observation well. The final loss includes this regularization term:
	\begin{equation}
	    \csLoss(\csLatent) = \| \csMatrix G(\csLatent) - \csObs \|_2^2 + \lambda \|\csLatent\|_2^2
	\end{equation}
	where $\lambda$ is a regularization coefficient.
	
	In summary, \ac{dnn}-aided \ac{cs} replaces the constrained optimization over the complex input signal with tractable optimization over the latent variable $\csLatent$, which follows a known simple distribution. This is achieved using a pre-trained \ac{dnn}-based prior $G$ to map it into the domain of interest. Inference is performed by minimizing $\csLoss$ in the latent space of $G$. {An illustration of the system operation is depicted in Fig.~\ref{fig:CSGM_flow}}.
	
	{\bf Quantitative Results:}
	To showcase the efficacy of the data-driven prior at capturing complex high-dimensional signal domains, we present the evaluation of its performance as reported in \cite{bora2017compressed}. The baseline model used for comparison is based on directly solving the LASSO loss \eqref{eqn:Lasso}. For CelebA, we formulate the LASSO objective in the discrete cosine transform (DCT) and the wavelet (WVT) basis, and minimize it via coordinate descent.
	
\begin{figure}
    	\centering
    	\includegraphics[width=0.9\columnwidth]{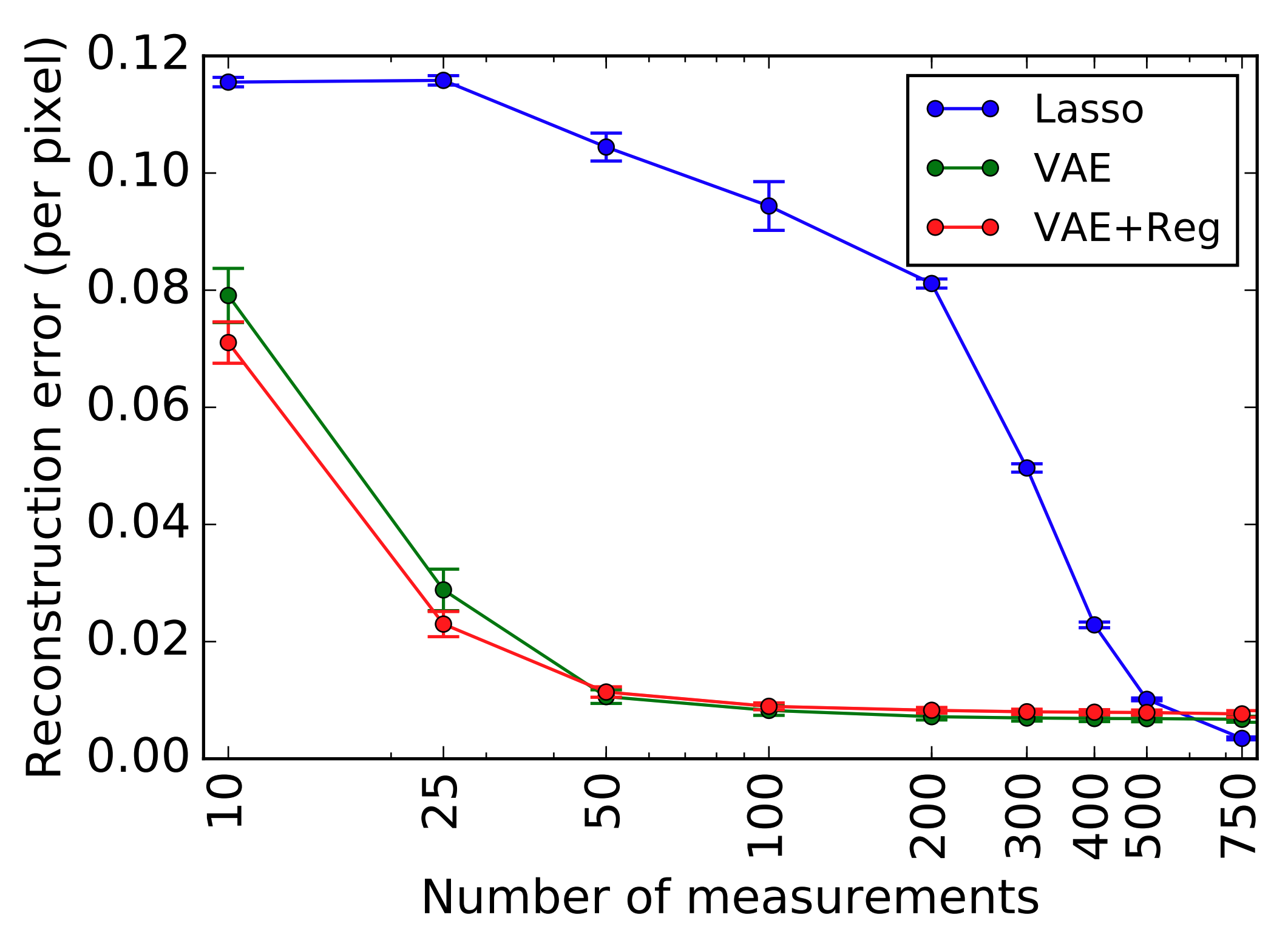} 
    	\caption{Experimental result  for noisy \ac{cs} on the MNIST data set.  Reproduced from \cite{bora2017compressed} with the authors' permission.}
    	\label{fig:CSGM_MNIST}	 
    \end{figure}

	The first task is the recovery of handwritten digit images from low-dimensional projections corrupted by additive  Gaussian noise. The reconstruction error is evaluated for various numbers of observations $M$. The results are depicted in Fig.~\ref{fig:CSGM_MNIST}.
    We  clearly see the benefit of using a data-driven deep prior in Fig. \ref{fig:CSGM_MNIST}, where the VAE-based methods (labeled \textsc{VAE} and \textsc{VAE+Reg}) show notable performance gain compared to the sparsity prior for small number of measurements. Implicitly imposing a sparsity prior via the LASSO objective outperforms the deep generative priors as the number of observations approaches the dimension of the signal. One explanation for this behavior is that the pre-trained generator $G$ does not perfectly model the MNIST digit distribution and may not actually contain the ground truth signal in its range. As such, its reconstruction error may never be exactly zero regardless of how many observations are given.  The LASSO objective, on the other hand, does not suffer from this issue and is able to make use of the extra observations available.
	
\begin{figure}
    	\centering
    	\includegraphics[width=0.9\columnwidth]{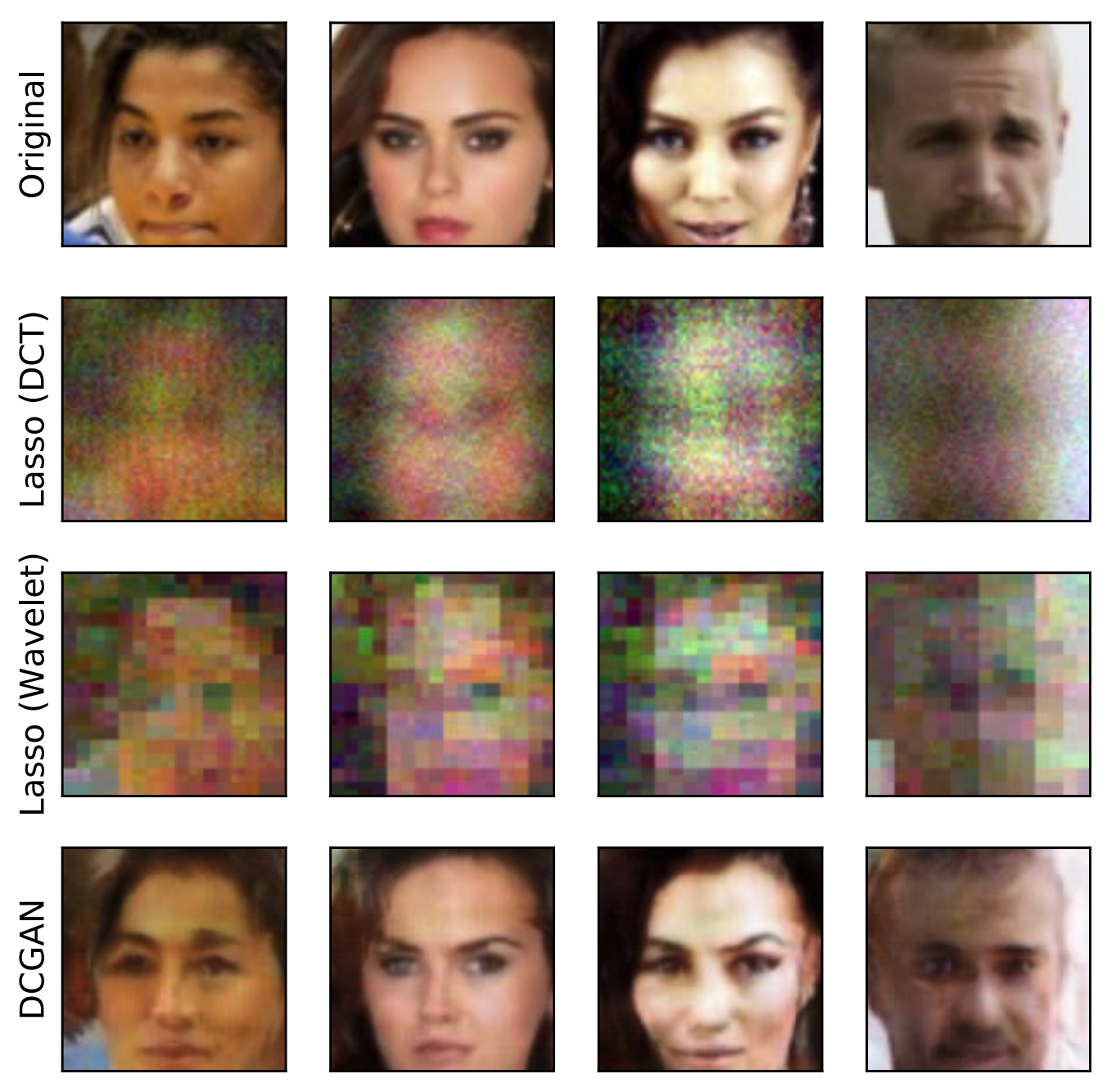} 
    	\caption{Visualization of the recovered signals from noisy \ac{cs} on the CelebA data set.  Reproduced from from \cite{bora2017compressed} with the authors' permission.}
    	\label{fig:CSGM_CelebA}	 
    \end{figure}

	The ability of deep generative priors to facilitate recovery from compressed measurements is also observed in Fig.~\ref{fig:CSGM_CelebA}, which  qualitatively evaluates  GAN-based \ac{cs} recovery on the CelebA data set. This experiment uses $M=500$ noisy measurements (out of $N=12288$ total dimensions). As shown in Fig. \ref{fig:CSGM_CelebA}, in this low-measurement regime, the data-driven prior again provides much more reasonable samples.
	
    \setcounter{paragraph}{0}
	\subsubsection*{Example 5: Plug-and-Play Networks for Image Restoration}
	The above example of \ac{dnn}-aided \ac{cs} allows to carry out regularized optimization over complex domains while using deep learning to avoid regularizing explicitly. This is achieved via deep priors, where the domain of interest is captured by a generative network. An alternative strategy, referred to as plug-and-play networks, applies deep denoisers as learned proximal mappings. Namely, instead of using \acp{dnn} to  evaluate the regularized objective as in \cite{bora2017compressed}, one uses \acp{dnn} to carry out an optimization procedure which relies on this objective without having to express the desired signal domain. In the following we exemplify the application of plug-and-play networks for image restoration using \ac{admm} optimization \cite{ahmad2020plug}.
	
	\paragraph{System Model}
	We again consider the linear inverse problem formulated in \eqref{eqn:CSModel} in which the additive noise $\myVec{w}$ is comprised of i.i.d. mutually independent Gaussian entries with zero mean and variance $\SigW$. However, unlike the setup considered in the previous example, the sensing matrix $\csMatrix$ is not assumed to be random, and can be any fixed matrix dictated by the underlying setup. 
	
	The recovery of the desired signal $\myVec{s}$ can be obtained via the \ac{map} rule, which is given by
	\begin{align}
	    \hat{\myVec{s}} &= \mathop{\arg\min}\limits_{\myVec{s}} -\log p(\myVec{s}|\myVec{x}) \notag \\
	    &= \mathop{\arg\min}\limits_{\myVec{s}} -\log p(\myVec{x}|\myVec{s}) -\log p(\myVec{s}) \notag \\
	    &= \mathop{\arg\min}\limits_{\myVec{s}} \frac{1}{2}\|\myVec{x}-\csMatrix \myVec{s}\|^2 +   \phi(\myVec{s})
	    \label{eqn:RegOpt}
	\end{align}
	where $\phi(\myVec{s})$ is a regularization term which equals $- \SigW\log p(\myVec{s})$, with possibly some additive constant that does not affect the minimization in \eqref{eqn:RegOpt}. 
	
\begin{figure*}
	    \centering
	    \includegraphics[width=0.75\linewidth]{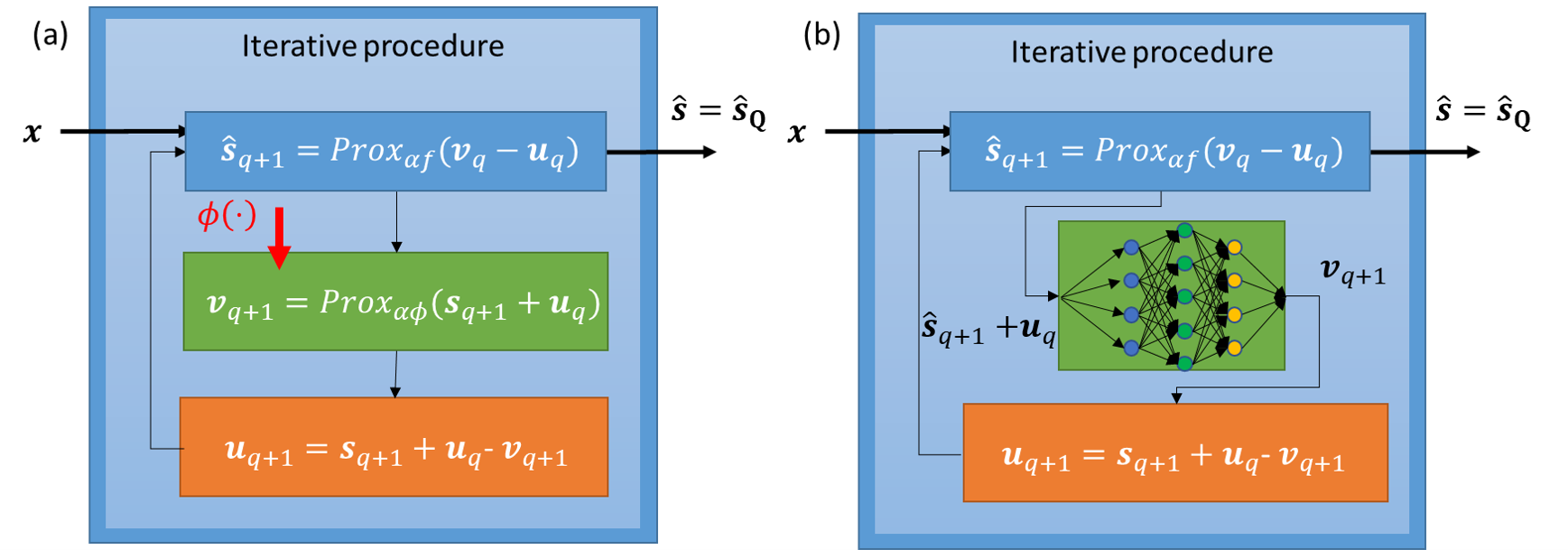}
	    \caption{Illustration of $(a)$ \ac{admm} algorithm compared to $(b)$ plug-and-play \ac{admm} network.  }
	    \label{fig:PnPADMM1}
	\end{figure*}

	\paragraph{Alternating Direction Method of Multipliers}
	\color{NewColor}
	The regularized optimization problem which stems from the \ac{map} rule in \eqref{eqn:RegOpt} can be solved using  \ac{admm} \cite{boyd2011distributed}. \ac{admm} introduces two auxiliary variables, denoted $\myVec{v}$ and $\myVec{u}$, and is given by the iterative procedure in Algorithm~\ref{alg:Algoadmm}, whose derivation is detailed in Appendix~\ref{app:ADMM_Exm}. In Step~\ref{stp:datamatch}, we defined $f(\myVec{v}) \triangleq \frac{1}{2}\|\myVec{x}-\csMatrix \myVec{v}\|^2 $, while the
	proximal mapping of some function $g(\cdot)$ used in Steps~\ref{stp:datamatch}-\ref{stp:prox} is defined as 
	\begin{equation}
		 \label{eqn:proxi}
		 {\rm prox}_g(\myVec{v}):= \arg\min_{\myVec{z}} \left( g(\myVec{z})+\frac{1}{2}\|\myVec{z}-\myVec{v}\|^2_2 \right).
	\end{equation}
	The \ac{admm} algorithm is illustrated in Fig.~\ref{fig:PnPADMM1}(a).
	
	\begin{algorithm}  
		\caption{\ac{admm}}
		\label{alg:Algoadmm}
		\KwData{Fix   $\alpha>0$. Initialize  $\myVec{u}^{(0)}$,  $\myVec{v}^{(0)}$ randomly} 
		\For{$q=0,1,\ldots$}{
	        Update  $ \hat{\myVec{s}}_{q+1} ={\rm prox}_{\alpha f}(\myVec{v}_{q} + \myVec{u}_{q})$.\label{stp:datamatch}\\
	    Update $
	    \myVec{v}_{q+1}  
	    =  {\rm prox}_{\alpha\phi}(\myVec{s}_{q+1} + \myVec{u}_{q})$. \label{stp:prox}\\
	    Update 	   $\myVec{u}_{q+1} = \myVec{u}_{q} +(\hat{\myVec{s}}_{q+1} - \myVec{v}_{q+1})$.
		}
		\KwOut{Estimate $\hat{\myVec{s}} = \hat{\myVec{s}}_q$.}
\end{algorithm}

	\paragraph{Plug-and-Play \ac{admm}} 
	The key challenge in implementing the \ac{admm} iterations stems from the computation of the proximal mapping in Step~\ref{stp:prox}. In particular, while one can evaluate Step~\ref{stp:datamatch} in closed-form, as shown in Appendix~\ref{app:ADMM_Exm},  computing Step~\ref{stp:prox} of Algorithm~\ref{alg:Algoadmm} requires explicit knowledge of the prior $\phi(\cdot)$, which is often not available. Furthermore, even when one has a good approximation of $\phi(\cdot)$, computing the proximal mapping in Step~\ref{stp:prox} may still be extremely challenging to carry out analytically.

		However, the proximal mapping in Step~\ref{stp:prox} of Algorithm~\ref{alg:Algoadmm} is invariant of the task and the data. In particular, it is the solution to the problem of \ac{map} denoising $\hat{\myVec{s}}_{q+1} + \myVec{u}_q$ assuming the noise-free signal has  prior $\phi(\cdot)$ and the noise is Gaussian with variance $\alpha$. Now, denoisers are common \ac{dnn} models, and are known to operate reliably on signal domains with intractable priors (e.g., natural images) \cite{zhang2017learning}. One can thus implement \ac{admm} optimization without having to specify the prior $\phi(\cdot)$ by replacing Step~\ref{stp:prox} of Algorithm~\ref{alg:Algoadmm} with a \ac{dnn} denoiser \cite{ahmad2020plug}, as illustrated in Fig.~\ref{fig:PnPADMM1}. Specifically, the proximal mapping is replaced with a \ac{dnn}-based denoiser $\dnnFunc$, such that
		\begin{equation}
		     \label{eqn:admm2bNew}
	    \myVec{v}_{q+1} =\dnnFunc\left(\hat{\myVec{s}}_{q+1} + \myVec{u}_q; \alpha_q \right)
		\end{equation}
		where $\alpha_q$ denotes the noise level to which the denoiser is tuned. This noise level can either be fixed to represent that used during training, or alternatively, one can use flexible \ac{dnn}-based denoiser in which, e.g., the noise level is provided as an additional input \cite{zhang2018ffdnet}.
	
\color{black}	
%	{\bf Architecture:}
	
\begin{figure}
	    \centering
	    \includegraphics[width=\columnwidth]{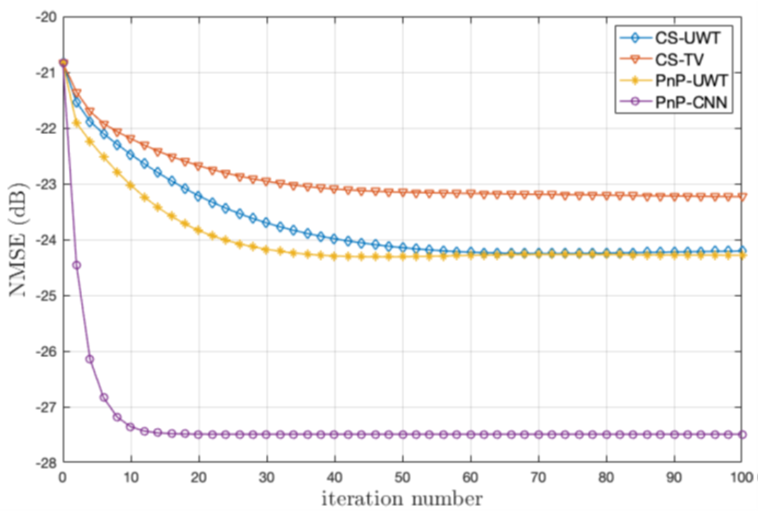}
	    \caption{Normalized \ac{mse} versus iteration for the recovery of cardiac MRI images. Here, plug-and-play networks using a \ac{cnn} denoiser (PnP-CNN) is compared to the model-based strategies of computing the proximal mapping by imposing as prior sparsity in the undecimated wavelet domain (PnP-UWT), as well as \ac{cs} with a similar constraint (CS-UWT) and with total-variation prior   (CS-TV). {Figure reproduced from \cite{ahmad2020plug} with authors' permission}.}
	    \label{fig:PnPMRI1}
	\end{figure}

	{\bf Quantitative Results:} 
	As an illustrative example of the quantitative gains on plug-and-play networks we consider the setup of cardiac \acl{mri} image reconstruction reported in \cite{ahmad2020plug}. The proximal mapping here is replaced with a five-layer \ac{cnn} with residual connection operating on spatiotemporal volumetric patches. The \ac{cnn} is trained offline to denoise clean images manually corrupted by Gaussian noise. The experimental results reported in Fig.~\ref{fig:PnPMRI1} demonstrate that the introduction of deep denoisers notably improves both the performance and the convergence rate of the iterative optimizer compared to utilizing model-based approaches for approximating the proximal mapping. 
	
\color{black}
	\subsubsection*{Discussion}	 Using deep learning to strengthen regularized optimization builds upon the model-agnostic nature of \acp{dnn}. Traditional optimization methods rely on mathematical expressions to capture the structure of the solution one is looking for, inevitably inducing model mismatch in domains which are extremely challenging to describe analytically. The ability of deep learning to learn complex mappings without relying on domain knowledge is exploited here to bypass the need for explicit regularization. %This application of \acp{dnn} can take the form of deep priors, where the domain of interest is captured by a generative network, as detailed in the above example. An alternative strategy is to apply deep denoisers as learned proximal mappings, as done in plug-and-play networks \cite{ahmad2020plug}. This approach avoids the need to mathematically formulate a proximal operator which faithfully matches the problem at hand, while relying on the qualities of established \ac{dnn} denoisers, such as  DnCNN  \cite{zhang2017beyond}.
	The need to learn to capture the domain of interest facilitates using pre-trained networks, thus reducing the dependency on massive amounts of labeled data. For instance,   deep generative priors  use \ac{dnn} architectures that are trained in an unsupervised manner, and thus rely only on unlabeled data, e.g., natural images. Such unlabeled samples are typically more accessible and easy to aggregate compared to labeled data. e.g., tagged natural images.  One can often utilize off-the-shelf pre-trained \acp{dnn} when such network exist for domains related to the ones over which optimization is carried out, with possible adjustments to account for the subtleties of the problem   by transfer learning. 
	
	Finally, while our description of \ac{dnn}-aided regularized optimization relies on model-based iterative optimizers which utilize a deep learning module, one can also incorporate deep learning into the optimization procedure. For instance, the iterative optimization steps can be unfolded into a \ac{dnn}, as in, e.g., \cite{gilton2019neumann}. This approach allows to benefit from both the ability of deep learning to implicitly represent complex domains, as well as the inference speed reduction  of deep unfolding along with its robustness to uncertainty and errors in the model parameters assumed to be known. Nonetheless, the fact that the iterative optimization must be learned from data in addition to the structure of the domain of interest implies that larger amounts of labeled data are required to train the system, compared to using the model-based optimizer.

% TODO NIR CONTINUE FROM HERE
     %----------------------------------------------------------------------------------------
    %	Structure-Oriented DNN-Aided Inference
    %----------------------------------------------------------------------------------------
    % \vspace{-0.2cm}
     \subsection{Structure-Oriented DNN-Aided Inference}
	\label{subsec:Inference_Oriented}
	% \vspace{-0.1cm}	
	The family of structure-oriented \ac{dnn}-aided inference algorithms utilize model-based methods designed to exploit an underlying statistical structure, while integrating \acp{dnn} to enable operation without additional explicit characterization of this model. The types of structures exploited in the literature can come in the form of an a-priori known factorizable distribution, such as causality and finite memory in communication channels \cite{shlezinger2019viterbinet, shlezinger2020data,farsad2020data}; it can follow from an established approximation of the statistical behavior, such as  modelling of images as conditional random fields \cite{arnab2018conditional,chandra2016fast, knobelreiter2020belief}; follow from physical knowledge of the system operation \cite{luijten2020adaptive, revach2022kalmannet, escoriza2021data}; or arise due to the distributed nature of the problem, as in   \cite{palangi2016distributed}. 
	
	The main advantage in accounting for such statistical structures stems from the availability of various model-based methods, tailored specifically to exploit these structures to facilitate accurate inference at reduced complexity. Many of these algorithms, such as the Kalman filter and its variants \cite[Ch. 7]{haykin2005adaptive}, which build upon an underlying state-space structure, or the Viterbi algorithm \cite{viterbi1967error}, which exploits the presence of a hidden Markov model, can be represented as special cases of the broad family of factor graph methods.  Consequently, our main example used for describing structure-oriented \ac{dnn}-aided inference focuses on the implementation of message passing over data-driven factor graphs. %, which encapsulate factorizable distributions. 

\subsubsection*{Design Outline} 
Structure-oriented \ac{dnn}-aided algorithms 
%exploit domain knowledge of statistical structures to carry out model-based inference methods in a data-driven fashion. These hybrid systems thus 
utilize deep learning not for the overall inference task, but for robustifying and relaxing the model-dependence of established model-based inference algorithms designed specifically for the structure induced by the specific  problem being solved. The design of such \ac{dnn}-aided hybrid inference systems consists of the following steps:
\begin{enumerate}
	\item A proper inference algorithm is chosen based on the available knowledge of the underlying statistical structure. The domain knowledge is  encapsulated in the selection of the algorithm which is learned from data. %For example, when the underlying distribution is known to take a factorizable form, accurate inference at linear complexity can be obtained by applying the \ac{sp} algorithm over a factor graph representation of the distribution. 
	\item Once a model-based algorithm is selected, we identify its model-specific computations, and replace them with dedicated compact \acp{dnn}.
	\item The resulting \acp{dnn} are either trained individually, or the overall system can be trained in an end-to-end manner.

\end{enumerate}

We next demonstrate how these steps are translated in a hybrid model-based/data-driven algorithm, using the example of learned factor graph inference for Markovian sequences proposed in \cite{shlezinger2020data,shlezinger2020inference}.

\subsubsection*{Example 6: Learned Factor Graphs} 
Factor graph methods, such as the \ac{sp} algorithm, exploit the factorization of a joint distribution to efficiently compute a desired quantity \cite{kschischang2001factor}. The application of the \ac{sp} algorithm for distributions which can be represented as non-cyclic factor graphs, such as Markovian models, allows computing the \ac{map} rule, an operation whose burden typically grows exponentially with the label space dimensionality, with complexity that only grows linearly with it. While the following description focuses on Markovian stationary time sequences, it can be extended to various forms of factorizable distributions.

\paragraph{System Model}
We consider the recovery of a  time series $\{s_i\}$ taking values in a finite set $\mySet{S}$ from an observed sequence $\{x_i\}$ taking values in a set $\mySet{X}$. The subscript $i$ denotes the time index. The joint distribution of $\{s_i\}$ and $\{x_i\}$ obeys an $\Mem$th-order Markovian stationary model, $\Mem \geq 1$.
\color{NewColor}
Consequently, when the initial state $\{s_i\}_{i=-\Mem}^{0}$ is given,  the joint distribution of  $\Input = [x_1, \ldots, x_\Blklen]^T$ and $\myVec{s}  = [s_1, \ldots, s_\Blklen]^T$  satisfies 
\begin{equation}
\label{eqn:MarkovModel}
\PdfNew{\myVec{X},\myVec{S}}(\Input,\myVec{s}) \!=\! %\Pdf{\myVec{X}^\Mem,\myVec{S}^\Mem}(\Input^\Mem,\myVec{s}^\Mem)
\prod_{i=1}^{\Blklen} \PdfNew{Y_i | \myVec{S}_{i-\Mem}^{i}}\!\left(x_i|\myVec{s}_{i-\Mem}^{i}\right)\! \PdfNew{S_i |  \myVec{S}_{i-\Mem}^{i-1}}\!\left(s_i | \myVec{s}_{i-\Mem}^{i-1}\right)
%	 \Pdf{Y_i, S_i | \myVec{X}_{i-\Mem}^{i-1}, \myVec{S}_{i-\Mem}^{i-1}}\left(x_i, s_i |\Input_{i-\Mem}^{i-1}, \myVec{s}_{i-\Mem}^{i-1}\right). %, \quad \forall \myVec{s}^\Blklen \in \mySet{S}^\Blklen, \Input^\Blklen\in\mySet{Y}^\Blklen.
\end{equation}
for any fixed sequence length $\Blklen >0$, where we write $\myVec{s}_i^j \triangleq [s_i, s_{i+1},\ldots,s_{j}]^T$ for $i <j$.

\paragraph{The Sum-Product Algorithm}
When the joint distribution of $\myVec{s}$ and $\Input$ is a-priori known and can be computed, the  inference rule that minimizes the error probability for each time instance is the \ac{map} detector,
\begin{align}
\hat{s}_i\left( \Input\right)  &= \mathop{\arg \max}\limits_{s_i \in \mySet{S}}\PdfNew{S_i|\myVec{X}}(s_i|\Input)
\label{eqn:MAP0}
\end{align}
for each $i\in\{1,\ldots,\Blklen\} \triangleq  \Blkset$.
This rule can be efficiently approached when \eqref{eqn:MarkovModel} holds using the \ac{sp} algorithm \cite{kschischang2001factor}. The \ac{sp} algorithm represents the joint distribution~\eqref{eqn:MarkovModel} and computes the posterior distribution by message passing over this graph, as illustrated in Fig.~\ref{fig:SumProduct2}(a). The resulting procedure, detailed further in Appendix~\ref{app:SP_Exm}, is summarized as Algorithm~\ref{alg:SP}, where we define  $\myVec{s}_i \triangleq \myVec{s}_{i-\Mem+1}^i \in \mySet{S}^\Mem$, and the function
\begin{align}
f\left(x_i, \myVec{s}_i, \myVec{s}_{i-1} \right) &\triangleq
\PdfNew{Y_i | \myVec{S}_{i-\Mem+1}^{i}, \myVec{S}_{i-\Mem}^{i-1}}\left(x_i|\myVec{s}_{i}, \myVec{s}_{i-1}\right)   \PdfNew{\myVec{S}_{i-\Mem+1}^{i} |  \myVec{S}_{i-\Mem}^{i-1}}\left(\myVec{s}_{i}| \myVec{s}_{i-1}\right). 
\label{eqn:FSC_funcNode}
\end{align}
Algorithm~\ref{alg:SP} approaches the \ac{map} detector in \eqref{eqn:MAP0} with complexity that only grows linearly with $\Blklen$.

\begin{algorithm}  
	\caption{ The \ac{sp} algorithm for system model~\eqref{eqn:MarkovModel}}
	\label{alg:SP}
	\KwData{Fix an initial forward message $\FwdMsg{f_{0}}{\myState_{0}}(\myStateR)= 1$ and a final backward message $\BwdMsg{f_{0}}{\myState_{\Blklen}}(\myStateR)\equiv 1$. } 
	\For{$i=\Blklen-1,\Blklen-2,\ldots,1$}{
		For each  $\myStateR_i \in \mySet{S}^\Mem$, compute backward message 
		\begin{equation*}
		    \BwdMsg{f_{i\!+\!1}}{ \myState_i}(\myStateR_i) = \sum_{\myStateR_{i\!+\!1}} f(x_{i\!+\!1}, \myStateR_{i\!+\!1}, \myStateR_{i})\BwdMsg{f_{i\!+\!2} }{ \myState_{i\!+\!1}}(\myStateR_{i\!+\!1}).
		\end{equation*}
            \label{stp:Recursion1Backwards}
	}
	\For{$i=1,2,\ldots,\Blklen$}{
		For each  $\myStateR_i \in \mySet{S}^\Mem$, compute forward message 
		\begin{equation*}
		    \FwdMsg{f_i}{\myState_i}(\myStateR_i) = \sum_{\myStateR_{i-1}} f(x_{i},  \myStateR_{i}, \myStateR_{i-1})\FwdMsg{f_{i-1}}{\myState_{i-1}}(\myStateR_{i-1}).
		\end{equation*}
		\label{stp:Recursion1Forwards} \\
		Estimate 
		\begin{align*}
    \hat{s}_i   
    \!=\!\mathop{\arg \max}\limits_{s_i \in \mySet{S}}\!\! \sum_{  \myVec{s}_{i\! - \!1}\in \mySet{S}^{\Mem}}& \FwdMsg{f_{i\! - \!1}}{\myVec{s}_{i\! - \!1}}(\myVec{s}_{i\! - \!1}) f(x_{i},  [s_{i\! - \!\Mem\!+\!1}, \ldots, s_{i} ],\myVec{s}_{i\! - \!1})  \notag \\
    &\times \BwdMsg{f_{i+1}}{\myVec{s}_{i}}([s_{i\! - \!\Mem\!+\!1}, \ldots, s_{i} ]).
    \end{align*}
	}
	\KwOut{$\hat{\myVec{s}}^\Blklen = [\hat{s}_1, \ldots, \hat{s}_\Blklen]^T$ %\tcp*{\ac{map} inference}
	}
\end{algorithm}

\color{black}

\begin{figure*}
	\centering
	\includegraphics[width=0.7\linewidth]{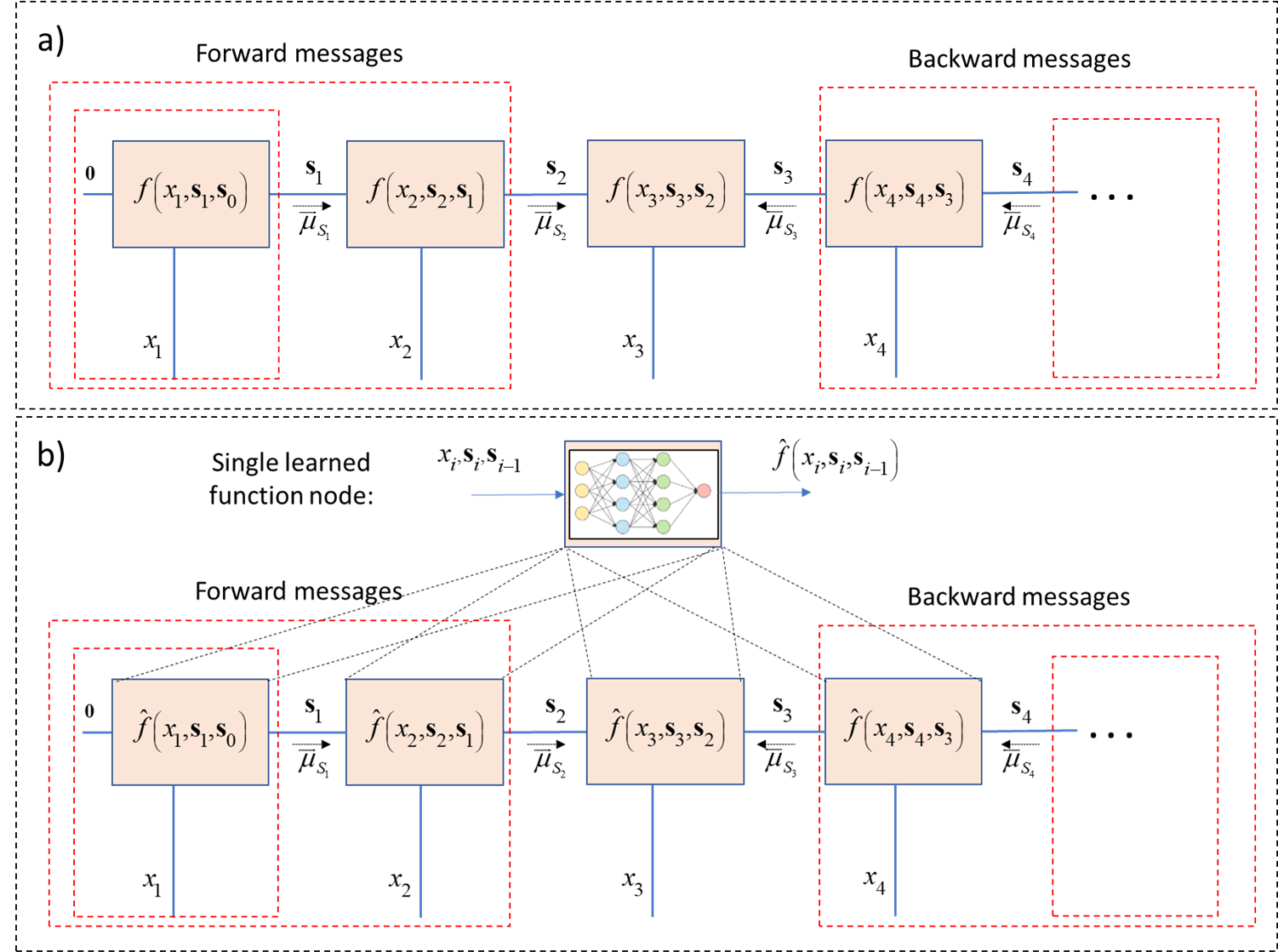} 
	\caption{Illustration of the \ac{sp} method for Markovian sequences using $a)$ the true factor graph; and $b)$ a learned factor graph.}
	\label{fig:SumProduct2}	 
\end{figure*}

% The BCJR algorithm is summarized as Algorithm~\ref{alg:Algo0}. 
% %

% \begin{algorithm}  
% 	\caption{ The BCJR algorithm}
% 	\label{alg:Algo0}
% 	\KwData{Fix an initial forward message $\FwdMsg{f_{0}}{\myState_{0}}(\myStateR)= 1$ and a final backward message $\BwdMsg{f_{0}}{\myState_{\Blklen}}(\myStateR)\equiv 1$. } 
% 	\For{$i=\Blklen-1,\Blklen-2,\ldots,1$}{
% 		For each  $\myStateR \in \mySet{S}^\Mem$, compute backward message $ \BwdMsg{f_{i}}{\myState_{i}}(\myStateR)$  via \eqref{eqn:Recursion1Backwards}
% 		%	\tcp*{backward messages} 
% 	}
% 	\For{$i=1,2,\ldots,\Blklen$}{
% 		For each  $\myStateR\in \mySet{S}^\Mem$, compute forward message $ \FwdMsg{f_{i}}{\myState_{i}}(\myStateR)$  via \eqref{eqn:Recursion1Forwards}
% 		%	\tcp*{forward messages} 
% 	}
% 	\KwOut{$\hat{\myVec{s}}^\Blklen = [\hat{\myVec{s}}_1, \ldots, \hat{\myVec{s}}_\Blklen]^T$, each obtained using \eqref{eqn:MAP2} %\tcp*{\ac{map} inference}
% 	}
% \end{algorithm}

\paragraph{Learned Factor Graphs}
Learned factor graphs enable learning to implement \ac{map} detection  from labeled data. It utilizes partial domain knowledge to determine the structure of the factor graph, while using deep learning to compute the function nodes without having to explicitly specify their computations. Finally, it carries out the \ac{sp} method for inference over the resulting learned factor graph.

{\bf Architecture:} %In order to implement the \ac{sp} algorithm, one must be able to specify the factor graph representing the underlying distribution. 
For Markovian relationships, the structure of the factor graph is that illustrated in Fig.~\ref{fig:SumProduct2}(a) regardless of the specific statistical model. Furthermore,  the stationarity assumption implies that the complete factor graph is  encapsulated in the single function $f(\cdot)$ \eqref{eqn:FSC_funcNode} {\em regardless of the block size $\Blklen$}. Building upon this insight,  \acp{dnn} can be utilized to learn the mapping carried out at the function node separately from the inference task.  
The resulting learned stationary factor graph is then used to recover  $\{\myS_i\}$ by message passing, as illustrated in Fig.~\ref{fig:SumProduct2}(b).
As learning a single function node is expected to be a simpler task compared to learning the overall inference method for recovering $\myVec{s}$ from $\Input$, this approach allows using relatively compact \acp{dnn}, which can be learned  from a relatively small data set. %Furthermore, the learned function node describes the factor graph for different values of $\Blklen$.  
%When the learned function node, denoted $\hat{f}_{\myVec{\theta}}(\cdot)$, is an accurate estimate of the true one, the \ac{map} detection rule \eqref{eqn:MAP0} is effectively implemented, and thus the inference performance approaches the minimal probability of error. 

{\bf Training:} %The function node which encapsulates the factor graph of stationary finite-memory channels is given in \eqref{eqn:FSC_funcNode}. 
In order to learn a stationary factor graph from samples, one must only learn its function node, which here boils down to learning  $\PdfNew{Y_i| \myVec{S}_{i-\Mem}^i }(x_i| \myVec{s}_{i-\Mem}^i)$ and $\PdfNew{S_i| \myVec{S}_{i-\Mem}^{i-1} }(s_i| \myVec{s}_{i-\Mem}^{i-1})$ by \eqref{eqn:FSC_funcNode}.  %Specifically, for stationary sequences, only a single function node must be learned, as the mapping $f(\cdot)$ does not depend on the time index $i$. 
Since  $\mySet{S}$ is finite, the transition probability $\PdfNew{S_i| \myVec{S}_{i-\Mem}^{i-1} }(s_i| \myVec{s}_{i-\Mem}^{i-1})$  can be learned   via a histogram. 

For learning the  distribution $\PdfNew{Y_i| \myVec{S}_{i-\Mem}^i }(x_i| \myVec{s}_{i-\Mem}^i)$, it is noted that 	\begin{equation}
	\label{eqn:Bayes}
\PdfNew{Y_i| \myVec{S}_{i-\Mem}^i }(x_i | \myVec{s}_i)  =
\PdfNew{\myVec{S}_{i-\Mem}^i |Y_i }\left(\myVec{s}_i|x_i \right) \PdfNew{Y_i  }\left(x_i   \right) {\big( \PdfNew{\myVec{S}_{i-\Mem}^i   }(\myVec{s}_i )\big)^{-1} }.
	\end{equation}
%	The resulting estimate consists of two parametric models: one for evaluating the conditional  $P_{{S}_i|Y}(s|x_i)$, and another for computing the marginal \ac{pdf} $P_{Y_i}(x_i)$.  
A parametric estimate of $\PdfNew{\myVec{S}_{i-\Mem}^i |Y_i }\left(\myVec{s}_i|x_i \right)$, denoted $\hat{P}_{\myVec{\theta}}(\myVec{s}_i|x_i)$,  is obtained  for each $\myVec{s}_i\in\mySet{S}^{\Mem+1}$ by training classification networks with softmax output layers to minimize the cross entropy loss. %For example, in our numerical study in Section \ref{sec:sims} we use a three-layer network trained with merely $5000$ training samples. 
As the \ac{sp} mapping  is invariant to scaling $f(\Input_i, \myStateR_{i}, \myStateR_{i-1})$ with some factor which does not depend on the $\myStateR_{i}, \myStateR_{i-1}$, one can set $ \PdfNew{Y_i }\left(x_i   \right) \equiv 1$  in \eqref{eqn:Bayes}, and use the result to obtain a scaled value of the function node, which, as discussed above, does not affect the inference mapping.

\begin{figure*}
	\centering
	\begin{subfigure}{0.42\textwidth}
		\centering
		{\includegraphics[width=\columnwidth]{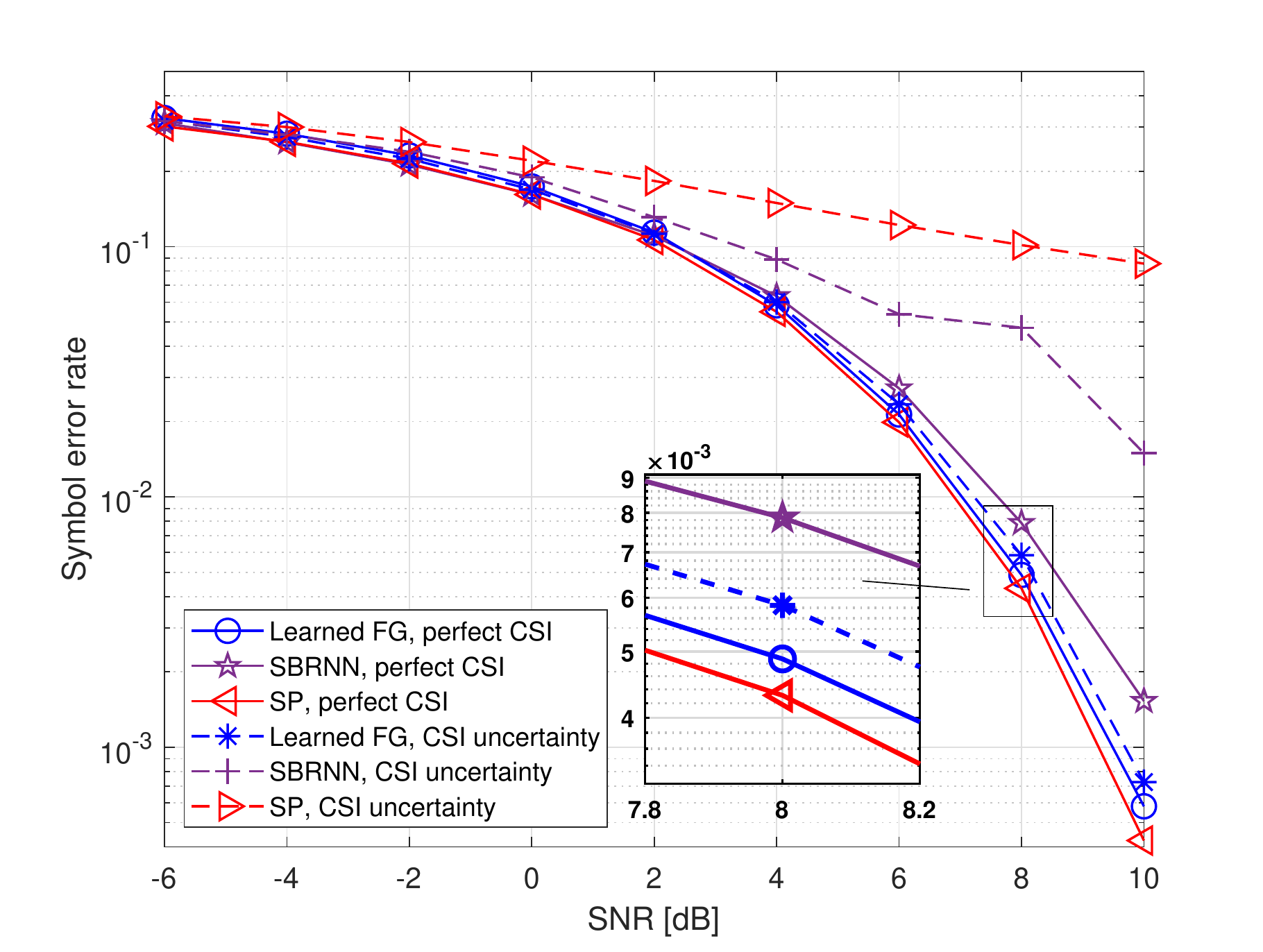}} 
		\caption{ Gaussian channel.
		}
		\label{fig:AWGN2} 	
	\end{subfigure}
	$\quad$
	\begin{subfigure}{0.42\textwidth}
		\centering
		{\includegraphics[width=\columnwidth]{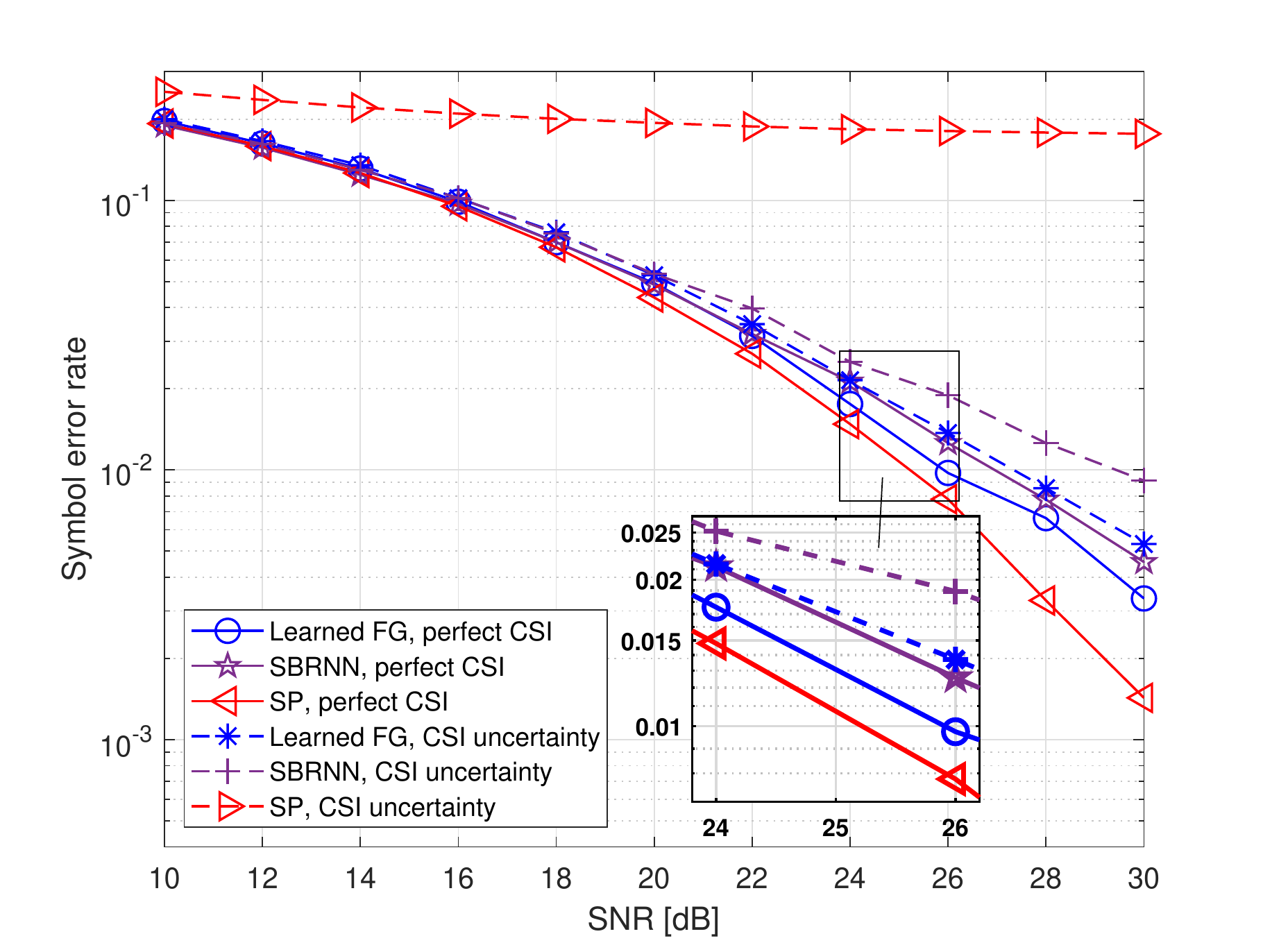}} 
		\caption{Poisson channel.
		}
		\label{fig:Poisson2} 
	\end{subfigure}
	\caption{Experimental results from \cite{shlezinger2020data} of learned factor graphs (Learned FG) compared to the model-based \ac{sp} algorithm and the data-driven sliding bidirectional \ac{rnn} (SBRNN) of \cite{farsad2018neural}. {\em Perfect CSI} implies that the system is trained and tested using
samples from the same channel, while under {\em CSI uncertainty} they are trained using samples from a set of different channels.}
    \label{fig:LearnedFGRes}
	%\vspace{-0.2cm}
\end{figure*}

{\bf Quantitative Results:} As a numerical example of learned factor graphs for Markovian models, we consider a scenario of symbol detection over causal stationary communication channels with finite memory, reproduced from \cite{shlezinger2020data}.  Fig.~\ref{fig:LearnedFGRes} depicts the numerically evaluated \ac{ser} achieved by applying the \ac{sp} algorithm over a factor graph learned from $\Ntraining=5000$ labeled samples, for channels with memory $\Mem=4$. The results are compared to the performance of  model-based \ac{sp}, which requires complete knowledge of the underlying statistical model, as well as the sliding bidirectional \ac{rnn} detector proposed in \cite{farsad2018neural} for such setups, which utilizes a conventional \ac{dnn} architecture that does not explicitly account for the Markovian structure. Fig.~\ref{fig:AWGN2} considers a Gaussian channel, while in Fig.~\ref{fig:Poisson2} the conditional distribution $\PdfNew{}(x_i|\myVec{s}_{i-\Mem}^{i})$ represents a Poisson distribution. Fig.~\ref{fig:LearnedFGRes} demonstrates the ability of learned factor graphs to enable accurate message passing inference in a data-driven manner, as the performance achieved using learned factor graphs approaches that of the \ac{sp} algorithm, which operates with full knowledge of the underlying statistical model. The numerical results also demonstrate that combining model-agnostic \acp{dnn} with model-aware inference notably improves robustness to model uncertainty compared to applying  \ac{sp}  with the inaccurate model. Furthermore, it also observed that explicitly accounting for the Markovian structure allows to achieve improved performance compared to utilizing black-box \ac{dnn} architectures such as the sliding bidirectional \ac{rnn} detector, with limited data sets for training.

\subsubsection*{Discussion} 
The integration of deep learning into structure-oriented model-based algorithms allows to exploit the model-agnostic nature of \acp{dnn} while explicitly accounting for available structural domain knowledge. Consequently, structure-oriented \ac{dnn}-aided inference is most suitable for setups in which  structured domain knowledge naturally follows from established models, while the subtleties of the complete statistical knowledge may be challenging to accurately capture analytically. Such structural knowledge is often present in various problems in signal processing and communications. For instance, modelling communication channels as causal finite-memory systems, as assumed in the above quantitative example, is a well-established representation of many physical channels. %, regardless of the specific characteristics of the setup and the communicating devices. 
%Another example is state-space models, which are widely-used in control scenarios. %State-space models represent transitions of an unknown state in a structured manner, while the unique properties of the problem at hand are encapsulated in the equations describing the state evolution and corresponding measurements. 
The availability of established structures in signal processing related setups makes structure-oriented \ac{dnn}-aided inference a candidate approach to facilitate inference in such scenarios in a manner which is ignorant of the possibly intractable subtleties of the problem, by learning to account for them implicitly from data. 

The fact that \acp{dnn} are used to learn an intermediate computation rather than the complete predication rule, facilitates the usage of relatively compact \acp{dnn}. This property can be exploited to implement learned inference on computationally limited devices, as was done in \cite{escoriza2021data} for \ac{dnn}-aided velocity tracking in autonomous racing cars. An additional consequence is   that the resulting system can be trained using scarce data sets. One can exploit the fact that the system can be trained using small training sets to, e.g., enable online adaptation to temporal variations in the  statistical model based on some  feedback on the correctness of the inference rule. This property was exploited in \cite{raviv2021meta} to facilitate online training of \ac{dnn}-aided receivers in coded communications. %Furthermore, while structure-oriented \ac{dnn}-aided systems are geared towards exploiting statistical structures, this approach can also incorporate additional forms of domain knowledge. For instance, the learned factor graphs example presented above utilize prior knowledge of an underlying stationarity to substantially reduce the number of learned parameters, capturing the complete learned factor graph in a single learned function node.  

A \ac{dnn} integrated into a structure-oriented model-based inference method  can be either trained individually, i.e., independently of the inference task, or in an end-to-end fashion. The first approach typically requires less training data, and the resulting trained \ac{dnn} can be combined with various inference algorithms. For instance, the learned function node used to carry out \ac{sp} inference in the above example can also be integrated into the Viterbi algorithm as done in \cite{shlezinger2019viterbinet}. Alternatively, the learned modules can be tuned  end-to-end  by formulating their objective as that of the overall inference algorithm, and backpropagating through the model-based computations, see, e.g., \cite{knobelreiter2020belief}. Learning in an end-to-end fashion facilitates overcoming inaccuracies in the assumed structures, possibly by incorporating learned methods to replace the generic computations of the model-based algorithm, at the cost of requiring larger volumes of data for training purposes.

	% TODO NIR CONTINUE FROM HERE
	%----------------------------------------------------------------------------------------
    %	Neural Augmentation
    %----------------------------------------------------------------------------------------
    % \vspace{-0.2cm}
     \subsection{Neural Augmentation}
	\label{subsec:Inference_Augmentation}
\begin{figure*}
	    \centering
	    \includegraphics[width=0.7\linewidth]{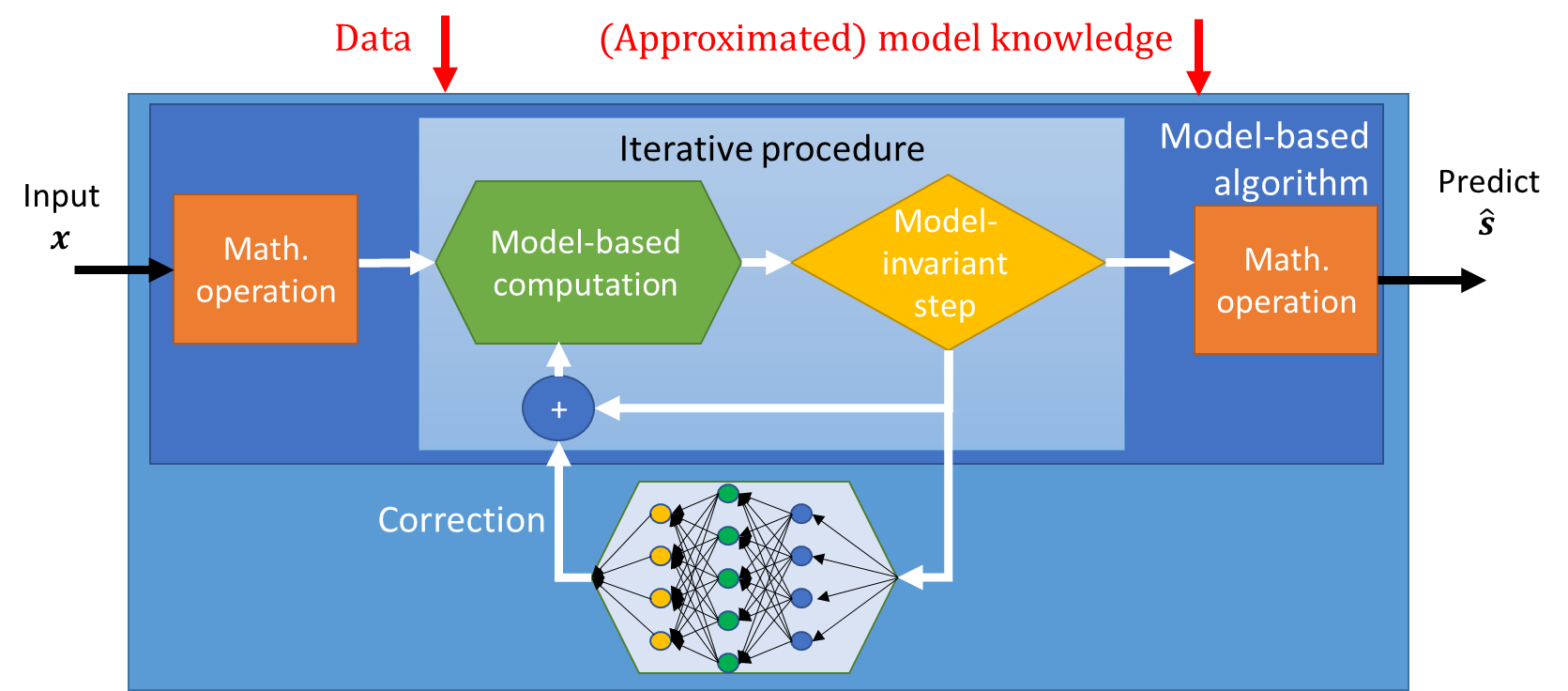}
	    \caption{Neural augmentation illustration.}
	    \label{fig:NeuralAug1}
	\end{figure*}

	% \vspace{-0.1cm}	
	The \ac{dnn}-aided inference strategies detailed in Subsections~\ref{subsec:Agnostic} and \ref{subsec:Inference_Oriented} utilize model-based algorithms to carry out inference, while replacing explicit domain-specific computations with dedicated \acp{dnn}. An alternative approach, referred to as {\em neural augmentation}, utilizes the complete model-based algorithm for inference, i.e., without embedding deep learning into its components, while using an external \ac{dnn} for correcting some of its intermediate computations 	\cite{satorras2019combining,satorras2020neural,pratik2020re, gao2022neural}.  An illustration of this approach is depicted in Fig.~\ref{fig:NeuralAug1}.
	
	The main advantage in utilizing an external \ac{dnn} for correcting internal computations %of a model-based algorithm follows 
	stems from its ability to notably improve the robustness of model-based methods to inaccurate knowledge of the underlying model parameters. Since the model-based algorithm is individually implemented, one must posses the complete domain knowledge it requires, and thus the external correction \ac{dnn} allows the resulting  system to overcome inaccuracies in this domain knowledge by learning to correct them from data. Furthermore, the learned correction term incorporated by neural augmentation can improve the performance of model-based algorithms in scenarios where they are sub-optimal, as detailed in the example in the sequel.
		
			\subsubsection*{Design Outline} %Neural augmentation is based on adding an external learned correction module to an iterative model-based optimization method, whose purpose is to provide a correction term to the information exchanged between the iterations of the algorithm.  
			The design of  neural-augmented inference systems is comprised of the following steps:
		\begin{enumerate}
		    \item Choose a suitable iterative optimization algorithm for the problem of interest, and  identify the information exchanged between the iterations, along with  the intermediate computations used to produce this information.
		    \item The information exchanged between the iterations is updated with a correction term learned by a \ac{dnn}. The \ac{dnn}  is designed to combine the same quantities used by the model-based algorithm, only in a learned fashion.
		    \item The overall hybrid model-based/data-driven system is trained in an end-to-end fashion, where one can consider not only the algorithm outputs in the loss function, but also the intermediate outputs of the internal iterations. 
		\end{enumerate}
		
		We next demonstrate how these steps are carried out in order to augment Kalman smoothing, as proposed in \cite{satorras2019combining}.

\subsubsection*{Example 7: Neural-Augmented Kalman Smoothing} 
	The \ac{dnn}-aided Kalman smoother proposed in \cite{satorras2019combining} implements state estimation in environments characterized by state-space models. Here, neural augmentation does not only to robustify the  smoother in the presence of inaccurate model knowledge, but also improves its performance in non-linear setups, where variants of the Kalman algorithm, such as the extended Kalman method, may be sub-optimal \cite[Ch. 7]{haykin2005adaptive}. %While the description below considers linear state-space models, the quantitative example taken from \cite{satorras2019combining} is taken from a non-linear case. 
		
	\paragraph{System Model} Consider a linear Gaussian state-space model. Here, one is interested in recovering a sequence of $\Blklen$ state \acp{rv} $\{\myS_i\}_{i=1}^\Blklen$ taking values in a continuous set from an observed sequence $\{\Input_i\}_{i=1}^{\Blklen}$. The observations are related to the desired state sequence via
	\begin{subequations}
	\label{eqn:KalmanSS}
		\begin{equation}
	\label{eqn:KalmanMeas}
	    \Input_i = \myMat{H} \myS_i + \myVec{r}_i
	\end{equation}
	while the state transition takes the form
	\begin{equation}
	\label{eqn:KalmanTrans}
	    \myS_i = \myMat{F} \myS_{i-1} + \myVec{w}_i.
	\end{equation}
	\end{subequations}
	In \eqref{eqn:KalmanSS}, $\myVec{r}_i$ and $\myVec{w}_i$ obey an i.i.d. zero-mean Gaussian distributions with covariance $\myMat{R}$ and $\myMat{W}$, respectively, while $\myMat{H}$ and $\myMat{F}$ are known linear mappings.
	
	We focus on scenarios where the state-space model in \eqref{eqn:KalmanSS} that is available to the inference system, is an inaccurate approximation of the true underlying dynamics. For such scenarios, one can apply Kalman smoothing, which is known to achieve minimal \ac{mse} recovery when  \eqref{eqn:KalmanSS} holds, while introducing a neural augmentation correction term  \cite{satorras2019combining}. 
	
	\paragraph{Kalman Smoothing} The Kalman smoother computes the minimal \ac{mse} estimate of each $s_i$ given a realization of  $\Input =[\Input_1,\ldots,\Input_{\Blklen}]^T$. Its procedure is comprised of forward and backward message passing, exploiting the Markovian structure of the state-space model to operate at complexity which only grows linearly with $\Blklen$. In particular, by writing  $\myVec{s} =[\myS_1,\ldots,\myS_{\Blklen}]^T$, one way to implement such smoothing to approach the minimal \ac{mse} estimate involves applying gradient descent optimization on the joint log likelihood function, i.e., by iterating over
	\begin{equation}
	\label{eqn:GradKal}
	    \myVec{s}^{(q+1)} = \myVec{s}^{(q)} + \eta \nabla_{\myVec{s}^{(q)} }\log \PdfNew{\myVec{X}, \myVec{S}}\left(\Input, \myVec{s}^{(q)} \right)
	\end{equation}
	where $\eta>0$ is a step-size. 
	\color{NewColor}
	Leveraging the state-space model \eqref{eqn:KalmanSS}, one can implement gradient descent iterations as message passing, via the procedure summarized in Algorithm~\ref{alg:Smoothing}, whose detailed formulation is given in Appendix~\ref{app:Kalman_Exm}.

	\begin{algorithm}  
		\caption{Smoothing via iterative gradient descent}
		\label{alg:Smoothing}
		\KwData{Fix step-size  $\eta>0$. Set initial guess $\hat{\myVec{s}}^{(0)}$} 
		\For{$q=0,1,\ldots$}{
		    \For{$i=1,\ldots,\Blklen$}{
		        Compute messages 
		        	\begin{align*}
	  \mu^{(q)}_{\myVec{S}_{i-1}\rightarrow \myVec{S}_i} & =  -\myMat{W}^{-1}\left( \myS_i^{(q)} - \myMat{F} \myS_{i-1}^{(q)}  \right),  \\
	  \mu^{(q)}_{\myVec{S}_{i+1}\rightarrow \myVec{S}_i} & =  \myMat{F}^T\myMat{W}^{-1}\left( \myS_{i+1}^{(q)} - \myMat{F} \myS_{i}^{(q)}  \right),  \\
	  \mu^{(q)}_{\myVec{X}_{i}\rightarrow \myVec{S}_i} & =  \myMat{H}^T\myMat{R}^{-1}\left( \Input_{i}- \myMat{H} \myS_{i}^{(q)}  \right).
	\end{align*}\label{stp:KalSmoothQuant1} \\
		        Update gradient step via
		        \begin{align*}
		             \hat{s}^{(q+1)}_i = \hat{s}^{(q)}_i + \eta &\Big(  \mu^{(q)}_{\myVec{S}_{i-1}\rightarrow \myVec{S}_i} \notag \\
		             &+  \mu^{(q)}_{\myVec{S}_{i+1}\rightarrow \myVec{S}_i} + \mu^{(q)}_{\myVec{X}_{i}\rightarrow \myVec{S}_i}\Big).
		        \end{align*}
		    }
		}
		\KwOut{Estimate $\hat{\myVec{s}} = \hat{\myVec{s}}^{(q)}$.}
\end{algorithm}
 
\color{black}

	\paragraph{Neural-Augmented Kalman Smoothing} The gradient descent formulation in \eqref{eqn:GradKal} is evaluated by the messages in Step~\ref{stp:KalSmoothQuant1} of Algorihtm~\ref{alg:Smoothing}, which in turn rely on accurate knowledge of the state-space model \eqref{eqn:KalmanSS}. To facilitate operation with inaccurate model knowledge due to, e.g., \eqref{eqn:KalmanSS} being a linear approximation of a non-linear setup, one can introduce neural augmentation to learn to correct inaccurate computations of the log-likelihood gradients. This is achieved by using an external \ac{dnn} to map the messages in Step~\ref{stp:KalSmoothQuant1} into a correction term, denoted $\myVec{\epsilon}^{(q+1)}$. 
	
	{\bf Architecture:} The learned mapping of the messages \eqref{eqn:KalmanSS} into a correction term operates in the form of a \ac{gnn} \cite{yoon2019inference}. This is implemented by maintaining an internal node variable for each variable in Step~\ref{stp:KalSmoothQuant1} of Algorithm~\ref{alg:Smoothing}, denoted $h_{\myS_i}^{(q)}$ for each $\myS_{i}^{(q)}$ and  $h_{\Input_i}$ for each $\Input_i$, as well as internal message variables $m_{\myVec{V}_n\rightarrow \myVec{S}_i}^{(q)}$ for each message computed by the model-based Algorithm~\ref{alg:Smoothing}. The node variables $h_{\myS_i}^{(q)}$ are updated along with the model-based smoothing algorithm iterations as estimates of their corresponding variables, while the variables $h_{\Input_i}$ are obtained once from $\Input$ via a neural network.  The \ac{gnn} then maps the messages produced by the model-based Kalman smoother into its internal messages via a neural network $f_{e}(\cdot)$ which operates on the corresponding node variables, i.e., 
	\begin{equation}
	    m_{\myVec{V}_n\rightarrow \myVec{S}_i}^{(q)} = f_{e}\left(h_{\myVec{v}_n}^{(q)}, h_{\myS_i}^{(q)}, \mu^{(q)}_{\myVec{V}_{n}\rightarrow \myVec{S}_i}  \right)
	    \label{eqn:MapEnc1}
	\end{equation}
	where $h_{\Input_n}^{(q)} \equiv h_{\Input_n}$ for each $q$. 
	These messages are then combined and forwarded into a \ac{gru}, which produces the refined estimate of the node variables $\{h_{\myS_i}^{(q+1)}\}$ based on their corresponding messages \eqref{eqn:MapEnc1}. Finally, each updated node variable $h_{\myS_i}^{(q+1)}$ is mapped into its corresponding error term $\myVec{\epsilon}_i^{(q+1)}$ via a fourth neural network, denoted $f_d(\cdot)$. 
	
\begin{figure*}
	    \centering
	    \includegraphics[width=0.7\linewidth]{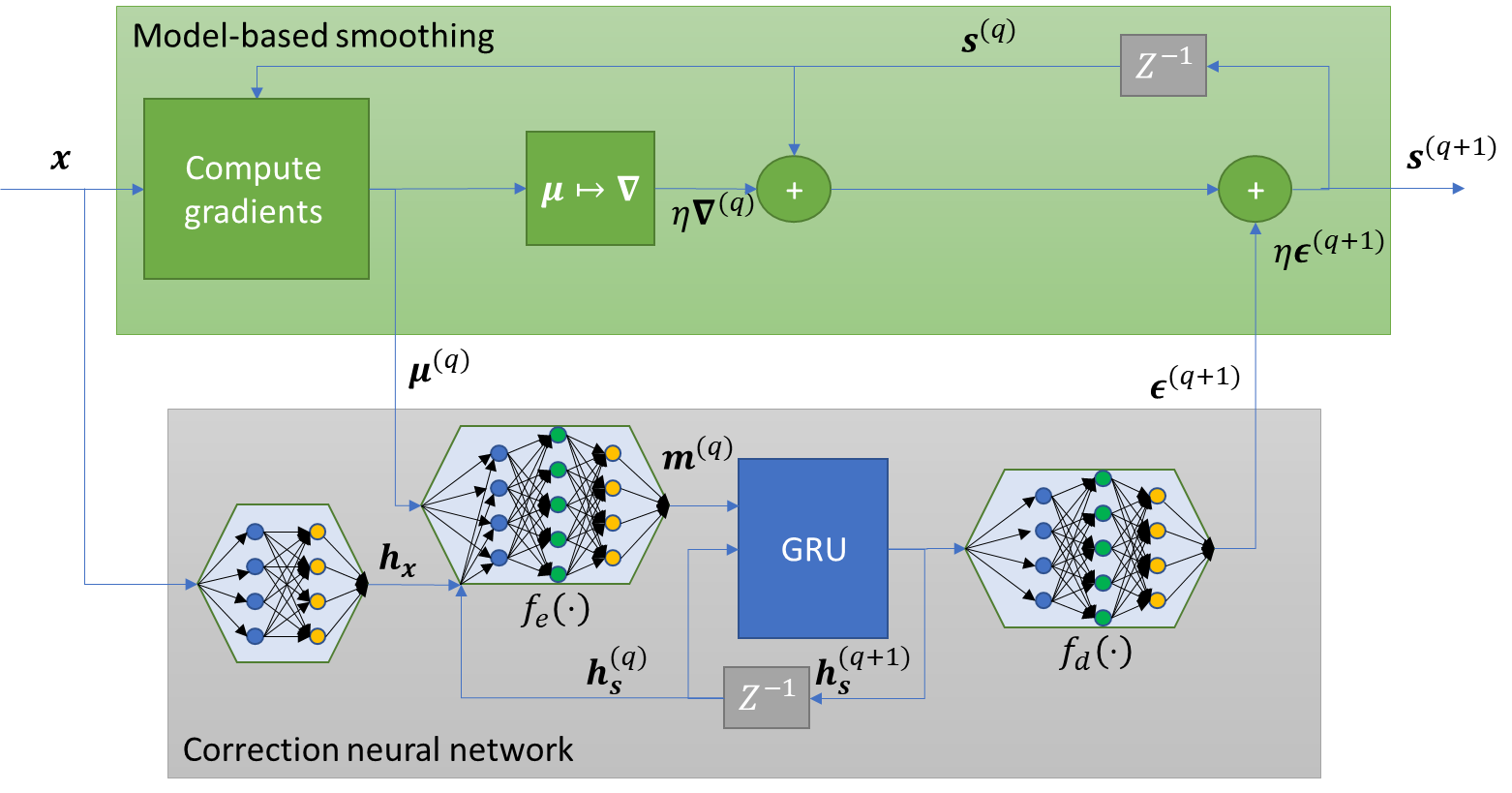}
	    \caption{Neural augmented Kalman smoother illustration. Blocks marked with $Z^{-1}$ represent a single iteration delay.}
	    \label{fig:NeuralKalman1}
	\end{figure*}

	The correction terms $\{\myVec{\epsilon}_i^{(q+1)}\}$ aggregated into the vector  $\myVec{\epsilon}^{(q+1)}$ are used  to update the log-likelihood gradients, resulting in the update equation \eqref{eqn:GradKal} replaced with
	\begin{equation}
	\label{eqn:GradKal2}
	    \myVec{s}^{(q+1)} = \myVec{s}^{(q)} + \eta \left( \nabla_{\myVec{s}^{(q)} }\log \PdfNew{\myVec{X}, \myVec{S}}\left(\Input, \myVec{s}^{(q)} \right) + \myVec{\epsilon}^{(q+1)} \right).
	\end{equation}
	The overall architecture is illustrated in Fig.~\ref{fig:NeuralKalman1}. 
	
	{\bf Training:} Let $\myVec{\theta}$ be the  parameters of the \ac{gnn} in  Fig.~\ref{fig:NeuralKalman1}. The hybrid system is trained  end-to-end  to minimize the empirical weighted $\ell_2$ norm loss over its intermediate layers, where the contribution of each iteration to the overall loss increases as the iterative procedure progresses. In particular,  letting $\{(\myVec{s}_t, \Input_t)\}_{t=1}^{\Ntraining}$ be the training set, the loss function used to train the neural-augmented Kalman smoother is given by
\begin{equation}
\label{eqn:LossKalman}
\mySet{L}(\myVec{\theta}) = \frac{1}{\Ntraining}\sum_{t=1}^{\Ntraining}\sum_{q=1}^{\Niter} \frac{q}{\Niter}\|\myVec{s}_t - \hat{\myVec{s}}_q(\Input_t; \myVec{\theta}) \|^2
\end{equation}
where $\hat{\myVec{s}}_q(\Input_t; \myVec{\theta})$ is the estimate produced by the $q$th iteration, i.e., via \eqref{eqn:GradKal2},  with parameters $\myVec{\theta}$ and input $\Input_t$.

\begin{figure}
	    \centering
	    \includegraphics[width=0.9\columnwidth]{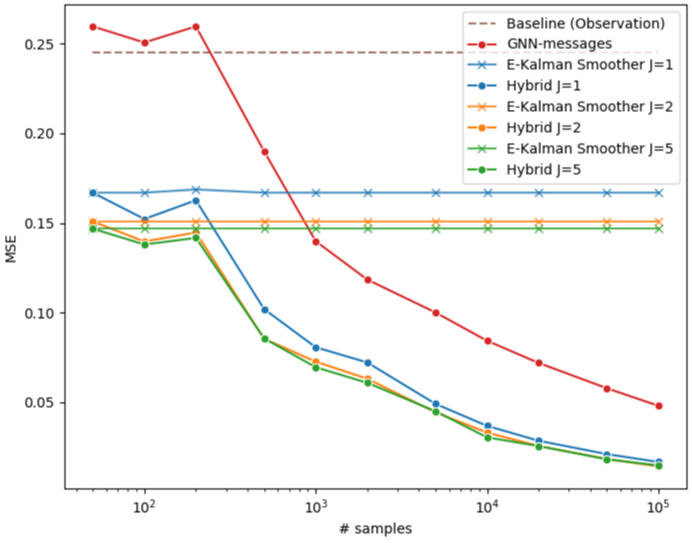}
	    \caption{MSE versus data set size for the neural-augmented Kalman smoother (Hybrid) compared to the model-based extended Kalman smoother (E-Kalman) and a solely data-driven \ac{gnn}, for various linearizations of state-space models (represented by the index $j$). Figure reproduced from \cite{satorras2019combining} with authors' permission.}
	    \label{fig:NKalRes1}
	\end{figure}

	{\bf Quantitative Results:}  The experiment whose results are depicted in Fig.~\ref{fig:NKalRes1} considers a non-linear state-space model described by the Lorenz attractor equations, which describe atmospheric convection via continuous-time differential equations. The state space model is approximated as a discrete-time linear one  by replacing the dynamics with their $j$th order Taylor series. %Thus the state-space equations assumed by the model-based inference system constitute an approximation of the true dynamics.
	Fig.~\ref{fig:NKalRes1} demonstrates the ability of neural augmentation to  improve model-based inference. It is observed that introducing the \ac{dnn}-based correction term allows the system to learn to overcome the model inaccuracy, and achieve an error which decreases with the amount of available training data. It is also observed that the hybrid approach of combining model-based inference and deep learning enables accurate inference with notably reduced volumes of training data, as the individual application of the \ac{gnn} for state estimation, which does not explicitly account for the available domain knowledge, requires much more training data to achieve similar accuracy as that of the neural-augmented Kalman smoother. 
	
	%The results presented here in Fig.~\ref{fig:NKalRes1} are based on \cite[Fig. 4]{satorras2019combining} with the authors' permission. 
	
\subsubsection*{Discussion} 
	Neural augmentation implements hybrid model-based/data-driven inference by utilizing two individual modules -- a model-based algorithm and a \ac{dnn} -- with each capable of inferring on its own. %For instance,  the neural-augmented Kalman smoother utilizes both an individual model-aware Kalman smoother which carries out inference, along with a \ac{gnn} that can produce its own state estimates via its internal node variables. 
	The rationale here is to benefit from  both approaches by interleaving the iterative operation of the modules, and specifically by utilizing the data-driven component to learn to correct the model-based algorithm, rather than produce individual estimates. This approach thus conceptually differs from the \ac{dnn}-aided inference strategies discussed in Subsections~\ref{subsec:Agnostic} and \ref{subsec:Inference_Oriented}, where a \ac{dnn} is integrated into a model-based algorithm. % and thus inference is achievable only by the combined efforts of the model-based and data-driven components. 
	
	The fact that neural augmentation utilizes individual model-based and data-driven modules reflects on its requirements and use cases. First, one must posses full domain knowledge, or at least an approximation of the true model, in order to implement model-based inference. For instance, the neural-augmented Kalman smoother requires full knowledge of the state-space model \eqref{eqn:KalmanSS}, or at least an approximation of this analytical closed-form model as used in the quantitative example, in order to compute the exchanged messages in Algorithm~\ref{alg:Smoothing}. Additionally, the presence of an individual \ac{dnn} module implies that relatively large amounts of data are required in order to train it. Nonetheless, the fact that this \ac{dnn} only produces a correction term which is interleaved with the model-based algorithm operation implies that the amount of training data required to achieve a given accuracy is notably smaller compared to that required when using solely the \ac{dnn} for inference. For instance, the quantitative example of the neural augmented Kalman smoother demonstrate that it requires $10-20$ times less samples compared to that required by the individual \ac{gnn} to achieve similar \ac{mse} results. 
	
%	In summary, the main benefits of neural augmentation over purely model-based inference thus lie in its ability to improve and robustify performance in the presence of inaccurate domain knowledge. Compared to inferring solely based on deep learning, neural augmentation enables to train with smaller data sets, or alternatively, to infer at improved accuracy for a given limited data set, due to its interaction with the model-based component. 

% 	    %----------------------------------------------------------------------------------------
%     %	Summary
%     %----------------------------------------------------------------------------------------
%     \vspace{-0.2cm}
%      \subsection{Summary}
% 	\label{subsec:Inference_Summary}
% 	\vspace{-0.1cm}	
% 	TODO - Nir

	%----------------------------------------------------------------------------------------
	%	CONCLUSIONS
	%----------------------------------------------------------------------------------------
	% \vspace{-0.2cm}
	\section{Conclusions and Future Challenges}
	\label{sec:Conclusions}
	% \vspace{-0.1cm}
	In this article, we presented a mapping of methods for combining domain knowledge and data-driven inference via model-based deep learning in a tutorial manner. We noted that hybrid model-based/data-driven systems can be categorized into model-aided networks, which utilize model-based algorithms to design \ac{dnn} architectures, and \ac{dnn}-aided inference, where deep learning is integrated into traditional model-based methods. We detailed representative design approaches for each strategy in a systematic manner, along with design guidelines and  concrete examples. To conclude this overview, we first summarize the key advantages of model-based deep learning in Subsection~\ref{subsec:Conc_Discussion}. Then, we present guidelines for selecting a design approach for a given application in Subsection~\ref{subsec:Conc_Choosing}, intended to facilitate the derivation of future hybrid data-driven/model-based systems. Finally, we review some future  research challenges in Subsection~\ref{subsec:Conc_Future}.

	%----------------------------------------------------------------------------------------
    %	Discussion
    %----------------------------------------------------------------------------------------
    % \vspace{-0.2cm}
     \subsection{Advantages of Model-Based Deep Learning}
	\label{subsec:Conc_Discussion}
	% \vspace{-0.1cm}	
	The combination of traditional handcrafted algorithms with emerging data-driven tools via model-based deep learning brings forth several key advantages. Compared to purely model-based schemes, the integration of deep learning facilitates inference in complex environments, where accurately capturing the underlying model in a closed-form mathematical expression may be infeasible. For instance, incorporating \ac{dnn}-based implicit regularization was shown to enable \ac{cs} beyond its traditional domain of sparse signals, as discussed in Subsection~\ref{subsec:Agnostic}, while the implementation of the \ac{sic} method as an interconnection of neural building blocks enables its operation in non-linear setups, as demonstrated in Subsection~\ref{subsec:Networks_Blocks}.  The model-agnostic nature of deep learning also allows hybrid model-based/data-driven inference to achieve improved resiliency to model uncertainty compared to inferring solely based on domain knowledge. % as the combination of \acp{dnn} allows the resulting system to learn to overcome such uncertainty from data. 
	For example, augmenting model-based Kalman smoothing with a \ac{gnn} was shown in Subsection~\ref{subsec:Inference_Augmentation} to notably improve its performance when the state-space model does not fully reflect the true dynamics, while the usage of learned factor graphs for \ac{sp} inference was demonstrated to result in improved robustness to model uncertainty in Subsection~\ref{subsec:Inference_Oriented}.    Finally, the fact that hybrid systems learn to carry out part of their inference based on data allows to infer with reduced delay compared to the corresponding fully model-based methods, as demonstrated by deep unfolding in Subsection~\ref{subsec:Networks_Unfolding}.

	Compared to utilizing conventional \ac{dnn} architectures for inference, the incorporation of domain knowledge via a hybrid model-based/data-driven design results in systems which are tailored for the problem at hand. As a result, model-based deep learning systems require notably less data in order to learn an accurate mapping, as demonstrated in the comparison of learned factor graphs and the sliding bidirectional \ac{rnn} system in the quantitative example in Subsection~\ref{subsec:Inference_Oriented}, as well as the  comparison between the neural augmented Kalman smoother and the \ac{gnn} state estimator in the corresponding example in Subsection~\ref{subsec:Inference_Augmentation}. This property of model-based deep learning systems enables quick adaptation to variations in the underlying statistical model, as shown in \cite{raviv2021meta}. {Finally, a system combining \acp{dnn} with model-based inference often provides the ability to analyze its resulting predictions, yielding interpretability and confidence which are commonly challenging to obtain with conventional black-box deep learning.}

	%----------------------------------------------------------------------------------------
    %	Choosing a Model-Based Deep Learning Strategy
    %----------------------------------------------------------------------------------------
    % \vspace{-0.2cm}
     \subsection{Choosing a Model-Based Deep Learning Strategy}
	\label{subsec:Conc_Choosing}
	% \vspace{-0.1cm}	
    The aforementioned gains of model-based deep learning are shared at some level by all the different approaches presented in Sections~\ref{sec:Networks}-\ref{sec:Inference}. However, each strategy is focused on exploiting a different advantage of hybrid model-based/data-driven inference, particularly in the context of signal processing oriented applications. Consequently, to complement the mapping of model-based deep learning strategies and facilitate the implementation of future application-specific hybrid systems, we next enlist the main considerations one should take into account when seeking to combine model-based methods with data-driven tools for a given problem. 	 %Once the suitable model-based deep learning approach is selected based on the guidelines detailed below, the specific implementation can follow the design outline detailed along with a concrete example in the corresponding part in Sections~\ref{sec:Networks}-\ref{sec:Inference}.
    
    {\bf Step 1: Domain knowledge and data  characterization:} First, one must ensure the availability of the two key ingredients in model-based deep learning, i.e., domain knowledge and data. The former corresponds to what is known \textit{a priori} about the problem at hand, in terms of statistical models and established assumptions, as well as what is unknown, or is based on some approximation that is likely to be inaccurate. The latter addresses the amount of labeled and unlabeled samples one posses in advance for the considered problem, as well as whether or not they reflect the scenario in which the system is requested to infer in practice.  
    
    {\bf Step 2: Identifying a model-based method:} Based on the available domain knowledge, the next step is to identify a suitable model-based algorithm for the problem. This choice should rely on the portion of the domain knowledge which is {\em available}, and not on what is {\em unknown}, as the latter can be compensated for by integration of deep learning tools. This stage must also consider the requirements of the inference system in terms of performance, complexity, and real-time operation, as these are encapsulated in the selection of the algorithm. 
    The identification of a model-based algorithm, combined with the availability of domain knowledge and data, should also indicate whether model-based deep learning mechanisms are required for the application of interest. %For instance, if a suitable model-based algorithm based on established accurate modelling is applicable for the problem at hand, then one may adopt the classical signal processing approach, namely, apply the model-based algorithm directly, while possibly exploiting the presence of data to estimate some missing parameters. Alternatively, if no suitable and reliable model-based method exists for the current application, while one has access to large volumes of data, it may be preferable to train a generic \ac{dnn} for inference, while possibly utilizing the availability of limited domain knowledge in the selection of the specific architecture.
    
    {\bf Step 3: Implementation challenges:} Having identified a suitable model-based algorithm, the selection of the approach to combine it with deep learning should be based on the understanding of its main implementation challenges. Some representative  issues  and their relationship with the recommended model-based deep learning approaches include:
    \begin{enumerate}
        \item Missing domain knowledge - model-based deep learning can implement the model-based inference algorithm when  parts of the underlying model are unknown, or alternatively, too complex to be captured analytically, by harnessing the model-agnostic nature of deep learning. In this case, the selection of the implementation approach depends on the format of the identified model-based algorithm: When it builds upon some known structures via, e.g., message passing based inference, structure-oriented \ac{dnn}-aided inference detailed in Subsection~\ref{subsec:Inference_Oriented} can be most suitable as means of integrating \acp{dnn} to enable operation with missing domain knowledge. Similarly, when the missing domain knowledge can be represented as some complex search domain, or alternatively, an unknown and possibly intractable regularization term,  structure-agnostic \ac{dnn}-aided inference detailed in Subsection~\ref{subsec:Agnostic} can typically facilitate optimization with implicitly learned regularizers.  Finally, when the algorithm can be represented as an interconnection of model-dependent building blocks, one can maintain the overall flow of the algorithm while operating in a model-agnostic manner via neural building blocks, as discussed in Subsection~\ref{subsec:Networks_Blocks}.
        \item Inaccurate domain knowledge - model-based algorithms are typically sensitive to inaccurate knowledge of the underlying model and its parameters. In such cases, where one has access to a complete description of the underlying model up to some uncertainty, model-based deep learning can robustify the model-based algorithm and learn to achieve improved accuracy. A candidate approach to robustify model-based processing is by adding a learned correction term via neural augmentation, as detailed in Subsection~\ref{subsec:Inference_Augmentation}. Alternatively, when the model-based algorithm takes an iterative form, improved resiliency can be obtained by unfolding the algorithm into a \ac{dnn}, as discussed in Subsection~\ref{subsec:Networks_Unfolding}, as well as use robust optimization in unfolding \cite{pu2021rest}.
        \item Inference speed - model-based deep learning can learn to implement iterative inference algorithms, which typically require a large amount of iterations to converge, with reduced inference speed. This is achieved by designing model-aided networks, typically via deep unfolding (see Subsection~\ref{subsec:Networks_Unfolding}) or neural building blocks (see Subsection~\ref{subsec:Networks_Blocks}).  
        The fact that model-aided networks learn their iterative computations from data allows the resulting system to infer reliably with a much smaller number of iteration-equivalent layers, compared to the iterations required by the model-based algorithm. 
        Alternatively, when the delaying aspect is an internal lengthy computation, one can improve run-time by replacing it with a fixed run-time \acp{dnn} via \ac{dnn}-aided inference, as shown in, e.g., \cite{ni2022rtsnet}. 
    \end{enumerate}

    The aforementioned implementation challenges constitute only a partial list of the considerations one should account for when selecting a model-based deep learning design approach. Additional considerations include computational capabilities during both training as well as inference; the  need to handle variations in the statistical model, which in turn translate to a possible requirement to periodically re-train the system; and the quantity and the type of available data. Nonetheless, the above division  provides systematic guidelines which one can utilize and possibly extend when seeking to implement an inference system relying on both data and domain knowledge.
    Finally, we note that some of the detailed model-based deep learning strategies can be combined, and thus one can select more than a single design approach. For instance, one can interleave \ac{dnn}-aided inference via implicitly learned regularization and/or priors, with deep unfolding of the iterative optimization algorithm, as discussed in Subsection~\ref{subsec:Agnostic}.  
 	
	    %----------------------------------------------------------------------------------------
    %	Future Challenges
    %----------------------------------------------------------------------------------------
    % \vspace{-0.2cm}
     \subsection{Future Research Directions}
	\label{subsec:Conc_Future}
	% \vspace{-0.1cm}	
    %The formulation of the model-based deep learning framework for combining domain knowledge in the form of model-based algorithms and data-driven inference in a systematic manner facilitates identifying some important research challenges in this emerging paradigm. In the following, we
    We end by discussing a few representative unexplored research aspects of model-based deep learning:
    
    {\bf Performance Guarantees:} One of the key strengths of model-based algorithms is their established theoretical performance guarantees. In particular, the analytical tractability of model-based methods implies that one can quantify their expected performance as a function of the parameters of underlying statistical or deterministic models. For conventional deep learning, such performance guarantees are very challenging to characterize, and deeper theoretical 
    understanding is a crucial missing component. The combination of deep learning with model-based structure increases interpretability thus possibly leading to theoretical guarantees.
    Theoretical guarantees improve the reliability of hybrid model-based/data-driven systems, as well as improve performance. 
    For example, some preliminary theoretical results were identified for specific model-based deep learning methods, such as the convergence analysis of the unfolded LISTA in \cite{chen2018theoretical} and of  plug-and-play networks in \cite{ryu2019plug}. 
    
    %Lacking theory - model-based algorithms have established theoretical guarantees. Meaningful theoretical results for deep learning are difficult to obtain. Can one advance deep learning theory by focusing on model-based deep learning? 
    
    {\bf Deep Learning Algorithms:}  Improving model interpretabilty and incorporating human knowledge is crucial for artificial intelligence development. Model-based deep learning can constitute a systematic framework to incorporate domain knowledge into data-driven systems, and can thus give rise to new forms of deep learning algorithms. \textcolor{NewColor}{For instance, while our description of the methodologies in Sections~\ref{sec:Networks}-\ref{sec:Inference} systematically commences with a model-based algorithm which is then augmented into a data-aided design via deep learning techniques, one can also envision algorithms in which model-based algorithms are utilized to improve upon an existing \ac{dnn} architecture. Alternatively, one can leverage model-based techniques to propose interpretable \ac{dnn} architectures which follow traditional model-based methods to account for domain knowledge.}

    {\bf Collaborative Model-Based Deep Learning:}  The increasing demands for accessible and personalized artificial intelligence give rise to the need to operate \acp{dnn} on edge devices such as smartphones, sensors, and autonomous cars~\cite{chen2019deep}. The limited computational and data resources of edge devices make model-based deep learning strategies particularly attractive for edge intelligence.
    Latency considerations and privacy constraints for mobile and sensitive data are further driving research in distributed training (e.g., through the framework of federated learning~\cite{li2020federated,gafni2022federated}) and collaborative inference~\cite{shlezinger2022collaborative}.
    Combining model-based structures with federated learning and  distributed inference remains as interesting research directions.

    {\bf Unexplored Applications:} 
    The increasing interest in hybrid model-based/data-driven deep learning methods is motivated by the need for robustness and structural understanding. 
Applications falling under the broad family of signal processing, communications, and control problems are natural candidates to benefit due to the proliferation of established model-based algorithms. We believe that model-based deep learning can contribute to the development of  technologies such as \ac{iot} networks, autonomous systems, and wireless communications. 

\color{NewColor}
\begin{appendix}
    \numberwithin{lemma}{subsection} 
    \numberwithin{corollary}{subsection} 
    \numberwithin{remark}{subsection} 
    \numberwithin{equation}{subsection}	

\subsection{Detailed Formulation of Project Gradient Descent (Example~1, Section~\ref{sec:Networks})}\label{app:DetNet_Exm}
Projected gradient descent  iteratively refines its estimate by taking a gradient step with respect the to unconstrained objective, followed by projection into the constrained set of the optimization variable. For the system model in \eqref{eqn:Gaussian}, this operation at iteration index $q+1$ is obtained recursively as
\begin{align}
\hat{\myVec{s}}_{q + 1} 
&= \mySet{P}_{\mySet{S}}\left(\hat{\myVec{s}}_{q} - \eta \left.\frac{\partial \|\Input-\myMat{H}\myVec{s}\|^2}{\partial \myVec{s}}\right|_{\myVec{s} = \hat{\myVec{s}}_q} \right) \notag \\
&=  \mySet{P}_{\mySet{S}}\left(\hat{\myVec{s}}_{q} - \eta\myMat{H}^T\Input + \eta\myMat{H}^T\myMat{H}\hat{\myVec{s}}_{q} \right)
\label{eqn:ProjGrad}
\end{align}
where $\eta$ is the step size, and $\hat{\myVec{s}}_0$ is an initial guess.

\subsection{Detailed Formulation of Proximal Gradient Method (Example~2, Section~\ref{sec:Networks})}\label{app:DCEA_Exm}
The recovery of the clean image $\myVec{\mu}$ which can be represented using a convolutional dictionary from the noisy observations $\Input$ can be formulated as a convolutional sparse coding problem: 
\begin{align}
  \!\!\!\!  \big(\hat{\myVec{s}},\hat{\myVec{H}} \big) &= \mathop{\arg \min}\limits_{\myVec{s}, {\myVec{H}} } -\log p_{\Input | \myVec{\mu}}(\Input | \myVec{\mu}=\myMat{H}\myVec{s})\! +\! \lambda \|\myVec{s}\|_1 \notag \\
    &= \mathop{\arg \min}\limits_{\myVec{s}, \hat{\myVec{H}}} \myVec{1}^T\!\exp\left(\myMat{H}\myVec{s}\right)\!-\!\Input^T \myMat{H}\myVec{s} \!+\!  \lambda \|\myVec{s}\|_1, 
    \label{eqn:ConvSparse} 
\end{align}
where the dictionary optimization variable is constrained to be block-Toeplitz. The clean image is then obtained as
\begin{align}
    \hat{\myVec{\mu}} &= \exp\left(\hat{\myMat{H}}\hat{\myVec{s}}\right).
    \label{eqn:ConvSparseDec}
\end{align}
Here, $\myVec{1}$ is the all ones vector, $\lambda$ is a regularizing term that controls the degree of sparsity, boosted by the usage of the $\ell_1$ norm.

Algorithm~\ref{alg:DictLearn} tackles \eqref{eqn:ConvSparse} via alternating optimization, where the update equations at iteration of index $l$ are given by
\begin{align}
\label{eqn:ProxGrad1}
    \hat{\myVec{s}}_{l+1} &= \mathop{\arg \min}\limits_{\myVec{s} } \myVec{1}^T\exp\left(\myMat{H}\myVec{s}\right)-\Input^T \myMat{H}\myVec{s} +  \lambda \|\myVec{s}\|_1,  \\ &\text{subject to } \myMat{H} = \hat{\myMat{H}}_l  \notag
    \end{align}
    and
    \begin{align}
    \hat{\myVec{H}}_{l+1} &= \mathop{\arg \min}\limits_{ \myVec{H}  } \myVec{1}^T\exp\left(\myVec{H} \myVec{s}\right)-\Input^T \myVec{H}\myVec{s},  \label{eqn:ProxGrad2} \\ 
   & \text{subject to } \myVec{s} = \hat{\myVec{s}}_{l+1}. \notag
\end{align}
The $\ell_1$ regularized optimization problem \eqref{eqn:ProxGrad1} can be tackled for a given $\myMat{H}$ and index $l$ via  proximal gradient descent iterations. This optimizer involves multiple iterations, indexed $q=0,1,2,\ldots$, of the form
\begin{equation}
     \hat{\myVec{s}}_{q+1} = \mySet{T}_{b}\left( \hat{\myVec{s}}_{q} + \eta \myMat{H}^T\left(\Input - \exp\left(\myVec{H} \hat{\myVec{s}}_{q}\right) \right)  \right). \label{eqn:Proxmapping}
\end{equation}
The threshold parameter $b$ is dictated by the regularization parameter $\lambda$.

\subsection{Detailed Formulation of Iterative Soft Interference Cancellation (Example 3, Section~\ref{sec:Networks})}\label{app:DeepSIC_Exm}
To formulate the iterative \ac{sic} algorithm, we consider the  Gaussian \ac{mimo} channel in \eqref{eqn:Gaussian}. 
Each iteration of the iterative \ac{sic} algorithm indexed  $q$ generates $\Nusers$ distribution vectors over the set of possible symbols $\mySet{S}$. The \acp{pmf} are denoted by the vectors  $\hat{\myVec{p}}_k^{(q)}$ of size $|\mySet{S}|\times 1$, where $k \in \NusersSet$. These vectors are computed from the observed $\Input$ as well as the distribution vectors obtained at the previous iteration, $\{ \hat{\myVec{p}}_k^{(q-1)}\}_{k=1}^{\Nusers}$. The entries of  $\hat{\myVec{p}}_k^{(q)}$ are estimates of the distribution of $s_k$ for each possible symbol in $\mathcal{S}$, given the observed $\Input$ and assuming that the interfering symbols $\{s_l \}_{l \neq k}$ are distributed via $\{ \hat{\myVec{p}}_l^{(q-1)}\}_{l \neq k}$. %Note that for binary constellations, i.e., $\CnstSize = 2$, $\hat{\myVec{p}}_k^{(q)}$ can be represented using a single scalar value, as $\big(\hat{\myVec{p}}_k^{(q)}\big)_2 = 1- \big(\hat{\myVec{p}}_k^{(q)}\big)_1$. 
Every iteration consists of two steps, carried out in parallel for each user: {\em Interference cancellation}, and {\em soft decoding}. 
Focusing on the $k$th user and the $q$th iteration, the interference cancellation stage first computes the expected values and variances of $\{s_l\}_{l \neq k}$ based on the estimated \ac{pmf}  $\{ \hat{\myVec{p}}_l^{(q-1)}\}_{l \neq k}$.  
The contribution of the interfering symbols from $\Input$ is then canceled by replacing them with $\{e_l^{(q-1)}\}$ and subtracting their resulting term. Letting $\myVec{h}_l$ be the $l$th column of $\myMat{H}$, the interference canceled channel output is given by
%	\begin{subequations} 
\begin{align}
\label{eqn:Cancelled1} 
\myVec{z}_k^{(q)}
&= \Input  \!- \!\sum\limits_{l \neq k} \myVec{h}_l e_l^{(q-1)}  . 
\end{align} 
%\end{subequations}
Substituting the channel output $\Input$ into \eqref{eqn:Cancelled1}, the realization of the interference canceled $\myVec{z}_k^{(q)} $  is obtained. 

To implement soft decoding, it is assumed that $\myVec{z}_k^{(q)} =\myVec{h}_k s_k  + \tilde{\myVec{w}}_k^{(q)}$, where the interference plus noise term $\tilde{\myVec{w}}_k^{(q)}$ obeys a zero-mean Gaussian distribution, independent of $s_k $, with covariance 
%	\begin{equation}
$\CovMat{k}^{(q)} = \SigW \myI_{\Nusers} + \sum_{l \neq k}v_l^{(q-1)}  \myVec{h}_l \myVec{h}_l^T$, where $\SigW$ is the noise variance. 
%\end{equation}
Combining this assumption with \eqref{eqn:Cancelled1}, while writing the set of possible symbols as $\mySet{S}=\{\alpha_j\}_{j=1}^{|\mySet{S}|}$, the conditional distribution of $\myVec{z}_k^{(q)}$ given $s_k  = \alpha_j$ is multivariate Gaussian with mean   $\myVec{h}_k \alpha_j$ and covariance $ \CovMat{k}^{(q)}$. The conditional \ac{pmf} of $s_k $ given $\Input $ is  approximated from the conditional distribution of $\myVec{z}_k^{(q)}$ given $s_k $ via Bayes theorem, assuming that the marginal \ac{pmf} of each  $s_k $ is uniform over $\mySet{S}$,  this estimated conditional distribution is computed as 
\begin{align*}
&\left( \hat{\myVec{p}}_k^{(q)}\right)_j %
%	&= \frac{\Pdf{ \myVec{Z}_k^{(q)} | S_k}( \myVec{z}_k^{(q)} | \alpha_j)}{\sum\limits_{j'\in\mySet{S} }\Pdf{ \myVec{Z}_k^{(q)} | S_k}( \myVec{z}_k^{(q)} | \alpha_{j'})} \notag \\
= \notag \\
&\frac{\exp \left\{-\frac{1}{2} \left( \myVec{z}_k^{(q)} - \myVec{h}_k\alpha_j \right)^T\left(\CovMat{k}^{(q)}\right) ^{-1} \left( \myVec{z}_k^{(q)} - \myVec{h}_k\alpha_j \right)   \right\} } { \sum\limits_{\alpha_{j'}\in\mySet{S} }\exp \left\{-\frac{1}{2} \left( \myVec{z}_k^{(q)} - \myVec{h}_k\alpha_{j'} \right)^T\left(\CovMat{k}^{(q)}\right) ^{-1} \left( \myVec{z}_k^{(q)} - \myVec{h}_k\alpha_{j'} \right)   \right\} }.
%\label{eqn:CondDist2} 
\end{align*}
After the final iteration, the symbols are decoded by maximizing the estimated \acp{pmf} for each $k \in \mySet{K}$, i.e., via 
\begin{equation}
    \hat{s}_k  = \alpha_{\hat{j}}, \qquad \hat{j} =  \mathop{\arg \max}\limits_{j }\left( \hat{\myVec{p}}_k^{(\Niter)}\right)_j,
    \label{eqn:DecisionDeepSIC}
\end{equation}
and the overall estimate is set to $\hat{\myVec{s}} = [\hat{s}_1,\ldots,\hat{s}_K]$.

\subsection{Detailed Formulation of Sparsity-Based \ac{cs} (Example 4, Section~\ref{sec:Inference})}\label{app:CS_Exm}
    Consider the case where
	$\csSignal^*$ is sparse in some dictionary $\myMat{B}$, e.g., in the wavelet domain, such that $\csSignal^* = \myMat{B} \myVec{c}^*$ where $\|\myVec{c}^*\|_0 = l$ with $l \ll N$.  In this case, the goal is to find the sparsest $\myVec{c}$ such that $\csSignal= \myMat{B} \myVec{c}$ agrees with the noisy observations:
	\begin{align*}
	    &\text{minimize } \| \myVec{c} \|_{0} \\
	    &\text{subject to } \| \csMatrix\myMat{B} \myVec{c} - \csObs \|_2 \le {\epsilon},
	\end{align*}
	where ${\epsilon}$ is a noise threshold. Since one can define $\tilde{\csMatrix} := \csMatrix\myMat{B}$, we henceforth  focus on the setting where $\myMat{B}$ is the identity matrix, and the optimization variable of the above $\ell_0$ norm optimization problem is $\csSignal$. 

	Although the above problem is NP-hard, \cite{candes2006robust,donoho2006compressed} showed that it suffices to minimize the  $\ell_1$ relaxed LASSO objective in \eqref{eqn:Lasso}.
	The formulation \eqref{eqn:Lasso} is convex, and for Gaussian $\myMat{A}$ with $l = \|\csSignal^*\|_{0}$ and $M = \Theta(l \log \frac{N}{l})$, the unique minimizer of $\csLasso$ is equal to $\csSignal^*$ with high probability.

\subsection{Detailed Formulation of \ac{admm} (Example 5, Section~\ref{sec:Inference})}\label{app:ADMM_Exm}
	\ac{admm} tackles the optimization problem in \eqref{eqn:RegOpt} by utilizing variable splitting. Namely, it introduces an additional auxiliary variable $\myVec{v}$ in order to decouple the regularizer $\phi(\myVec{s})$ from the likelihood term $\|\myVec{x}-\csMatrix \myVec{s}\|^2$. The resulting formulation of \eqref{eqn:RegOpt} is expressed as
	\begin{align}
	\label{eqn:admm1}
	    \hat{\myVec{s}} &= \mathop{\arg\min}\limits_{\myVec{s}}\mathop{\min}\limits_{\myVec{v}} \frac{1}{2}\|\myVec{x}-\csMatrix \myVec{s}\|^2 +   \phi(\myVec{v}),  \\
	    &\text{subject to } \myVec{v}=\myVec{s}.
	\end{align}
	The problem \eqref{eqn:admm1} is then solved by formulating the augmented Lagrangian (which introduces an additional optimization variable $\myVec{u}$) and solving it in an alternating fashion. This results in the following update equations for the $q$th iteration \cite{ryu2019plug}
	\begin{subequations}
	\label{eqn:admm2}
	\begin{align}
	\label{eqn:admm2a}
	    \hat{\myVec{s}}_{q+1} &=\mathop{\arg\min}\limits_{\myVec{s}}\frac{\alpha}{2}\|\myVec{x}\!-\!\csMatrix \myVec{s}\|^2 \! +\!\frac{1}{2}\| \myVec{s} \!-\! (\myVec{v}_q \!-\! \myVec{u}_q)\|^2, \\
	    \label{eqn:admm2b}
	    \myVec{v}_{q+1} &=\mathop{\arg\min}\limits_{\myVec{v}}\alpha  \phi(\myVec{v})    + \frac{1}{2}\| \myVec{v} - (\hat{\myVec{s}}_{q+1} + \myVec{u}_q)\|^2, \\
	    \label{eqn:admm2c}
	     \myVec{u}_{q+1} &= \myVec{u}_{q} +(\hat{\myVec{s}}_{q+1} - \myVec{v}_{q+1}).
	\end{align}
	\end{subequations}
	Here, $\alpha>0$ is an optimization hyperparameter. Steps \eqref{eqn:admm2a} and \eqref{eqn:admm2b} are the proximal mappings with respect to the functions $\alpha \phi(\cdot)$ and $\alpha f(\cdot)$, respectively, with $f(\myVec{v}) \triangleq \frac{1}{2}\|\myVec{x}-\csMatrix \myVec{v}\|^2 $. Step \eqref{eqn:admm2c} represents a gradient ascent iteration.
	
	For brevity, in Algorithm~\ref{alg:Algoadmm} we  write \eqref{eqn:admm2a} as $ \hat{\myVec{s}}_{q+1} = {\rm Prox}_{\alpha f}(\myVec{v}_q - \myVec{u}_q)$ and \eqref{eqn:admm2b} as  $ \myVec{v}_{q+1} = {\rm Prox}_{\alpha \phi}(\hat{\myVec{s}}_{q+1} + \myVec{u}_q)$ in Algorithm~\ref{alg:Algoadmm}. In particular, it is noted that \eqref{eqn:admm2a} equals $
    \myVec{s}_{q+1} = (\alpha \csMatrix^T\csMatrix  +   \myMat{I})^{-1}(\alpha \csMatrix^T\myVec{x} + (\myVec{v}_q - \myVec{u}_q))$.

\subsection{Detailed Formulation of Sum-Product Method (Example 6, Section~\ref{sec:Inference})}\label{app:SP_Exm}
To formulate the \ac{sp} method,  the factorizable distribution \eqref{eqn:MarkovModel} is first represented as a factor graph. To that aim, we recall the definitions of the vector variable  $\myVec{s}_i = \myVec{s}_{i-\Mem+1}^i \in \mySet{S}^\Mem$, and the function $f\left(x_i, \myVec{s}_i, \myVec{s}_{i-1} \right)$ in \eqref{eqn:FSC_funcNode}. 
When  $ \myVec{s}_i$ is a shifted version of   $\myVec{s}_{i-1}$,  \eqref{eqn:FSC_funcNode} coincides with  $\PdfNew{Y_i | \myVec{S}_{i-\Mem}^{i}}\left(x_i|\myVec{s}_{i-\Mem}^{i}\right) \PdfNew{S_i |  \myVec{S}_{i-\Mem}^{i-1}}\left(s_i | \myVec{s}_{i-\Mem}^{i-1}\right)$,  and  equals zero otherwise.
Using \eqref{eqn:FSC_funcNode}, the joint distribution $\PdfNew{\myVec{X},\myVec{S}}(\Input,\myS)$ in \eqref{eqn:MarkovModel}  can be written as
\begin{align}
\PdfNew{\Input , \myS}\left(\Input , \myVec{s} \right)  
&=   \prod\limits_{i\!=\!1}^{\Blklen} f\left(x_i, \myVec{s}_i, \myVec{s}_{i-1} \right).
\label{eqn:ChModel2}
\end{align}
The factorizable expression of the joint distribution \eqref{eqn:ChModel2} implies that it can be represented as a factor graph with $\Blklen$ function nodes $\{	f\left(x_i, \myStateR_i, \myStateR_{i-1} \right) \}$, in which $\{\myStateR_i\}_{i=2}^{\Blklen-1}$ are edges while the remaining variables are half-edges. % as illustrated in Fig. \ref{fig:FGFinite}.

%	
%	\begin{figure}
%		\centering
%		{\includefig{fig/FG_state3.png}} 
%		\caption{Factor graph of finite-memory channels.}
%		\label{fig:FGFinite}	 
%	\end{figure}

%
Using its factor graph representation, one can compute the joint distribution of $\myVec{s}$ and $\Input$  by recursive message passing along its factor graph as  illustrated in Fig.~\ref{fig:SumProduct2}(a). 
In particular, 
\begin{align}
\!\!\PdfNew{\myState_k,\myState_{k\!+\!1}, \myVec{X}}(\myStateR_k,\myStateR_{k\!+\!1}, \Input)\! &=\! \FwdMsg{f_k}{\myState_k}(\myStateR_k) f(x_{k\!+\!1},   \myStateR_{k\!+\!1}, \myStateR_k) 
% \notag \\
%&\times 
\BwdMsg{f_{k+2}}{\myState_{k+1}}(\myStateR_{k\!+\!1})
\label{eqn:Recursion1}
\end{align}
where the forward path messages satisfy 
\begin{equation}
\FwdMsg{f_i}{\myState_i}(\myStateR_i) = \sum_{\myStateR_{i-1}} f(x_{i},  \myStateR_{i}, \myStateR_{i-1})\FwdMsg{f_{i-1}}{\myState_{i-1}}(\myStateR_{i-1})
\label{eqn:Recursion1Forwards}
\end{equation}
for $i = 1, 2,\ldots, k$. Similarly, the backward messages are 
\begin{equation}
\BwdMsg{f_{i\!+\!1}}{ \myState_i}(\myStateR_i) = \sum_{\myStateR_{i\!+\!1}} f(x_{i\!+\!1}, \myStateR_{i\!+\!1}, \myStateR_{i})\BwdMsg{f_{i\!+\!2} }{ \myState_{i\!+\!1}}(\myStateR_{i\!+\!1})
\label{eqn:Recursion1Backwards}
\end{equation} 	
for $i = \Blklen-1, \Blklen -2, \ldots, k+1$.

The ability to compute the joint distribution in \eqref{eqn:Recursion1} via message passing allows to obtain the \ac{map} detector in \eqref{eqn:MAP0} with complexity that only grows linearly with $\Blklen$. This is achieved by noting that the \ac{map} estimate satisfies 
\begin{align}
\hat{\myVec{s}}_i\left( \Input\right)  
%\mathop{\arg \max}\limits_{s \in \mySet{S}} \Pdf{S_i,\myVec{X}^\Blklen}(s,\Input^\Blklen) 
%	\notag \\
%	&=
% &=	\mathop{\arg \max}\limits_{s_i \in \mySet{S}} \sum_{\{s_{i\! - \!\Mem}, \ldots, s_{i\! - \!1} \}\in \mySet{S}^{\Mem}} 	P_{\myVec{S}_{i\! - \!1},\myVec{S}_{i}, \myVec{X}^{\Blklen}}([s_{i\! - \!\Mem}, \ldots, s_{i\! - \!1} ], [s_{i\! - \!\Mem+1}, \ldots, s_{i} ], \Input^{\Blklen}) \notag \\
\!=\!\mathop{\arg \max}\limits_{s_i \in \mySet{S}} \sum_{  \myVec{s}_{i\! - \!1}\in \mySet{S}^{\Mem}}& \FwdMsg{f_{i\! - \!1}}{\myVec{s}_{i\! - \!1}}(\myVec{s}_{i\! - \!1}) f(x_{i},  [s_{i\! - \!\Mem\!+\!1}, \ldots, s_{i} ],\myVec{s}_{i\! - \!1})  \notag \\
&\times \BwdMsg{f_{i+1}}{\myVec{s}_{i}}([s_{i\! - \!\Mem\!+\!1}, \ldots, s_{i} ])
\label{eqn:MAP2}
\end{align}
for each $i \in \Blkset$, where the summands can be computed recursively, resultin in Algorithm~\ref{alg:SP}. It is noted that when the block size $\Blklen$ is large, the messages may tend to zero, and are thus commonly scaled \cite{loeliger2004introduction}, e.g., $\BwdMsg{f_{i+1}}{\myState_i}(\myStateR)$ is replaced with $\gamma_i \BwdMsg{f_{i+1}}{\myState_i}(\myStateR)$ for some scale factor  which does not depend on $\myStateR$, and thus does not affect the \ac{map} rule.

\subsection{Detailed Formulation of Iterative Kalman Smoother (Example 7, Section~\ref{sec:Inference})}\label{app:Kalman_Exm}
	The state-space model \eqref{eqn:KalmanSS} implies that the joint distribution of the state and observations satisfies
	\begin{align}
	     p\left(\Input, \myVec{s}\right) &= p\left(\Input| \myVec{s}\right)p\left(\myVec{s}\right) = \prod_t p(\Input_t | \Label_t) p(\Label_t | \Label_{t-1}).
	\end{align}
	Consequently, it holds that 
	\begin{align}
	&\frac{\partial}{\partial \Label_t} \log p\left(\Input, \myVec{s}\right) 
	= \frac{\partial}{\partial \Label_t} \sum_\tau \log p(\Input_\tau | \Label_\tau) +\sum_\tau \log p(\Label_\tau | \Label_{\tau-1}) \notag \\
	&= \frac{\partial}{\partial \Label_t} \log p(\Input_t | \Label_t) + \frac{\partial}{\partial \Label_t}  \log p(\Label_t | \Label_{t-1})  + \frac{\partial}{\partial \Label_t} \log  p(\Label_{t+1} | \Label_{t}) \notag \\
	&= \frac{\partial}{\partial \Label_t} (\Input_t - \myMat{H}\Label_t)^T\myMat{R}^{-1} (\Input_t - \myMat{H}\Label_t) \notag \\
	&\qquad+ \frac{\partial}{\partial \Label_t} (\Label_t - \myMat{F}\Label_{t-1})^T\myMat{W}^{-1}(\Label_t - \myMat{F}\Label_{t-1}) \notag \\
	&\qquad
	+ \frac{\partial}{\partial \Label_t} (\Label_{t+1} - \myMat{F}\Label_{t})^T\myMat{W}^{-1}(\Label_{t+1} - \myMat{F}\Label_{t}) \notag \\
	&= \myMat{H}^T\myMat{R}^{-1}\left( \Input_{t}- \myMat{H} \myS_{t} \right) + -\myMat{W}^{-1}\left( \myS_t - \myMat{F} \myS_{t-1}  \right) \notag \\ 
	&\qquad + \myMat{F}^T\myMat{W}^{-1}\left( \myS_{t+1} - \myMat{F} \myS_{t}  \right).
	\end{align}
		Therefore, the $t$th entry of the log likelihood gradient in \eqref{eqn:GradKal}, abbreviated henceforth as $\nabla^{(q)}_t$, can be obtained as 
	$\nabla^{(q)}_t = \mu^{(q)}_{\myVec{S}_{t-1}\rightarrow \myVec{S}_t} +  \mu^{(q)}_{\myVec{S}_{t+1}\rightarrow \myVec{S}_t} + \mu^{(q)}_{\myVec{X}_{t}\rightarrow \myVec{S}_t}$, where the summands, referred to as messages, are given by
	\begin{subequations}
	\label{eqn:KalSmoothQuant1}
	\begin{align}
	  \mu^{(q)}_{\myVec{S}_{t-1}\rightarrow \myVec{S}_t} & =  -\myMat{W}^{-1}\left( \myS_t^{(q)} - \myMat{F} \myS_{t-1}^{(q)}  \right),  \\
	  \mu^{(q)}_{\myVec{S}_{t+1}\rightarrow \myVec{S}_t} & =  \myMat{F}^T\myMat{W}^{-1}\left( \myS_{t+1}^{(q)} - \myMat{F} \myS_{t}^{(q)}  \right),  \\
	  \mu^{(q)}_{\myVec{X}_{t}\rightarrow \myVec{S}_t} & =  \myMat{H}^T\myMat{R}^{-1}\left( \Input_{t}- \myMat{H} \myS_{t}^{(q)}  \right).
	\end{align}
	\end{subequations}
	The iterative procedure in \eqref{eqn:GradKal}, is repeated until convergence, as stated in Algorithm~\ref{alg:Smoothing} and the resulting $\myVec{s}^{(q)}$ is used as the estimate.

% 	The state-space model \eqref{eqn:KalmanSS} implies that the $i$th entry of the log likelihood gradient in \eqref{eqn:GradKal}, abbreviated henceforth as $\nabla^{(q)}_i$, can be obtained as 
% 	$\nabla^{(q)}_i = \mu^{(q)}_{\myVec{S}_{i-1}\rightarrow \myVec{S}_i} +  \mu^{(q)}_{\myVec{S}_{i+1}\rightarrow \myVec{S}_i} + \mu^{(q)}_{\myVec{X}_{i}\rightarrow \myVec{S}_i}$, where the summands, referred to as messages, are given by
% 	\begin{subequations}
% 	\label{eqn:KalSmoothQuant1}
% 	\begin{align}
% 	  \mu^{(q)}_{\myVec{S}_{i-1}\rightarrow \myVec{S}_i} & =  -\myMat{W}^{-1}\left( \myS_i^{(q)} - \myMat{F} \myS_{i-1}^{(q)}  \right),  \\
% 	  \mu^{(q)}_{\myVec{S}_{i+1}\rightarrow \myVec{S}_i} & =  \myMat{F}^T\myMat{W}^{-1}\left( \myS_{i+1}^{(q)} - \myMat{F} \myS_{i}^{(q)}  \right),  \\
% 	  \mu^{(q)}_{\myVec{X}_{i}\rightarrow \myVec{S}_i} & =  \myMat{H}^T\myMat{R}^{-1}\left( \Input_{i}- \myMat{H} \myS_{i}^{(q)}  \right).
% 	\end{align}
% 	\end{subequations}
% 	The iterative procedure in \eqref{eqn:GradKal}, is repeated until convergence, and the resulting $\myVec{s}^{(q)}$ is used as the estimate. 

\end{appendix}
\color{black}

	%----------------------------------------------------------------------------------------
	%	BIBLIOGRAPHY
	%----------------------------------------------------------------------------------------
	\bibliographystyle{IEEEtran}
	\bibliography{IEEEabrv,refs}

\end{document}